\documentclass[]{article}

\title{The Scattering Map on Oppenheimer--Snyder Space-time}
\author{Fred Alford}
\date{Department of Applied Mathematics and Theoretical Physics,\\
	University of Cambridge\\[2ex]
	This work was funded by EPSRC DTP\\
	$[1936235]$\\
	[2ex]\today}
\newcommand{\p}{\partial}
\newcommand{\R}{\mathbb{R}}
\usepackage[british]{babel}
\usepackage{amsmath,amssymb,amsthm}
\usepackage{geometry}
\usepackage{slashed}
\usepackage{array}
\usepackage{tikz}
\usepackage{wrapfig}
\usepackage{graphicx}
\usepackage{float}
\usepackage{cutwin}
\usepackage{bbm}
\usepackage{textcomp}
\usetikzlibrary{decorations.pathmorphing}
\newtheorem{Lemma}{Lemma}[section]

\newtheorem{Theorem}{Theorem}[section]
\newtheorem{thm}{Theorem}
\newtheorem*{Theorem*}{Theorem}
\newtheorem{Corollary}{Corollary}[section]

\newtheorem{Proposition}{Proposition}[section]

\newtheorem{Remark}{Remark}[section]
\usetikzlibrary{decorations.pathmorphing}
\tikzset{snake it/.style={decorate, decoration=snake}}
\usepackage{xpatch}
\makeatletter
\xpatchcmd{\@thm}{\thm@headpunct{.}}{\thm@headpunct{}}{}{}
\makeatother
\DeclareMathOperator{\Ima}{Im}
\begin{document}

\maketitle

\begin{abstract}
In this paper we analyse the boundedness of solutions $\phi$ of the wave equation in the Oppenheimer--Snyder model of gravitational collapse in both the case of a reflective dust cloud and a permeating dust cloud. We then proceed to define the scattering map on this space-time and look at the implications of our boundedness results on this scattering map. 

Specifically, it is shown that the energy of $\phi$ remains uniformly bounded going forwards in time and going backwards in time for both the reflective and the permeating cases. It is then shown that the scattering map is bounded going forwards, but not backwards. Therefore the scattering map is not surjective onto the space of finite energy on $\mathcal{I}^+\cup\mathcal{H}^+$. Thus there does not exist a backwards scattering map from finite energy radiation fields on $\mathcal{I}^+\cup\mathcal{H}^+$ to finite energy radiation fields on $\mathcal{I}^-$. We will then contrast this with the situation for scattering in pure Schwarzschild.
\end{abstract}

\section{Overview}

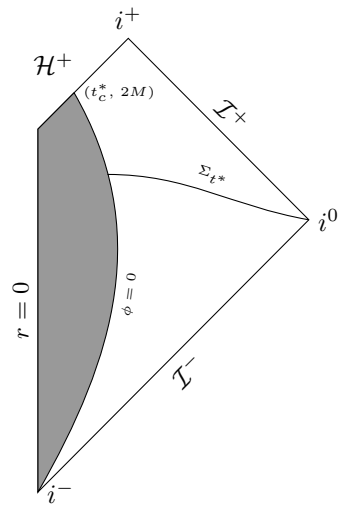
\begin{wrapfigure}{r}{5cm}
	\vspace{-10mm}
	\begin{tikzpicture}[scale =1.2]
	\node (I)    at ( 0,0) {};
	
	\path 
	(I) +(90:2)  coordinate[label=90:$i^+$]  (Itop)
	+(-90:2) coordinate (Imid)
	+(0:2)   coordinate[label=0:$i^0$] (Iright)
	+(-1,1) coordinate (Ileft)
	+(-0.6,1.4) coordinate[label=0:\tiny ($t^*_c$\text{, }$2M$)] (BHH)
	+(-1,-3) coordinate[label=0:$i^-$] (Ibot)
	;
	\draw (Ileft) -- 
	node[midway, above left]    {$\mathcal{H}^+$}
	(Itop) --
	node[midway, above, sloped] {$\mathcal{I}^+$}
	(Iright) -- 
	node[midway, below, sloped] {$\mathcal{I}^-$}
	(Ibot) --
	node[midway, above, sloped]    {\small $r=0$}    
	(Ileft) -- cycle;
	\draw[fill=gray!80] (Ibot) to[out=60, in=-60]
	node[midway, below, sloped] {\tiny $\phi=0$} (BHH)--(Ileft)--cycle;
	\draw (Iright) to[out=170, in=0] node[midway, above, sloped] {\tiny $\Sigma_{t^*}$} (-0.23,0.5);
	\end{tikzpicture}
	\caption{Penrose diagram of Oppenheimer--Snyder space-time with reflective boundary conditions, with spacelike hypersurface $\Sigma_{t^*}$.}\label{fig:PenRef}
\end{wrapfigure}

In this paper we will be studying energy boundedness of solutions to the linear wave equation
\begin{equation}\label{eq:wave}
\Box_g \phi=\frac{1}{\sqrt{-g}}\p_a(\sqrt{-g}g^{ab}\p_b\phi)=0
\end{equation}
on Oppenheimer--Snyder space-time $(\mathcal{M},g)$ \cite{S-O}. This is one of the simplest models of gravitational collapse.
We will further be considering two different sets of boundary conditions:~\emph{reflective}, where we will impose the condition $\phi=0$ on the surface of the star (in a trace sense), and \emph{permeating}, where we will be solving the linear wave equation throughout the whole space-time, including the interior of the star. We will then be using these results to define a scattering theory for this space-time. 

\begin{wrapfigure}{r}{5cm}
	\vspace{-10mm}
	\begin{tikzpicture}[scale =1.2] 
	\node (I)    at ( 0,0) {};
	
	\path 
	(I) +(90:2)  coordinate[label=90:$i^+$]  (Itop)
	+(-90:2) coordinate (Imid)
	+(0:2)   coordinate[label=0:$i^0$] (Iright)
	+(-1,1) coordinate (Ileft)
	+(-0.6,1.4) coordinate[label=0:\tiny ($\tau_c$\text{, }$2M$)] (BHH)
	+(-1,-3) coordinate[label=0:$i^-$] (Ibot)
	;
	\draw (Ileft) -- 
	node[midway, above left]    {$\mathcal{H}^+$}
	(Itop) --
	node[midway, above, sloped] {$\mathcal{I}^+$}
	(Iright) -- 
	node[midway, below, sloped] {$\mathcal{I}^-$}
	(Ibot) --
	node[midway, below, sloped]    {\small $r=0$}    
	(Ileft) -- cycle;
	\draw[dotted] (Ibot) to[out=60, in=-60]
	node[midway, above, sloped] {\tiny $r=r_b(\tau)=R^*(t^*)$} (BHH);
	\draw (Iright) to[out=170, in=0] node[midway, above, sloped] {\tiny $\Sigma_{\tau}$} (-1,0.5);
	\end{tikzpicture}
	\caption{Penrose Diagram of Oppenheimer--Snyder space-time with permeating boundary conditions, with spacelike hypersurface $\Sigma_{\tau}$.}\label{fig:PenPerm}
	
\end{wrapfigure}
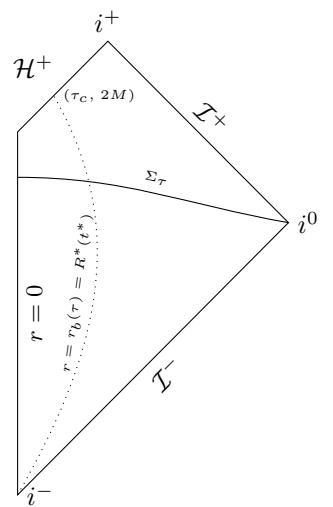

The first main theorem dealing with solutions of \eqref{eq:wave} in the bulk of the space-time is informally stated below:

\begin{thm}[Non-degenerate Energy ($N$-energy) boundedness]\label{Thm:VagueBound}
	In Oppenheimer--Snyder space-time, let the map $\mathcal{F}_{(t^*_0,t^*_1)}$ take the solution of \eqref{eq:wave} on a time slice $\Sigma_{t^*_0}$ (or $\Sigma_{\tau_0}$), forward to the same solution on a later time slice, $\Sigma_{t^*_1}\cup(\mathcal{H}^+\cap\{t^*\in [t^*_0,t^*_1]\})$ (or $\Sigma_{\tau_1}\cup(\mathcal{H}^+\cap\{\tau\in[\tau_0,\tau_1]\})$). Then $\mathcal{F}_{(t^*_0,t^*_1)}$ is uniformly bounded in time with respect to the non-degenerate energy, in both the reflective and permeating cases. Furthermore, for $t^*_1\leq t^*_c$ (or $\tau_1\leq\tau_c$), its inverse is also bounded with respect to this non-degenerate energy.
	
	The contents of this theorem are stated more precisely across Theorems \ref{Thm:ForwardReflBound}, \ref{Thm:BackReflBound}, \ref{Thm:ForwardPermBound} and \ref{Thm:BackPermBound}.
\end{thm}

The sphere $(t^*_c,2M)$ and the time slice $\Sigma_{t^*}$ (for $t^*<t^*_c$) are shown in Figure \ref{fig:PenRef}. The sphere $(\tau_c,2M)$ and the time slice $\Sigma_{\tau}$ ($\tau<\tau_c$) are shown in Figure \ref{fig:PenPerm}.

Non-degenerate energy means the energy with respect to an everywhere timelike vector field (including on the horizon $\mathcal{H}^+$) which coincides with the timelike Killing vector in a neighbourhood of null infinity $\mathcal{I}^\pm$. This energy controls the $L^2$ norm of each $1^{st}$ derivative of the field, $\phi$. 

In the reflective case we also go on to show forwards and backwards boundedness of higher order derivatives, see Theorem \ref{Thm:NForwardBound} and \ref{Thm:NBackwardBound}. In the permeating case we go on to show forwards and backwards boundedness of $2^{nd}$ order derivatives, see Theorem \ref{Thm:2nd order energy}.

We then consider the limiting process to look at the radiation field on past null infinity $\mathcal{I}^-$, and obtain the following result.

\begin{thm}[Existence and Non-degenerate Energy Boundedness of the Past Radiation Field]\label{Thm:VaguePastRad}
	In Oppenheimer--Snyder space-time, we define the map $\mathcal{F}^-$ as taking the solution of \eqref{eq:wave} on $\Sigma_{t^*_0}$, $t^*\leq t^*_c$ (or $\Sigma_{\tau_0}\cup(\mathcal{H}^+\cap\{\tau\leq\tau_0\})$) to the radiation field on $\mathcal{I}^-$. $\mathcal{F}^-$ is well-defined and bounded with respect to the non-degenerate energy, for both reflective and permeating boundary conditions.

	This theorem is stated more precisely as Theorem \ref{Thm:RadBound}.
\end{thm}

On Schwarzschild, we know that the future radiation field exists, so the map $\mathcal{G}^+$ from data on $\Sigma_{t^*}$ to $\mathcal{I}^+\cup\mathcal{H}^+$ exists (see \cite{Moschidis} for example). It is also bounded in terms of the $N$-energy, \cite{redshift}. It is, however, unbounded, going backwards, in terms of the $N$-energy (see for example \cite{Blue}). This is stated more precisely as Proposition \ref{prop:SwarzBound}. This result immediately applies to Oppenheimer--Snyder space-time. Together with Theorem \ref{Thm:VaguePastRad} and a new result about decay towards the past on asymptotically null foliations (see Proposition \ref{prop:decay}), this allows us to define the inverse of $\mathcal{F}^-$, $\mathcal{F}^+$ (see Theorem \ref{Thm:RadBackBound}). This combination also gives us the final theorem:

\begin{thm}[Boundedness but non-surjectivity of the scattering map]\label{Thm:VagueScat}
	We define the scattering map, 
	\begin{align}\label{eq:scatmap}
	\mathcal{S}^+&:\mathcal{E}^{\p_{t^*}}_{\mathcal{I}^-}\to\mathcal{E}^{\p_{t^*}}_{\mathcal{I}^+}\times \mathcal{E}^N_{\mathcal{H}^+}\\\nonumber
	\mathcal{S}^+&:=\mathcal{G}^+\circ\mathcal{F}^+
	\end{align}
	on Oppenheimer--Snyder space-time from data on $\mathcal{I}^-$ to data on $\mathcal{I}^+\cup\mathcal{H}^+$. $\mathcal{S}^+$ is injective and bounded going forwards, with respect to the non-degenerate energy ($L^2$ norms of $\p_v(r\phi)$ on $\mathcal{I}^-$ and $\mathcal{H}^+$ and $\p_u(r\phi)$ on $\mathcal{I}^+$). One can then define the inverse, $\mathcal{S}^-$, of \eqref{eq:scatmap}, going backwards from $\mathcal{S}^+(\mathcal{E}^{\p_{t^*}}_{\mathcal{I}^-})$, in either the reflective or permeating case. However, $\mathcal{S}^-$ is not bounded with respect to the non-degenerate energy. It follows that $\mathcal{S}^+$ is not surjective. Moreover, $\mathcal{E}^{\p_{t^*}}_{\mathcal{I}^+}\times \{0\}_{\mathcal{H}^+}$ is not a subset of $\Ima(\mathcal{S}^+)$.
	
	This Theorem is stated more precisely as Theorem \ref{Thm:O-SScat}.
\end{thm}

In proving Proposition \ref{prop:decay}, we obtain a result on the rate at which our solution decays (towards $i^-$, with respect to this asymptotically null  foliation) for data decaying sufficiently quickly towards spatial infinity. However we do not look at optimising this rate, as only very weak decay is required for Theorem \ref{Thm:VagueScat}.

The non-invertibility of $\mathcal{S}^+$ is inherited from that of $\mathcal{G}^+$. This ultimately arises from the red-shift effect along $\mathcal{H}^+$, which for backwards time evolution corresponds to a blue-shift instability. It is the existence of the map $\mathcal{F}^+$ mapping into the space of non-degenerate energy however, that extends this non-invertibility to data on $\mathcal{I}^-$. Note that for $\mathcal{I}^-$, the notion of energy is completely canonical. This is in contrast to the pure Schwarzschild case, where no such $\mathcal{F}^+$ exists.

It remains an open problem to precisely characterise the image of the scattering map $\mathcal{S}^+$.

\subsection{Acknowledgements}
We would like to thank Mihalis Dafermos for many insightful discussions. We would also like
to thank Christoph Kehle for reading through the manuscript, and Owain Salter Fitz-Gibbon for many insightful comments.

\section{Previous Work}
\begin{wrapfigure}{r}{6cm}
	\begin{tikzpicture}[scale =1.2]
	\node (I)    at ( 0,0) {};
	
	\path 
	(I) +(90:2)  coordinate[label=90:$i^+$]  (Itop)
	+(-90:2) coordinate (Imid)
	+(0:2)   coordinate[label=0:$i^0$] (Iright)
	+(135:1.414213613) coordinate (Ileft)
	+(120:1.464101615) coordinate[label=180:\small $t^*_c$\text{, }$2M$] (BHH)
	+(-108.434948823:3.16227766) coordinate[label=0:$i^-$] (Ibot)
	;
	\draw (Ileft) -- 
	node[midway, above left]    {$\mathcal{H}^+$}
	(Itop) --
	node[midway, above, sloped] {$\mathcal{I}^+$}
	(Iright) -- 
	node[midway, below, sloped] {$\mathcal{I}^-$}
	(Ibot) --
	node[midway, below, sloped]    {\small $r=0$}    
	(Ileft) -- cycle;
	\draw (Ibot) to[out=60, in=-60]
	node[midway, above, sloped] {\tiny $r=R^*(t^*)$} (BHH);
	\draw[-latex] (0.5,-1.5) to[out=135, in=-90] (0,0);
	\draw[-latex] (0,0) to[out=90, in=-135] (1,1);
	\draw[-latex] (0,0) to[out=90, in=-45] (-0.5,1.5);
	\end{tikzpicture}
	\caption{Penrose Diagram of Oppenheimer--Snyder space-time with scattering map}\label{fig:Pen}
\end{wrapfigure}
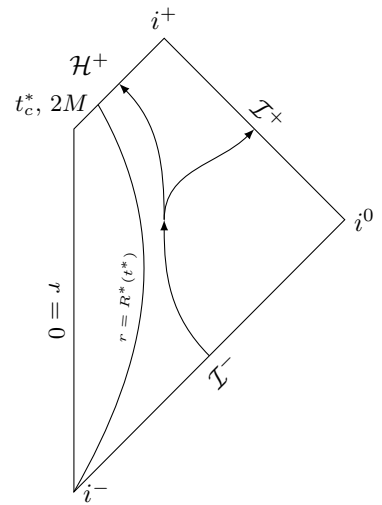
There has been a substantial amount of work done concerning the scattering map on Schwarzschild. However there has been considerably less concerning the scattering map for collapsing space-times such as Oppenheimer--Snyder. The exterior of the star is a vacuum spherically symmetric space-time and therefore has the Schwarzschild metric by Birkhoff's Theorem, \cite{ReallBH}. We will thus be using a couple of results in this region from previous papers. However, we will not be discussing the scattering map on Schwarzschild very much beyond this. For a more complete discussion of the wave equation on Schwarzschild, see \cite{BHL}.

Most previous works on scattering in gravitational collapse, such as \cite{Bachelot}, \cite{DiracScatter}, \cite{BachelotHawking}, \cite{Melnyk2004}, assume that the star/dust cloud is at a finite radius from infinite past up to a certain time and then proceed to let this cloud collapse. Thus these models are stationary in all but a compact region of space-time. This model allows these previous works avoid the difficulty of allowing the star to tend to infinite radius towards the past, as happens in the original Oppenheimer--Snyder model that we will be studying here. Also, dynamics on the interior of the star have not been examined, and so only the case of reflective boundary conditions has been studied previously. The energy current techniques we will be using here can, with relatively little difficulty, also be applied to these finite-radius models. These energy current methods are also more easily generalisable to other space-time models: for example, to obtain boundedness of the forward scattering map, all that is required to apply these techniques is that the star is collapsing. Nonetheless, in this paper we will specifically restrict to the Oppenheimer--Snyder model. 

In this paper, we look at defining the scattering map $\mathcal{S}^+$ geometrically as a map from data on $\mathcal{I}^-$ to data on $\mathcal{H}^+\cup\mathcal{I}^+$ (equation \eqref{eq:scatmap}). This is treating scattering in terms of the Friedlander radiation formalism (as in \cite{friedlander_1980}). In the above papers (\cite{Bachelot}, \cite{DiracScatter}, \cite{BachelotHawking}, \cite{Melnyk2004}), their solution is evolved a finite time, then evolved back to $t=0$ with respect to either Schwarzschild metric (for the horizon radiation field) or Minkowski metric (for the null infinity radiation field). Then the authors show that the limit as we let that time tend to infinity exists. All this is done using the language of wave operators. For a comparison of these two approaches to scattering theory, the reader may wish to refer to Section 4 of \cite{Conformal}.

Let us discuss two related works in more detail. The work \cite{Bachelot} studies the Klein--Gordon equation (\eqref{eq:wave} is the massless Klein--Gordon equation, thus is studied as a special case) on the finite-radius model discussed above. In this context, the author obtains what can be viewed as a partial result towards the analogue of Theorem \ref{Thm:VagueBound} for each individual spherical harmonic. However they do not find a bound independent of angular frequency.

Again in the finite-radius model, \cite{DiracScatter} studies the Dirac equation for spinors. However, as this has a $0^{th}$ order conserved current, this allows a Hilbert space to be defined such that the propagator through time is a unitary operator. Thus there is no need for the (first order) energy currents we will be using. This also allows questions of surjectivity to be answered with relative ease.

There have also been some papers discussing the Hawking effect, such as \cite{hawking1975} and \cite{BachelotHawking}, where a collapsing background is the set up for this. We hope that having a theory of the scattering map will be useful for applications in this direction.

\section{Oppenheimer--Snyder Space-time}
The Oppenheimer--Snyder space-time \cite{S-O} $(\mathcal{M},g)$ is that of a homogeneous spherically symmetric collapsing dust star. That is to say, a spherically symmetric solution of the Einstein equations:
\begin{equation}\label{eq:EE}
R_{\mu\nu}-\frac{1}{2}Rg_{\mu\nu}=8\pi\boldsymbol{T}_{\mu\nu}
\end{equation}
where for dust, we have 
\begin{equation}\label{eq:DustEMtensor}
\boldsymbol{T}_{\mu\nu}=\rho u_\mu u_\nu.
\end{equation}
Here the vector $u^\mu$ is the $4$-velocity of the dust, and $\rho$ is the density of the dust. On our initial timelike hypersurface, this density is a positive constant inside the star, but $0$ outside the star. The case of the non-homogeneous dust cloud was studied by Christodoulou in \cite{Christodoulou1984}.

As this density is not continuous across the boundary of the star, the Oppenheimer--Snyder model is only a global solution of the Einstein equations in a weak sense. However, it is a classical solution on both the interior and the exterior of the star.

We  therefore have two specific regions of the space-time to consider: inside the star (section \ref{sec:int}), and outside the star (section \ref{sec:ext}). We will finally give the definition of our manifold and global coordinates in section \ref{sec:GlobCoord}. If the reader is uninterested in the derivation of the metric, they may want to skip to that section. Finally in section \ref{sec:Pen}, we discuss the Penrose diagram for this space-time.

\subsection{Exterior}\label{sec:ext}

We will first consider the exterior of the star. This region is a spherically symmetric vacuum space-time, thus by Birkhoff's theorem, \cite{ReallBH}, this is a region of Schwarzschild space-time. It is bounded by the timelike hypersurface $r= R(t)=R^*(t^*)$, where $R(t)=R^*(t^*(t,R(t)))$. This hypersurface will be referred to in this paper as the boundary of the star. We will be using the following two coordinate systems in the exterior of the star:
\begin{equation}\label{eq:metric}
g=-\left(1-\frac{2M}{r}\right)dt^2+\left(1-\frac{2M}{r}\right)^{-1}dr^2+r^2g_{S^2} \qquad (t,r,\theta,\varphi)\in\R\times [R(t),\infty)\times S^2
\end{equation}
\begin{equation}\label{eq:metric*}
g=-\left(1-\frac{2M}{r}\right)dt^{*2}+\frac{4M}{r}dt^*dr+\left(1+\frac{2M}{r}\right)dr^2+r^2g_{S^2}\qquad (t^*,r,\theta,\varphi)\in\R\times [R^*(t^*),\infty)\times S^2
\end{equation}
where $g_{S^2}$ is the usual metric on the unit sphere, and $t^*$ is defined by
\begin{equation}
t^*=t+2M\log\left(\frac{r}{2M}-1\right).
\end{equation} 
Note, the first coordinate system \eqref{eq:metric} becomes degenerate on $r=2M$, so we will have to use the second \eqref{eq:metric*} when considering the horizon itself.

As the surface of the star is itself free-falling and massive, we may assume that the surface of the star follows timelike geodesics (and so is smooth). This assumption is true in the Oppenheimer--Snyder model, but also generalises to other models, provided the matter remains well behaved. Thus if a particle on the surface has space-time coordinates $x^\alpha(\tau)$, then these coordinates satisfy

\begin{align}\nonumber
-1=g_{ab}\bigg(\frac{dx}{d\tau}^a\bigg)\bigg(\frac{dx}{d\tau}^b\bigg)&=-\left(1-\frac{2M}{r}\right)\left(\frac{dt^*}{d\tau}\right)^2+\frac{4M}{r}\frac{dt^*}{d\tau}\frac{dr}{d\tau}+\left(1-\frac{2M}{r}\right)\left(\frac{dr}{d\tau}\right)^2\\\label{eq:timelike}
&=\left(-\left(1-\frac{2M}{r}\right)+\frac{4M}{r}{\dot{R}^*}+\left(1-\frac{2M}{r}\right)\dot{R}^{*2}\right)\left(\frac{dt^*}{d\tau}\right)^2.
\end{align}
Here we are using the $t$ and $r$ coordinates in equation \eqref{eq:metric*}, and using the fact that this space-time is spherically symmetric to ignore $\frac{d\theta}{d\tau}$ and $\frac{d\varphi}{d\tau}$ terms. Note that $\dot{R}^*=\frac{dR^*}{dt^*}$.

Now, as $R^*(t^*)$ is to be timelike and is the surface of a collapsing star, we assume $\dot{R}^*<0$, and that the surface emanates from past timelike infinity. Again, this is true in Oppenheimer--Snyder space-time, but also in many other models of gravitational collapse. At some time, $t^*_c$, we have $R^*(t^*_c)=2M$ (note that $R(t)$ does not cross $r=2M$ in $t$ coordinates, as $t$ becomes degenerate at the horizon). For $t^*>t^*_c$ and $r\geq 2M$, we have that the space-time is standard exterior Schwarzschild space-time, with event horizon at $r=2M$.

In the exterior region, we define our outgoing and ingoing null coordinates as follows:
\begin{align}\label{eq:extuv}
v& = t^*+r\\
u& = t^*-r-4M\log\left(\frac{r}{2M}-1\right)\\
g&=-\left(1-\frac{2M}{r}\right)dudv+r(u,v)^2g_{S^2}.
\end{align}

\subsection{Interior}\label{sec:int}
We now move on to considering the interior of the star. One thing that is important to note here is that as we go from considering the exterior of the star to considering the interior, i.e.~as our coordinates cross the boundary of our star, our metric changes from solving the vacuum Einstein equations to solving the Einstein equations with matter. Thus across the boundary, our metric will not be smooth, so we must be careful when wishing to take derivatives of the metric. This will have implications on the regularity of our solutions of \eqref{eq:wave} for the permeating case.

This derivation will closely follow the original Oppenheimer--Snyder paper, \cite{S-O}.

We first consider taking a spatial hypersurface in our space time, which is preserved under the spherical symmetry $SO_3$ action. We can therefore parametrise this by some $R$, $\theta$, $\varphi$, where $\theta$ and $\varphi$ are our spherical angles. Then we locally extend this coordinate system to the space-time off this surface by constructing the radial geodesics through each point with initial direction normal to the surface. In these coordinates, our metric must be of the form
\begin{equation}\label{eq:Rtau}
g=-d\tau^2+e^{\bar\omega}dR^2+e^\omega g_{S^2}.
\end{equation} 
for $\omega=\omega(\tau,R)$ and $\bar{\omega}=\bar{\omega}(\tau,R)$.

Now our matter is moving along lines of constant $R$, $\theta$ and $\phi$, so in these coordinates the dust's velocity $u^\mu$ is proportional to $\p_{\tau}$. Thus, we have from equation \eqref{eq:DustEMtensor} that $\boldsymbol{T}^\tau_\tau=-\rho$, for density $\rho$. We also have that all other components of the energy momentum tensor $\boldsymbol{T}$ vanish. Then the Einstein equations \eqref{eq:EE} imply that the following is a solution:
\begin{equation}
e^{\bar\omega}=\frac{1}{4}{\omega'}^2e^\omega
\end{equation}
\begin{equation}
e^{\omega}=(F\tau+G)^\frac{4}{3},
\end{equation}
where $'$ denotes derivative with respect to $R$, and $F$, $G$ are arbitrary functions of $R$. Then we can rescale $R$ to choose $G=R^\frac{3}{2}$. We now assume that at $\tau=0$, $\rho$ is a constant density $\rho_0$ inside the star, and vacuum outside the star, i.e.\begin{equation}
\rho(0,R)=\begin{cases} \rho_0 & R\leq R_b\\0 & R>R_b\end{cases},
\end{equation} for $R_b>0$ constant. Then the equation for $\boldsymbol{T}^\tau_\tau$ gives:
\begin{equation}
FF'=\begin{cases} 9\pi \rho_0 R^2 & R\leq R_b \\ 0 & R>R_b \end{cases}
\end{equation}
where, in these coordinates, $\{R=R_b\}$ is the boundary of the star. This has the particular solution
\begin{equation}
F=\begin{cases}
-\frac{3}{2}\sqrt{2M}\left(\frac{R}{R_b}\right)^\frac{3}{2} & R\leq R_b\\-\frac{3}{2}\sqrt{2M} &R>R_b
\end{cases},
\end{equation}
for $M=4\pi\rho_0 R_b^3/3$. This gives us a range for which our coordinate system is valid, as the angular part of the metric, $e^\omega$ has to be greater than or equal to $0$. Thus we obtain $\tau\leq \frac{2R}{3\sqrt{2M}}$. Now, if we transform to a new radial coordinate, $r=e^\frac{\omega}{2}$, then we obtain a metric of the form:
\begin{equation}\label{eq:metricint0}
g=\begin{cases}
-\left(1-\frac{2Mr^2}{r_b^3}\right)d\tau^2+2\sqrt{\frac{2Mr^2}{r_b^3}}drd\tau+dr^2+r^2g_{S^2} & r< r_b\\-\left(1-\frac{2M}{r}\right)d\tau^2+2\sqrt{\frac{2M}{r}}drd\tau+dr^2+r^2g_{S^2} & r\geq r_b
\end{cases}
\end{equation}
where
\begin{equation}\label{eq:rbdef}
r_b(\tau)^\frac{3}{2}=R_b^\frac{3}{2}-\frac{3\tau}{2}\sqrt{2M}.
\end{equation}

Once $r_b(\tau)\leq 2M$, i.e.~$\tau\geq\tau_c=\frac{4M}{3}\left(\left(\frac{R_b}{2M}\right)^\frac{3}{2}-1\right)$, we have $r=2M$ is the surface of an event horizon, and the $r\geq 2M$ section of our space-time is exterior Schwarzschild space-time.

Thus any point which can be connected by a future directed null geodesic to a point outside $r=2M$ at $\tau\geq\tau_c$ is outside our black hole, and any point which cannot reach $r>2M$ at $\tau\geq\tau_c$ is inside our black hole. The future directed, outgoing radial, null geodesic which passes through $r=2M$, $\tau=\tau_c$ is given by:
\begin{equation}\label{eq:H^+int}
r=r_b(\tau)\left(3-2\sqrt{\frac{r_b(\tau)}{2M}}\right).
\end{equation}
Thus the set of points obeying \eqref{eq:H^+int} intersect $\tau\in[\tau_{c^-},\tau_c]$ is then part of the boundary of our black hole for
\begin{equation}\label{eq:tauc-}
\tau_{c^-}=2M\left(\frac{2}{3}\left(\frac{R_b}{2M}\right)^\frac{3}{2}-\frac{9}{4}\right).
\end{equation} 
Before $\tau_{c^-}$, no part of the star is within a black hole, and for $\tau>\tau_c$, all of the collapsing star is inside the black hole region.

Thus for the permeating case, we define our ingoing and outgoing null geodesics defining their derivative:
\begin{align}\label{eq:unullint}
du=\begin{cases}
d\tau-(1-\sqrt{2M/r})^{-1}dr& r\geq r_b\\
\alpha(d\tau-(1-\sqrt{2Mr^2/r_b^3})^{-1}dr)& r<r_b
\end{cases}\\\label{eq:vnullint}
dv=\begin{cases}
d\tau+(1+\sqrt{2M/r})^{-1}dr& r\geq r_b\\
\beta(d\tau+(1+\sqrt{2Mr^2/r_b^3})^{-1}dr)& r<r_b
\end{cases}.
\end{align}
These coordinates exist, thanks to Frobenius' theorem (see for example \cite{ReallBH}) with $\alpha$ and $\beta$ bounded above and away from 0. However, we may not be able to write $\alpha$ and $\beta$ explicitly.

\begin{Remark}
	Note that when using different coordinates across the boundary of the star, $r=r_b(\tau)$, such as in \eqref{eq:unullint} and \eqref{eq:vnullint} compared to \eqref{eq:metricint0}, one should be concerned that these coordinates may define different smooth structures on $\mathcal{M}$. For example, the function $f(\tau,r,\theta,\varphi)=r-r_b(\tau)$ is smooth on $r=r_b(\tau)$ with respect to $(\tau,r,\theta,\varphi)$, but is not smooth with respect to coordinates $(\tau,x:=(r-r_b(\tau))^3,\theta,\varphi)$.
	
	However, when considering (in the exterior) the coordinates in \eqref{eq:metric*} compared to \eqref{eq:metricint0}, the change of coordinates is smooth with bounded (above and away from $0$)  Jacobian. Thus a function is smooth with respect to \eqref{eq:metric*} if and only if it is smooth with respect to \eqref{eq:metricint0}, so this is not a concern in this case.
\end{Remark}

\subsection{Global Coordinates and the Definition of Our Manifold}\label{sec:GlobCoord}
We summarise the work of the previous sections by defining our manifold and metric with respect to global coordinates. Fix $M>0,R_b\geq 0$, let $\tau_{c^-}=\sqrt{\frac{2R_b^3}{3M}}$, and consider $\R^4=\R\times\R^3$. Here $\R$ is parametrised by $\tau$ and $\R^3$ is parametrised by the usual spherical polar coordinates. We then define $\mathcal{M}$ by:
\begin{equation}\label{eq:ManifoldRange}
\mathcal{M}:=\R^4\backslash \{\tau\in[\tau_{c^-},\infty), r=0\}.
\end{equation}
In these coordinates, we then have the metric:
\begin{equation}\label{eq:metricint}
g_{M,R_b}=\begin{cases}
-\left(1-\frac{2Mr^2}{r_b^3}\right)d\tau^2+2\sqrt{\frac{2Mr^2}{r_b^3}}drd\tau+dr^2+r^2g_{S^2} & r< r_b(\tau)\\-\left(1-\frac{2M}{r}\right)d\tau^2+2\sqrt{\frac{2M}{r}}drd\tau+dr^2+r^2g_{S^2} & r\geq r_b(\tau)
\end{cases}
\end{equation}
where $r_b(\tau)$ is defined by
\begin{equation}
r_b(\tau)=\left(R_b^{3/2}-\frac{3\tau}{2}\sqrt{2M}\right)^{2/3}.
\end{equation}
Note that choice of $R_b$ is equivalent to choosing when $\tau=0$. Also note the $r=0$ line (as a subset of $\R^4$) ceases to be part of the manifold $\mathcal{M}$ when the singularity ``forms" at $\tau_{c^-}$, where $r_b=0$. For $\tau<\tau_{c^-}$, $r=0$ is included in the manifold, as the metric is perfectly regular on this line.

We define our future event horizon by:
\begin{equation}
\mathcal{H}^+=\left\{r=r_b(\tau)\left(3-2\sqrt{\frac{r_b(\tau)}{2M}}\right),\tau\in[\tau_{c^-},\tau_c]\right\}\cup\{r=2M,\tau\geq\tau_c\}.
\end{equation}

Note that geometrically, this family of space-times ($H^1_{loc}$ Lorentzian manifolds), $(\mathcal{M},g_{M,R_b})$, is a one parameter family of space-times. The geometry depends only on $M$, as $R_b$ just corresponds to the coordinate choice of where $\tau=0$. Thus constants which only depend on the overall geometry of the space-time only depend on $M$.

We can also explicitly calculate $\rho$ in these coordinates for $r<r_b(\tau)$:
\begin{equation}
\rho(\tau)=\frac{3M}{4\pi r_b^3(\tau)}=\frac{3M}{4\pi \left(R_b^{3/2}-\frac{3\tau}{2}\sqrt{2M}\right)^2}=\frac{R_b^3}{r_b^3(\tau)}\rho_0.
\end{equation}

In the exterior of the space-time, we have one timelike Killing field, $\p_{t^*}=\p_\tau$, which is not Killing in the interior. Throughout the whole space-time, we have 3 angular Killing fields, $\{\Omega_i\}_{i=1}^3$, which between them span all angular derivatives. When given in the usual $\theta,\varphi$ coordinates, these take the form:
\begin{align}\nonumber
\Omega_1&=\p_\varphi\\\label{eq:AngularKillingFields}
\Omega_2&=\cos\varphi\p_\theta-\sin\varphi\cot\theta\p_\varphi\\\nonumber
\Omega_3&=-\sin\varphi\p_\theta-\cos\varphi\cot\theta\p_\varphi
\end{align}

\subsection{Penrose Diagram of $(\mathcal{M},g)$}\label{sec:Pen}

\begin{wrapfigure}{r}{7cm}
	\vspace{-5mm}
	\begin{tikzpicture}[scale=0.7]
	\node (I)    at ( 0,0) {};
	
	\path 
	(I) +(90:3)  coordinate[label=90:$i^+$]  (Itop)
	+(0:3)   coordinate[label=0:$i^0$] (Iright)
	+(180:3) coordinate (Ileft)
	+(120:2.92820323) coordinate (BHH)
	+(-116.565051177:6.708203932) coordinate[label=0:$i^-$] (Ibot)
	+(135:4.242640687) coordinate (R0+)
	+(116.565051177:3.345) coordinate (tc)
	+(95:1.5) coordinate[label=\small $t^*_c$\text{, }$2M$]
	;
	\draw (Ileft) -- 
	node[midway, above left]    {$\mathcal{H}^+$}
	(Itop) --
	node[midway, above, sloped] {$\mathcal{I}^+$}
	(Iright) -- 
	node[midway, below, sloped] {$\mathcal{I}^-$}
	(Ibot) --
	node[midway, below, sloped]    {\small $r=0$}    
	(Ileft) -- cycle;
	\draw (Ibot) to[out=60, in=-60]
	node[midway, above, sloped] {\tiny $r=R^*(t^*)$} (tc);
	\draw (Ileft) -- (R0+);
	\draw[snake it] (R0+) -- 
	node[midway, above, sloped]    {\small $r=0$}
	(Itop);
	\end{tikzpicture}
	\caption{Penrose diagram of Oppenheimer--Snyder space-time}\label{fig:PenFull}
\end{wrapfigure}
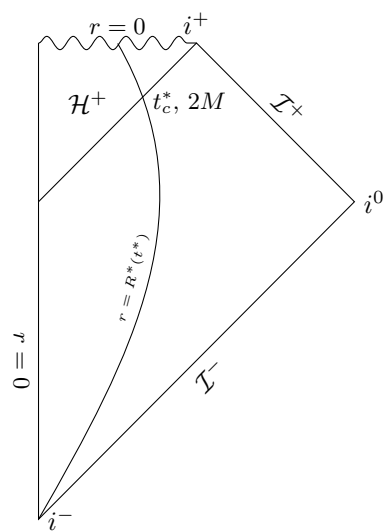

We now look to derive the Penrose diagram for the space-time $(\mathcal{M},g)$. Recall that the Penrose diagram corresponds to the range of globally defined radial double null coordinates. Using the original $R$ and $\tau$ coordinates in \eqref{eq:Rtau}, we obtain that the interior of the dust cloud has metric
\begin{equation}
g=-d\tau^2+\left(1-\frac{3\sqrt{2M}\tau}{2{R_b}^{\frac{3}{2}}}\right)^\frac{4}{3}\left(dR^2+R^2g_{S^2}\right)
\end{equation}
for $R\leq R_b$ and $\tau\leq\tau_c=\frac{2{R_b}^\frac{3}{2}}{3\sqrt{2M}}$. We then choose a new time coordinate, $\eta$ such that 
\begin{equation}
\eta(\tau)=\int_{\tau'=0}^{\tau}\left(1-\frac{3\sqrt{2M}\tau'}{2{R_b}^{\frac{3}{2}}}\right)^{-\frac{2}{3}}d\tau'.
\end{equation}
Then we change to coordinates $u=\eta-R$, and $v=\eta+R$. Thus we obtain the metric to be of the form
\begin{equation}
g=\left(1-\frac{3\sqrt{2M}\tau(u,v)}{2{R_b}^{\frac{3}{2}}}\right)^\frac{4}{3}(-du dv+R(u,v)^2g_{S^2}).
\end{equation}

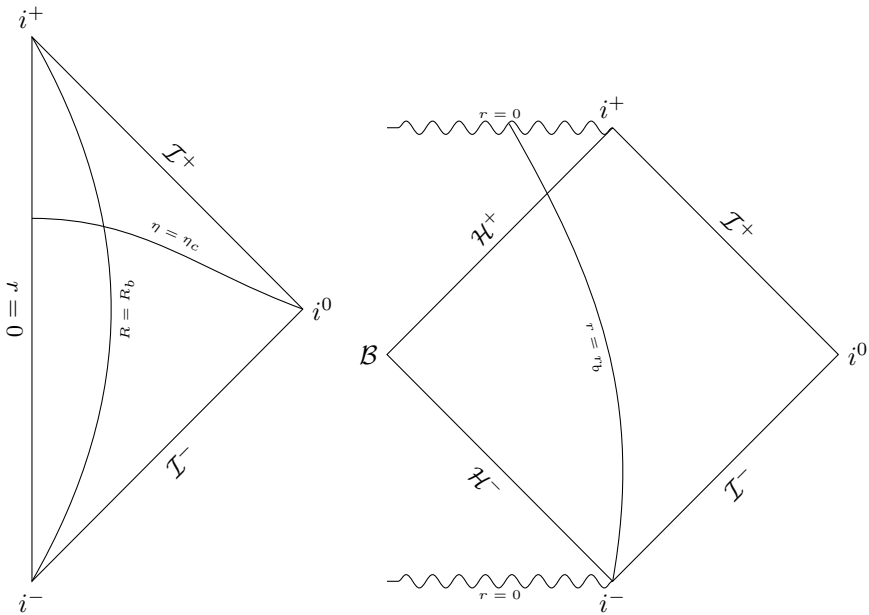
\begin{figure}
	\begin{tikzpicture}[scale=1.2]
	\node (I)    at ( 0,0) {};
	
	\path 
	(I) +(90:3)  coordinate[label=90:$i^+$]  (Itop)
	+(-90:3) coordinate[label=-90:$i^-$] (Ibot)
	+(0:3)   coordinate[label=0:$i^0$] (Iright)
	+(90:1)   coordinate (tc);
	\draw (Itop) -- 
	node[midway, below, sloped]    {\small $r=0$}
	(Ibot) --
	node[midway, below, sloped] {$\mathcal{I}^-$}
	(Iright) -- 
	node[midway, above, sloped] {$\mathcal{I}^+$}
	(Itop) -- cycle;
	\draw (Ibot) to[out=60, in=-60]
	node[midway, below, sloped] {\tiny $R=R_b$} (Itop);
	\draw (tc) to[out=0, in=160]
	node[midway, above, sloped] {\tiny $\eta=\eta_c$} (Iright);
	\end{tikzpicture}
	\begin{tikzpicture}[scale=1]
	\node (I)    at ( 0,0) {};
	
	\path 
	(I) +(45:4.242640687)  coordinate[label=90:$i^+$]  (Itop)
	+(-45:4.242640687) coordinate[label=-90:$i^-$] (Ibot)
	+(0:6)   coordinate[label=0:$i^0$] (Iright)
	+(0:0) coordinate[label=180:$\mathcal{B}$] (BH)
	+(90:3) coordinate (R0+)
	+(-90:3) coordinate (R0-)
	+(62:3.458) coordinate (R0);
	\draw (Itop) -- 
	node[midway, above, sloped]    {\small $\mathcal{H}^+$}
	(BH) -- 
	node[midway, below, sloped]    {\small $\mathcal{H}^-$}
	(Ibot) --
	node[midway, below, sloped] {$\mathcal{I}^-$}
	(Iright) -- 
	node[midway, above, sloped] {$\mathcal{I}^+$}
	(Itop);
	\draw[snake it] (Itop) -- 
	node[midway, above, sloped] {\tiny $r=0$}
	(R0+);
	\draw[snake it] (Ibot) -- 
	node[midway, below, sloped] {\tiny $r=0$}
	(R0-);
	\draw (Ibot) to[out=80, in=-60]
	node[midway, below, sloped] {\tiny $r=r_b$} (R0);
	
	\end{tikzpicture}
	\caption{Penrose diagram of Minkowski (left) and Schwarzschild (right) space-times, with appropriate boundaries.}
\end{figure}
In this coordinate system, the range of $u$ and $v$ is given by $u+v\leq B$ and $0\leq v-u\leq A$. Thus the interior of the star is conformally flat. Hence the Penrose diagram for the interior is that of Minkowski space-time, subject to the above ranges of $u+v$ and $v-u$. We also note that we have that $R_{abcd}R^{abcd}$ blows up as $\eta$ approaches $\eta_c=\eta(\tau_c)$, so this corresponds to a singular boundary of space-time.

On the exterior of the dust cloud, our solution is a subregion of Schwarzschild space-time. The boundary of this region is given by a timelike curve going from past timelike infinity to $r=0$. Matching these two diagrams across the relevant boundary, we obtain the Penrose diagram shown in Figure \ref{fig:PenFull}. Again, remember the metric is only a piecewise smooth and $H^1_{loc}$ function of $u$ and $v$.

\section{Notation}
We will be studying solutions of the wave equation \eqref{eq:wave}. 
Generally we will be considering the solutions to arise from initial data. Initial data consists of the values of our function $\phi$ on a hypersurface of constant $t$ or $\tau$, and the values of the normal derivative of $\phi$ to this surface. For the permeating case, we will be considering initial data to be $H^2_{loc}$ (with normal derivative in $H^1_{loc}$) and compactly supported on $\Sigma_{\tau}$. For the reflective case, we will consider smooth and compactly supported  initial data on $\Sigma_{t^*}$ in the region $r\in[R^*(t^*),\infty)$, with $\phi=0$ on $r=R^*(t^*)$. Functions which obey these conditions for all $\Sigma_{\tau}$ or $\Sigma_{t^*}$ will be said to be in $H^2_{c\forall \tau}$ or $C^\infty_{c\forall t^*}$ respectively. In Theorems \ref{Thm:H1ReflExist} and \ref{Thm:H1PermExist} we obtain an existence result for more general solutions, which compatible theorems then generalise to by density arguments. We may use either of the coordinate systems in equations \eqref{eq:metric} and \eqref{eq:metric*} when discussing the manifold $(\mathcal{M},g)$ away from $r=2M$. However, as \eqref{eq:metric} becomes problematic on the sphere $r=2M$, we will mainly be using the coordinates given by \eqref{eq:metric*} throughout this paper.

The main norms that we will be using are the $H^n$ norm, the $\dot{H}^n$ norm, the $\ddot{H}^n$ norm, and the $L^2$ norm. These are defined on either spacelike submanifolds $\Sigma$ of $\mathcal{M}$ (or on $\mathcal{M}$ itself) in the reflective case by
\begin{equation}
\Vert f\Vert_{L^2(\Sigma)}^2=\int_{\Sigma} \vert f\vert^2dV
\end{equation}
\begin{equation}\label{eq:Hdotdef}
\Vert f\Vert_{\ddot{H}^n(\Sigma)}^2=\sum_{\substack{n_1,n_2, n_3\\n_1+n_2+n_3=n\\n_1,n_2,n_3\geq0}}\int_\Sigma \frac{1}{r^{2n_1}}\Vert\mathring{\slashed\nabla}{}^{n_1}\p_{t^*}^{n_2}\p_r^{n_3}f\Vert^2dV
\end{equation}
\begin{equation}
\Vert f\Vert_{H^n(\Sigma)}^2=\sum_{m=0}^n\Vert f\Vert_{\ddot{H}^m(\Sigma)}^2
\end{equation}
\begin{equation}
\Vert f\Vert_{\dot{H}^n(\Sigma)}^2=\sum_{m=1}^n\Vert f\Vert_{\ddot{H}^m(\Sigma)}^2
\end{equation}
where $\mathring{\slashed\nabla}$ is the induced gradient on the unit sphere. For hypersurfaces, $dV$ is the volume form on $\Sigma$ induced from generalised Stokes' theorem with respect to the normals given below. In the permeating case, we will have all $\p_{t^*}$ derivatives replaced by $\p_{\tau}$ derivatives. Note that in the above, $f$ is a suitable function on space-time, and the norms are for now not associated with normed spaces. (We shall discuss various genuine normed spaces later.)

In \eqref{eq:Hdotdef}, the norm on the right hand side is on a tensor on the unit sphere. This norm is defined by:
\begin{equation}
\Vert T_{a_1a_2...a_m}\Vert^2=\sum_{a_1,...,a_m=1}^{n}\vert T_{a_1a_2...a_m}\vert^2
\end{equation}
for $T$ an $m$ tensor on $S_n$, in any orthonormal basis tangent to the sphere at that point.

The case of null hypersurfaces, $\mathcal{N}$, is defined similarly, but only with derivatives contained in the surface itself. Let $x$ denote either $u$ or $v$, whichever is a valid parameter for our null surface. Then we define the $\ddot{H}^n$ norm by

\begin{equation}
\Vert f\Vert_{\ddot{H}^n(\mathcal{N})}^2=\sum_{m=0}^{n}\int_\mathcal{N} \frac{1}{r^{2m}}\Vert\mathring{\slashed\nabla}{}^{m}\p_{x}^{n-m}f\Vert^2dV.
\end{equation}

All other norm definitions follow as in the spacelike case.

The above functional norms will be used on $\Sigma$s, with respective volume forms and (not necessarily unit!) normals:
\begin{align}\label{eq:voldef}
\Sigma_{t^*_0}:=\{(t^*,r,\theta,\varphi):r\geq R^*(t^*), t^*=t^*_0\}\qquad &dV=r^2drd\omega^2\qquad dn=-dt^*\\
\Sigma_{\tau_0}:=\{(\tau,r,\theta,\varphi):\tau=\tau_0\}\qquad &dV=r^2drd\omega^2\qquad dn=-d\tau\\
\Sigma_{v_0}:=\{(v,r,\theta,\phi):v=v_0, r\geq R^*(t^*)\}\qquad &dV=\frac{1}{2}\left(1-\frac{2M}{r}\right)r^2dud\omega^2\qquad dn=-dv\\
\Sigma_{u_0}:=\{(u,v,\theta,\phi):u=u_0, r\geq R^*(t^*)\}\qquad &dV=\frac{1}{2}\left(1-\frac{2M}{r}\right)r^2dvd\omega^2\qquad dn=-du\\
S_{[t^*_0,t^*_1]}:=\{(t^*,R^*(t^*),\theta,\varphi):t^*\in[t^*_0,t^*_1]\}\qquad &dV=r^2dt^*d\omega^2\qquad dn=dr-{\dot{R}^*}dt^*\\
S_{[\tau_0,\tau_1]}:=\{(\tau,r_b(\tau),\theta,\varphi):\tau\in[\tau_0,\tau_1]\}\qquad &dV=r^2d\tau d\omega^2\qquad dn=dr-\dot{r_b}d\tau
\label{eq:voldef2}
\end{align}
where $d\omega^2$ is the Euclidean metric on the unit sphere.

\begin{wrapfigure}{r}{6cm}
	\vspace{-9mm}
	\begin{tikzpicture}[scale=1.5]
	\node (I)    at ( 0,0) {};
	
	\path
	(I) +(-90:2) coordinate (Imid)
	+(0:2)   coordinate[label=0:$i^0$] (Iright)
	+(120:1.154700538) coordinate[label=90:\small $t^*_c$\text{, }$2M$] (Ileft)
	+(150:1.154700538) coordinate (C)
	+(-108.434948823:3.16227766) coordinate[label=0:$i^-$] (Ibot)
	+(-120:0.34) coordinate (R)
	;
	\draw[color=white] (R) --
	node[midway, above, sloped] {\tiny \textcolor{black}{$S_{[t^*,t^*_c]}$}} (Ileft);
	\draw (Ileft) to[out=0, in=150] 
	node[midway, above, sloped] {$\Sigma_{t^*_c}$}
	(Iright) -- 
	node[midway, below, sloped] {$\mathcal{I}^-$}
	(Ibot) --
	node[midway, below, sloped]    {\small $r=0$}    
	(C);
	\draw (Ibot) to[out=60, in=-60]
	node[midway, above, sloped] {\tiny $r=R^*(t^*)$} (Ileft);
	\draw (R) to[out=20, in=-160]
	node[midway, above, sloped] {$\Sigma_{t^*}$} (Iright);
	\draw (C) -- (Ileft);
	\draw (-0.76,-2.56) to node[midway, above, sloped] {$\Sigma_{u_0}$} (1.87,0.07);
	\end{tikzpicture}
\end{wrapfigure}

We define future/past null infinity $\mathcal{I}^\pm$ by
\begin{align}
\mathcal{I}^+:=\R\times S^2 \qquad dV=dud\omega^2\qquad
\mathcal{I}^-:=\R\times S^2 \qquad dV=dvd\omega^2\qquad
\end{align}
Past null infinity is viewed as the limiting surface as we take $u$ to $-\infty$, keeping $v$ fixed. For appropriate space-time functions (or integrands) $f(u,v,\theta, \varphi)$, we will write the function evaluated on $\mathcal{I}^-$ to mean the limit
\begin{equation}
f(v,\theta,\varphi)\vert_{\mathcal{I}^-}=\lim_{u\to-\infty}f(u,v,\theta,\varphi),
\end{equation}
when this limit exists in an appropriate sense (see Section \ref{Sec:TheScatMap}).

\begin{wrapfigure}{r}{6cm}
	\vspace{-5mm}
	\begin{tikzpicture}[scale=1.5]
	\node (I)    at ( 0,0) {};
	
	\path
	(I) +(-90:2) coordinate (Imid)
	+(0:2)   coordinate[label=0:$i^0$] (Iright)
	+(120:1.154700538) coordinate[label=90:\small $\tau_c$\text{, }$2M$] (Ileft)
	+(150:1.154700538) coordinate (C)
	+(150:1.25) coordinate[label=90:\tiny $\tau_{c^-}$\text{, }$0$]
	+(-108.434948823:3.16227766) coordinate[label=0:$i^-$] (Ibot)
	+(180:0.21) coordinate (R)
	+(-1,0) coordinate (R1)
	;
	\draw[color=white] (R) --
	node[midway, above, sloped] {\tiny \textcolor{black}{$S_{[\tau,\tau_c]}$}} (Ileft);
	\draw (Ileft) to[out=0, in=150] 
	node[midway, above, sloped] {$\Sigma_{\tau_c}$}
	(Iright) -- 
	node[midway, below, sloped] {$\mathcal{I}^-$}
	(Ibot) --
	node[midway, below, sloped]    {\small $r=0$}    
	(C);
	\draw[dotted] (Ibot) to[out=60, in=-60]
	node[midway, above, sloped] {\tiny $r=r_b(\tau)$} (Ileft);
	\draw (R1) to[out=20, in=-160]
	node[midway, above, sloped] {$\Sigma_{\tau}$} (Iright);
	\draw (C) -- (Ileft);
	\draw (-1,-2.56) to (-0.76,-2.56) to node[midway, above, sloped] {$\tilde{\Sigma}_{\tau_0}$} (1.87,0.07);
	\end{tikzpicture}
\end{wrapfigure}

We define $\mathcal{I}^+$ similarly, with
\begin{equation}
f(u,\theta,\varphi)\vert_{\mathcal{I}^+}=\lim_{v\to\infty}f(u,v,\theta,\varphi).
\end{equation}

We will also later be using, for the permeating case, the eventually null foliation, $\tilde{\Sigma}_{\tau_0}$. These are the set of points with $\tau=\tau_0$ for $r< r_b$, and $v=v_0$ for $r\geq r_b$. Here $v_0$ is the value of $v$ at $(\tau_0,r_b(\tau_0))$.
\begin{equation}
\tilde{\Sigma}_{\tau_0}=\left(\Sigma_{\tau_0}\cap\{r<r_b(\tau_0)\}\right)\cup\left(\Sigma_{v_0}\cap\{r\geq r_b(\tau_0)\}\right).
\end{equation}
This will have the same volume form as $\Sigma_{\tau_0}$ for $r<r_b(\tau_0)$ and the same volume form as $\Sigma_{v_0}$ for $r\geq r_b(\tau_0)$.

Any surface integrals from this point on that do not have a volume form stated are to be understood as using the volume forms stated in \eqref{eq:voldef} through to \eqref{eq:voldef2}. Any space-time integrals with no volume form stated are using the volume form given by $\sqrt{-\det g}$.

We will then define the $H^1(\Sigma)$ norm (or $\dot{H}^1(\Sigma)$ norm) of a pair of functions $(\phi_0,\phi_1)\in H^1_{loc}(\Sigma)\times L^2_{loc}(\Sigma)$ on any spacelike $\Sigma$, as follows:

\begin{equation}
\Vert(\phi_0,\phi_1)\Vert^2_{H^1(\Sigma)}:=\Vert\phi\Vert^2_{H^1(\Sigma)}\quad \text{ for any }\phi\quad s.t.\quad (\phi|_\Sigma,\p_{t^*,\tau}\phi|_\Sigma)=(\phi_0,\phi_1).
\end{equation}
As $\phi_0, \phi_1$ may not be smooth, we will take $(\phi|_\Sigma,\p_{t^*,\tau}\phi|_\Sigma)=(\phi_0,\phi_1)$ in a trace sense. 

We may also just take $\phi=\phi_0+t^*\phi_1$ (or $\tau\phi_1$). An explicit calculation then gives:
\begin{equation}
\int_{\Sigma}(\p_r\phi_0)^2+\frac{1}{r^2}\vert\mathring{\slashed\nabla}\phi_0\vert^2+\phi_1^2.
\end{equation}

We then define the Hilbert space $H^1(\Sigma)$ to be the space of all pairs of functions in $ H^1_{loc}(\Sigma)\times L^2_{loc}(\Sigma)$ with finite $H^1(\Sigma)$ norm.

Similarly, we define the norm on an $(n+1)$-tuple of functions $(\phi_0,\phi_1,...,\phi_n)\in H^n_{loc}(\Sigma)\times H^{n-1}_{loc}(\Sigma)\times...\times L^2_{loc}(\Sigma)$ to be
\begin{align}
\Vert(\phi_0,\phi_1,...,\phi_n)\Vert^2_{H^1(\Sigma)}:&=\Vert\phi\Vert^2_{H^1(\Sigma)}\quad \\\nonumber\text{ for any }\phi\quad s.t.\quad (\phi,\p_{t^*}\phi,...,\p_{t^*}^n\phi)|_\Sigma&=(\phi_0,\phi_1,...,\phi_n) \text{ in a trace sense}.
\end{align}

Again, we may just take $\phi=\phi_0+t^*\phi_1+...+t^{*n}\phi_n$ (with $t^*$ replaced by $\tau$ where necessary).

The Hilbert space $H^n(\Sigma)$ is defined to be that of $(n+1)$-tuples $(\phi_0,\phi_1,...,\phi_n)\in H^n_{loc}(\Sigma)\times H^{n-1}_{loc}(\Sigma)\times...\times L^2_{loc}(\Sigma)$ with finite $H^n(\Sigma)$ norm.

We define the $H^n(\mathcal{N})$ norm on a function, $\phi_0$ on $\mathcal{N}$ for null surface $\mathcal{N}$ by

\begin{equation}
\Vert\phi_0\Vert^2_{H^1(\mathcal{N})}:=\Vert\phi\Vert^2_{H^1(\mathcal{N})}\quad \text{ for any }\phi\quad s.t.\quad \phi|_\mathcal{N}=\phi_0 \text{ in a trace sense}.
\end{equation}
Note this definition applies to $\mathcal{I}^\pm$.

We next introduce the energy momentum tensor of our wave, $\phi$ (note this is unrelated to the energy momentum tensor $\boldsymbol{T}$ in \eqref{eq:EE}). We will also introduce the notion of energy currents, modified energy currents, and energy through a surface, $S$. For a given vector field $X$ and scalar function $w$, we define:
\begin{align}
T_{\mu\nu}(\phi)&=\nabla_\mu\phi\nabla_\nu\phi-\frac{1}{2}g_{\mu\nu}\nabla^\rho\phi\nabla_\rho\phi\\\label{eq:energycurrent}
J^X_{\mu}(\phi)&=X^\nu T_{\mu\nu}(\phi)\\
K^X(\phi)&=\nabla^\mu J^X_\mu(\phi)\\
\label{eq:modcurrent}
J^{X,w}_\mu(\phi)&=X^\nu T_{\mu\nu}(\phi)+w\nabla_\mu(\phi^2)-\phi^2\nabla_\mu w\\
K^{X,w}(\phi)&=\nabla^\nu J^{X,w}_\nu(\phi)=K^X(\phi)+2w\nabla_\mu\phi\nabla^\mu\phi-\phi^2\Box_gw\\
\label{eq:energydef}X\text{-energy}(\phi,S)&=\int_Sdn(J^X(\phi))
\end{align}
where $dn$ is the normal to $S$ which is given for each surface above. Note that for $S$  spacelike or null, we have chosen $dn$ to be future pointing. We have that if $X$ is future pointing and causal, this energy gives a norm on timelike or null surfaces (by the dominant energy condition). We will often choose vectors such that their energy is an equivalent norm to the $\dot{H}^1$ norm defined in \eqref{eq:Hdotdef}.

When we later discuss the forwards, backwards and scattering maps, we will need to use the notion of function spaces on our different surfaces. For this, we will be using similar notation to \cite{KerrScatter}. We define the space of finite $X$-energy pairs of functions on a spacelike surface, $S$ by first defining the $X$ norm:
\begin{equation}\label{eq:energyspacedef}
\Vert(\phi_0,\phi_1)\Vert^2_X=X\text{-energy}(\phi,S)\text{ for any }\phi\quad s.t.\quad (\phi|_S,\p_{t^*,\tau}\phi|_S)=(\phi_0,\phi_1) \text{ in a trace sense}.
\end{equation}

An explicit calculation shows that for a given $(\phi_0,\phi_1)\in H^1(S)$, we have that $\Vert(\phi_0,\phi_1)\Vert_X$ is independent of the choice of $\phi$. Thus we may take $\phi=\phi+t^*\phi_1$ (or $\tau\phi_1$).

Then we define the space of finite energy pairs of functions, $\mathcal{E}_S^X$ by
\begin{equation}
(\phi_0,\phi_1)\in\mathcal{E}_S^X\subset H^1(S) \iff \Vert(\phi_0,\phi_1)\Vert^2_X<\infty.
\end{equation}

Note this requires $X$ to be future pointing and causal, and that in the reflective case, any $\phi_0$ function with $(\phi_0,\phi_1) \in\mathcal{E}^X_{S}$ is zero when restricted to $r=R^*(t^*)$ (in a trace sense).

We similarly define the space of finite $X$-energy functions on a null surface, $S$ by:
\begin{equation}
\Vert\phi_0\Vert_X^2=X\text{-energy}(\phi,S)\text{ for any }\phi\in H_{loc}^1(\mathcal{M}))\quad s.t.\quad \phi|_S=\phi_0 \text{ in a trace sense},
\end{equation}
\begin{equation}
\phi_0\in\mathcal{E}_S^X \iff \phi_0\in H^1(S), \Vert\phi_0\Vert^2_X<\infty.
\end{equation}

Note this space of functions is complete.

The final function space and norm we define is $\mathcal{E}^{\p_{t^*,\tau}}_{\mathcal{I}^\pm}$, which have norms given by:
\begin{align}
\Vert\psi^-(v,\theta,\varphi)\Vert_{\p_{t^*,\tau}}^2=\int_{\mathcal{I}^-}(\p_v\psi^-)^2dvd\omega^2\\
\Vert\psi^+(u,\theta,\varphi)\Vert_{\p_{t^*,\tau}}^2=\int_{\mathcal{I}^+}(\p_u\psi^+)^2dud\omega^2.
\end{align}
We similarly define the space of finite $\p_{t^*,\tau}$-energy functions on $\mathcal{I}^\pm$ as
\begin{equation}
\psi\in\mathcal{E}_{\mathcal{I}^\pm}^{\p_{t^*,\tau}} \iff \psi\in L^2(\mathcal{I}^\pm), \p_v\psi\in L^2(\mathcal{I}^\pm), \Vert\psi\Vert^2_{\p_{t^*,\tau}}<\infty,
\end{equation}
where $\p_v\psi$ is a weak derivative of $\psi$ in the $v$ direction.

\section{Existence and Uniqueness of Solutions to the Wave Equation}\label{Sec:Existence}
Consider initial data given by $\phi=\phi_0$ and $\p_{t^*} \phi=\phi_1$ on the spacelike hypersurface $\Sigma_{t^*_0}$ (or $\p_{\tau} \phi=\phi_1$ on $\Sigma_{\tau_0}$). For the reflective case, we also impose vanishing Dirichlet conditions on the surface of the star, $\phi =0$ on $r=R(t)$. We first show existence of a solution to the forced wave equation,
\begin{equation}\label{eq:forcedwave}
\Box_g \phi=\frac{1}{\sqrt{-g}}\p_a(\sqrt{-g}g^{ab}\p_b\phi)=F
\end{equation}
for $g$ as given in \eqref{eq:metric*} and \eqref{eq:metricint}.

These are standard results which can be taken from literature, but there is no elementary reference. For completeness we will write a proof out here.

\subsection{Existence and Uniqueness for the Reflective Case}

We initially prove existence and uniqueness for smooth, compactly supported initial data in the reflective case. We will be proving this up until the surface of the star passes through the horizon. For later $t^*$ times, we are then in exterior Schwarzschild space-time with the usual boundaries, so can refer to standard existing proofs of existence and uniqueness (see for example proposition 3.1.1 in \cite{BHL}). The proof below closely follows that of Theorems 4.6 and 5.3 of Jonathan Luk's notes on Nonlinear Wave Equations \cite{Luk}, and comes in two parts:

We proceed by first proving uniqueness via the following lemma: 
\begin{Lemma}[Uniqueness of Solution to the Forced Wave Equation]\label{lm:bdd}
Let $\phi\in C^\infty_{c\forall t^*}$ be a solution to equation \eqref{eq:forcedwave} in some region ${r\geq R^*(t^*)\geq 2M}$, $t^*_{-1}\leq t^*_0\leq t^*_1$, with
\begin{multline}\label{eq:initial}\\
\phi=0 \text{ on } r=R^*(t^*)\\
\left.
\begin{array}{c}
\phi=\phi_0\\
\p_{t^*}\phi=\phi_1
\end{array}\right\}\text{ on } \Sigma_{t^*_0}\\
\end{multline}
with $g_{ab}$ given by $\eqref{eq:metric}$.

It follows that $\exists A,C>0$ s.t.
\begin{equation}
\sup_{t^*\in [t^*_{-1},t^*_1]}\Vert \phi\Vert_{\dot{H}^1(\Sigma_{t^*})}
\leq C\left(\Vert(\phi_0,\phi_1)\Vert_{\dot{H}^1(\Sigma_{t^*_0})}+\int_{t^*_{-1}}^{t^*_1}{\Vert F\Vert_{L^2(\Sigma_{t^*})}(t^*)dt^*}\right)\exp\left(A\vert t^*_1-t^*_{-1}\vert\right).
\end{equation}

In particular, if $\phi, \phi'\in C^\infty_{c\forall t^*}$ are both solutions to the above problem, then consider $\zeta=\phi-\phi'$. We have that $\zeta$ solves equation \eqref{eq:forcedwave} with $F=0$ and has $0$ initial data. Thus from this lemma, $\zeta=\phi-\phi'=0$ everywhere, and  we have uniqueness.
\end{Lemma}
\begin{proof}
	We first consider coordinates $(t^*,\rho,\theta,\varphi)$, where $\rho=r-R^*(t^*)+2M$. This causes $\p_{t^*}$ to be tangent to the boundary $r=R^*(t^*)$. The metric then takes the form
	\begin{equation}\label{eq:RhoCoords}
	g=-\left(\left(1-\frac{2M}{r}\right)-\frac{4M}{r}{\dot{R}^*}-\left(1+\frac{2M}{r}\right)\dot{R}^{*2}\right)dt^{*2}+\left(\frac{4M}{r}+2\left(1+\frac{2M}{r}\right){\dot{R}^*}\right)d\rho dt^*+\left(1+\frac{2M}{r}\right)d\rho^2+r(t^*,\rho)^2g_{S^2}.
	\end{equation}
	We integrate the following identity:
	\begin{equation}
	\p_0\phi(\p_a(\bar{g}^{ab}\p_b\phi)-\bar{F}))=0
	\end{equation}
	for \begin{equation}
	\bar{g}^{ab}=\sqrt{-g}g^{ab}
	\end{equation}
	\begin{equation}
	\bar{F}=\sqrt{-g}F.
	\end{equation}
	
	We look at the cases $a=b=0$, $a=i,b=j$, and $\{a,b\}=\{0,i\}$ separately, where $i,j\in \{1,2,3\}$.
	\begin{equation}\label{eq:unique00}
	\int_{t^*_0}^{t^*_1}\int_{\Sigma_{t^*}}\p_0\phi\p_0(\bar{g}^{00}\p_0\phi)d\rho d\omega^2dt^*=\frac{1}{2}\left(\int_{t^*_0}^{t^*_1}\int_{\Sigma_{t^*}}(\p_0\bar{g}^{00})(\p_0\phi)^2d\rho d\omega^2dt^*+\left(\int_{\Sigma_{t^*_1}}-\int_{\Sigma_{t^*_0}}\right)\bar{g}^{00}(\p_0\phi)^2d\rho d\omega^2\right).
	\end{equation}
	\begin{equation}\label{eq:unique0i}
	\int_{t^*_0}^{t^*_1}\int_{\Sigma_{t^*}}\p_0\phi\p_i(\bar{g}^{ij}\p_j\phi)d\rho d\omega^2dt^*=\frac{1}{2}\left(\int_{t^*_0}^{t^*_1}\int_{\Sigma_{t^*}}(\p_0\bar{g}^{ij})(\p_i\phi\p_j\phi) d\rho d\omega^2dt^*-\left(\int_{\Sigma_{t^*_1}}-\int_{\Sigma_{t^*_0}}\right)\bar{g}^{ij}(\p_i\phi\p_j\phi)d\rho d\omega^2\right).
	\end{equation}
	\begin{equation}\label{eq:uniqueij}
	\int_{t^*_0}^{t^*_1}\int_{\Sigma_{t^*}}\p_0\phi(\p_i(\bar{g}^{i0}\p_0\phi)+\p_0(\bar{g}^{i0}\p_i\phi))d\rho d\omega^2dt^*=\int_{t^*_0}^{t^*_1}\int_{\Sigma_{t^*}}(\p_0\bar{g}^{i0})(\p_0\phi\p_i\phi)d\rho d\omega^2dt^*.
	\end{equation}
	Here we have integrated by parts. Using the fact that as $\phi=0$ on our boundary and $\p_0$ is tangent to our boundary, we can see that $\p_0\phi=0$ on our boundary. We have used this to simplify the above boundary terms. Using
	\begin{equation}
	\bar{g}^{ab}=\left(\begin{array}{ccc}
	-\left(1+\frac{2M}{r}\right)&\frac{2M}{r}+\left(1+\frac{2M}{r}\right)\dot{R}^*&0\\ \frac{2M}{r}+\left(1+\frac{2M}{r}\right)\dot{R}^*& \left(1-\frac{2M}{r}\right)-\frac{4M}{r}\dot{R}^*-\left(1+\frac{2M}{r}\right)\dot{R}^{*2} & 0 \\
	0& 0& \frac{1}{r^2}g^{-1}_{S^2} 
	\end{array}\right)r^2\sin\theta
	\end{equation}
	gives us that
	\begin{equation}
	\vert\p_0\bar{g}^{ab}\vert\begin{cases}
	\leq A'r^2 & {a,b}={0,1}\\
	=0 & {a,b}={2,3}.
	\end{cases}
	\end{equation}
	(Note that coordinate singularities have been removed from $g^{ab}$ by multiplying by $\sqrt{-g}$.)
	
	We then have, by summing \eqref{eq:unique00}, \eqref{eq:unique0i} and \eqref{eq:uniqueij} together, that
	\begin{align}\nonumber
	\int_{\Sigma_{t^*_1}}\bar{g}^{ij}\p_i\phi\p_j\phi-\bar{g}^{00}(\p_0\phi)^2d\rho d\omega^2
	=&\int_{\Sigma_{t^*_0}}\bar{g}^{ij}\p_i\phi\p_j\phi-\bar{g}^{00}(\p_0\phi)^2d\rho d\omega^2 +\frac{1}{2}\int_{t^*_0}^{t^*_1}(\p_0\bar{g}^{ab})\p_a\phi\p_b\phi-\bar{F}\p_0\phi d\rho d\omega^2dt^*\\\label{eq:uniquebound}
	\leq& \int_{\Sigma_{t^*_0}}\bar{g}^{ij}\p_i\phi\p_j\phi-\bar{g}^{00}(\p_0\phi)^2d\rho d\omega^2 +\frac{1}{2}\int_{t^*_0}^{t^*_1}\Vert\phi\Vert_{\dot{H}^1(\Sigma_{t^*})}\Vert F\Vert_{L^2(\Sigma_{t^*})}+A'\Vert\phi\Vert^2_{\dot{H}^1(\Sigma_{t^*})}dt^*.
	\end{align}
	Note that the bar is removed from $F$ as the factor of $\sqrt{-g}$ is absorbed into the volume form in the norms of $F$ and $\phi$.
	
	Then we define 
	\begin{equation}
	E(t^*_1)=\int_{\Sigma_{t^*_1}}\bar{g}^{ij}\p_i\phi\p_j\phi-\bar{g}^{00}(\p_0\phi)^2d\rho d\omega^2.
	\end{equation}
	
	As the surface of the star is timelike, we have that $-g_{00}=g^{11}$ is bounded above and below by positive constants independent of time (from equation \eqref{eq:timelike}). We note that the $r^2$ term from using $\bar{g}$ instead of $g$ is identical to the volume form in $\Vert\phi\Vert^2_{\dot{H}^1(\Sigma_{t^*_1})}$. This implies $E(t^*_1)\sim\Vert\phi\Vert^2_{\dot{H}^1(\Sigma_{t^*_1})}$. Thus, using the fact that the RHS of \eqref{eq:uniquebound} is increasing in $t^*_1$, we have
	\begin{align}\nonumber
	f({t^*_1}):&=\sup_{t^*\in [{t^*_0},t^*_1]}\Vert\phi\Vert^2_{\dot{H}^1(\Sigma_{t^*})}\leq C\sup_{t^*\in [{t^*_0},t^*_1]}E(t^*)\leq C'\Vert\phi\Vert^2_{\dot{H}^1(\Sigma_0)}+C\int_{t^*_0}^{t^*_1}\Vert\phi\Vert_{\dot{H}^1(\Sigma_{t^*})}\Vert F\Vert_{L^2(\Sigma_{t^*})}+A'\Vert\phi\Vert^2_{\dot{H}^1(\Sigma_{t^*})}dt^*\\\nonumber
	&\leq C'f({t^*_0})+C\sqrt{f(t^*_1)}\int_{t^*_0}^{t^*_1}\Vert F\Vert_{L^2(\Sigma_{t^*})}dt^*+C\int_{t^*_0}^{t^*_1}A'f(t^*)dt^*\\
	&\leq C'f({t^*_0})+\frac{C^2}{2}\left(\int_{t^*_0}^{t^*_1}\Vert F\Vert_{L^2(\Sigma_{t^*})}dt^*\right)^2+\frac{f(t^*_1)}{2}+C\int_{t^*_0}^{t^*_1}A'f(t^*)dt^*.
	\end{align}
We can then subtract the $f(t^*_1)/2$ term from both sides to end up with an inequality of the form
\begin{equation}
f({t^*_1})\leq A(t^*_0,t^*_1) +\int_{t^*_0}^{t^*_1}f(t^*)h(t^*)dt^*.
\end{equation}
	An application of Gronwall's inequality gives our result, but with $t^*_{-1}$ replaced with $t^*_0$. We then repeat the same argument with time reversed to obtain the final result.
\end{proof}

Note we have written out the above argument explicitly in coordinates. It could be written out using the energy momentum tensor and a suitable vector field multiplier, as we have done in Section \ref{Sec:Bound}.

Next we need to deal with existence. To do this, we prove the following theorem:
\begin{Theorem}[Existence of Reflective Solutions]\label{Thm:ReflExist}
	Let $F \in C^k([t^*_{-1},t^*_1];C^\infty_0(\Sigma_{t^*}))$ $\forall k\in\mathbb{N}$ and $g_{ab}$ as above. Let also $\phi_0$ and $\phi_1$ smooth, compactly supported functions on $\Sigma_{t^*_0}$ such that $\phi_0(R^*(t^*_0),\theta,\varphi)=0$. There exists a $C^\infty_{c\forall t^*}$ solution to equation \eqref{eq:forcedwave} subject to \eqref{eq:initial}.
\end{Theorem} 

\begin{proof}
	We begin the proof with the case $(\phi_0,\phi_1)=(0,0)$.
	Let the set $C_0\subset C^\infty_{c\forall t^*}$ be the image under the map $\Box_g$ of $C_0^\infty(\mathcal{M})$. We define the map $W$ by:
	\begin{align*}
	W:\ &C_0\rightarrow \R\\
	&\Box_g\psi \mapsto \int_{t^*_{-1}}^{t^*_1}\int_{\Sigma_{t^*}}\psi F \sqrt{-g}d\rho d\theta d\varphi dt^*=:\langle F,\psi\rangle
	\end{align*}
	This is well defined by our previous uniqueness lemma: suppose two functions $\psi_1,\psi_2\in C_0$ have $\Box_g\psi_1=\Box_g\psi_2$. Then we can choose $t^*_0$ to be far back enough that $\Sigma_{t^*_0}$ does not intersect the support of either $\psi_1$ or $\psi_2$. Thus $\psi_1-\psi_2$ solves \eqref{eq:wave} with vanishing initial data. Lemma \ref{lm:bdd} then gives $\psi_1-\psi_2=0$ everywhere, i.e.~they are equal. 
	
	We then proceed by quoting Lemma 5.2 in \cite{Luk}, which relies on definitions of $H^{-k}$ spaces. The space $H^{-k}(\Sigma_{t^*})$ is defined to be the dual of $H^k(\Sigma_{t^*})$ (the space of bounded linear maps from $H^k(\Sigma_{t^*})$ to $\R$). Note also that, as a Hilbert space, $H^k(\Sigma_{t^*})$ is reflexive, i.e.~the dual of $H^{-k}(\Sigma_{t^*})$ is $H^k(\Sigma_{t^*})$. In the permeating case, we define $H^{-k}(\Sigma_{\tau})$ in an identical manner.
	\begin{Lemma}\label{lm:exist}
		Suppose $\psi \in C_0^\infty((-\infty,t^*_1)\times\Sigma_{t^*})$, supported away from $r=R^*(t^*)$, and $g$ as above. Fix $t^*_0\in(-\infty,t^*_1)$. Then for any $m\in\mathbb{Z}$, $\exists C=C(m,t^*_0,t^*_1,g)>0$ s.t.
		\begin{equation}
		\Vert\psi\Vert_{H^m(\Sigma_{t^*_2})}\leq C\int_{t^*_0}^{t^*_1}\Vert \Box_g \psi\Vert_{H^{m-1}(\Sigma_{t^*})}(s)dt^*\quad \forall t^*_2\in[t^*_0,t^*_1].
		\end{equation}
	\end{Lemma}

	\begin{Remark}
		To see this from \cite{Luk}, one must first  ``Euclideanise'', i.e.~replace angular and $r$ coordinates with some $x,y,z$ in order for these coordinates to be everywhere regular. We can then extend our metric smoothly to inside the star. Using the result of Lemma \ref{lm:bdd} allows the proof to proceed exactly as in \cite{Luk}. Note that linear maps on the space extended inside the star are also linear maps when restricted to functions on the outside of the star.
	\end{Remark}

	Lemma \ref{lm:exist} then gives the bound
	\begin{align}\nonumber
	\vert W(\Box_g\psi)\vert&=\left\vert\int_{t^*_0}^{t^*_1}\int_{\Sigma_{t^*}}\psi F\sqrt{-g}d\rho d\theta d\varphi dt^*\right\vert\\
	&\leq C\left(\int_{t^*_{-1}}^{t^*_1}\Vert F\Vert_{H^{k-1}(\Sigma_{t^*})}(t^*)dt^*\right)(\sup_{t^*\in [t^*_{-1},t^*_1]}\Vert\psi\Vert_{H^{-k+1}(\Sigma_{t^*})})
	\leq C\int_{t^*_{-1}}^{t^*_1}\Vert \Box_g\psi\Vert_{H^{-k}(\Sigma_s)}(s)ds,
	\end{align}
	for smooth and compactly supported functions away from the horizon. We then take the closure of such functions with respect to the $H^k$ norm, for which $W$ is linear and bounded. Thus by Hahn--Banach (Theorem 5.1, \cite{Luk}), there exists a function $\phi\in (L^1((-\infty,T);H^{-k}(\Sigma_{t^*})))^*=L^\infty((-\infty,T);H^k(\Sigma_{t^*}))\text{ } \forall k$, which extends $W$ as a linear map. This means 
	\begin{equation}\label{eq:weaksoln}
	\langle F,\psi\rangle=\langle\phi,\Box_g\psi\rangle \qquad\forall \psi\in C_0^\infty(\mathcal{M}).
	\end{equation} 
	Now $\Box_g$ obeys
	\begin{equation}
	\langle \Box_g\psi_1,\psi_2\rangle=\langle\psi_1,\Box_g\psi_2\rangle\quad \forall \psi_{1,2}\in C_0^\infty((-\infty,t^*_1)\times\Sigma_{t^*}).
	\end{equation}
	Thus equation \eqref{eq:weaksoln} means that $\phi$ is a solution of \eqref{eq:forcedwave} in the sense of distributions. 
	
	We then consider the following equation which $\p_{t^*}\phi$ solves, in a distributional sense:
	\begin{equation}\label{eq:phidot}
	v^\mu\nabla_\mu(\dot{\phi})=h\dot{\phi}+F',
	\end{equation}
	for 
	\begin{align}
	v^\mu&=(-g^{00},-2g^{01},-2g^{02},-2g^{03})\\
	h&=\frac{1}{\sqrt{-g}}\p_{\nu}\left(g^{\nu 0}\sqrt{-g}\right)\\
	F'&=\frac{1}{\sqrt{-g}}\p_{\nu}\left(g^{\nu i}\sqrt{-g}\right)\p_i\phi+g^{ij}\p_i\p_j\phi-F.
	\end{align}
	We explicitly have $h$ and $F'$, as we have $\phi$ and its spacelike derivatives. We can then easily solve this along integral curves of $v^\mu$ to obtain that $\dot{\phi}$ exists as a function and is continuous.
	
	We then look at the difference between equation \eqref{eq:phidot} and the wave equation \eqref{eq:forcedwave}. Here we are considering everything as distributions rather than functions. This gives us that 
	\begin{equation}
	v^\mu\nabla_\mu\left(\dot{\phi}-\p_{t^*}\phi\right)-h\left(\dot{\phi}-\p_{t^*}\phi\right)=0.
	\end{equation}
	Applying the zero distribution is the same as integrating against the zero function. We also know $\dot{\phi}-\p_{t^*}\phi$ is zero on the initial surface. It is then zero along all integral curves of $v^\mu$, and is therefore the zero function everywhere. Thus $\p_{t^*}\phi$ exists everywhere and is continuous.
	
	Then, by considering equation \eqref{eq:forcedwave} and its derivatives, we can determine further weak derivatives with respect to time. If $F\in C^k([t^*_{-1},t^*_1];C^\infty_0(\Sigma_{t^*}))\ \forall k$, then our final solution has finite $H^{k}(\Sigma_{t^*})$ norm for all $k$ and all $t^*\in[t^*_{-1},t^*_1]$, This means it is smooth. Due to finite speed of propagation of the wave equation  it is also compactly supported on each $\Sigma_{t^*}$. 
	
	Finally, we show $\phi$ is a classical solution. Let $\psi$ be an arbitrary function in $C_0^\infty(\mathcal{M})$, supported away from the boundary. We can then integrate \eqref{eq:weaksoln} by parts. Using the fact $\phi$ is smooth, we can see that $\Box_g\phi=F$. 
	
	By choosing $k=1$, we note that $\phi$ extends $W$ to the closure of $C^\infty_0(\mathcal{M})$ under the $H^1$ norm. In particular, this includes functions which are smooth with non-vanishing derivative at the horizon. From this set, we can choose any arbitrary smooth compactly supported function $\psi$. Let us chose one which is zero at the boundary, but with non-zero normal derivative at the boundary. The boundary term we obtain when integrating \eqref{eq:weaksoln} by parts gives that $\phi=0$ on $r=R^*(t^*)$, as required.
	
	Now let $(\phi_0,\phi_1)$ be smooth, as in the statement of the theorem. Let $u\in C^\infty_0\left([0,{t^*_1}]\times \Sigma_{t^*}\right)$ be any function with $(u,\p_tu)=(\phi_0,\phi_1)$ on $t={t^*_0}$. Then if we solve
	\begin{align}
	\Box_g\nu&=F-\Box_g u\\
	(\nu,\p_t\nu)&=(0,0) \text{ on } t={t^*_0},
	\end{align}
	then $\phi:=\nu+u$ is our required solution.
\end{proof}

\begin{Remark}\label{Rmk:H1ReflExist}
	Theorem \ref{Thm:ReflExist} will allow us to extend other results. Suppose we obtain any result on boundedness between times slices $\Sigma_{t^*}$ in the $H^1(\Sigma_{t^*})$ norm (not necessarily uniform in time). We can use a density argument to obtain that given initial data in $H^1(\Sigma_{t^*_0})$, there exists an $H^1(\Sigma_{t^*})\text{ }\forall t^*$ solution. Again, this would be a solution in the sense of distributions (see already Theorem \ref{Thm:H1ReflExist}).
\end{Remark}

\subsection{The Permeating Case}
The proof for the permeating case follows almost identical lines to that of the reflective case. There are fewer concerns about the boundary, but the solution itself cannot be shown to be smooth for smooth initial data.

We still have Lemma \ref{lm:bdd} applying in this case, with almost no change to the proof. Lemma \ref{lm:exist} also remains the same for all $m\leq 2$. This can be seen by considering $(\tau, R,\theta,\varphi)$ coordinates. We can then commute with both angular derivatives and $\p_{\tau}$, and also rearranging $\eqref{eq:forcedwave}$ for $\p_R^2\phi$. This just leaves the analogue of Theorem \ref{Thm:ReflExist}:

\begin{Proposition}[Existence of Permeating Solutions with Initial Data Constraints]\label{Prop:PermExist}
	Suppose $F \in C^1([\tau_{-1},\tau_1];H^{1}(\Sigma_{\tau}))$, and $g_{ab}$ as above. Suppose also we are given $(\phi_0, \phi_1) \in H^1(\Sigma_{\tau})$ such that there exists a function $u\in \Box_g^{-1}(H^1(\mathcal{M}))\cap H^2(\mathcal{M})$ with $(\phi_0,\phi_1)=(u,\p_\tau u)$ on $\Sigma_{\tau_0}$. Then there exists an $H^2([\tau_{-1},\tau_1]\times\Sigma_{\tau})$ weak solution to equation \eqref{eq:forcedwave}, subject to:
	\begin{equation}\label{eq:initial2}
	\left.
	\begin{array}{c}
	\phi=\phi_0\\
	\p_\tau\phi=\phi_1
	\end{array}\right\}\text{ on } \Sigma_{\tau_0}.
	\end{equation}
\end{Proposition}
\begin{proof}
	Again, we begin with the $(\phi_0,\phi_1)=(0,0)$ case. We define the map $W$ exactly as in the reflecting case. We define it on $C'_0$, the image under the map $\Box_g$ on $C^\infty_0(\mathcal{M})$. Note the components of $g$ are $H^1_{loc}$ functions, so have weak derivatives in $L^2_{loc}$. Thus this operator still exists. As before, this operator is well defined, is linear, and is bounded.
	
	Thus, again by Hahn--Banach, there exists a function $\phi\in (L^1((-\infty,\tau_1);H^{-k}(\Sigma_{t^*})))^*=L^\infty((-\infty,\tau_1);H^k(\Sigma_{t^*}))$ $\forall k\leq 2$ such that
	\begin{equation}\label{eq:weaksolnperm}
	\langle F,\psi\rangle=\langle\phi,\Box_g\psi\rangle \qquad\forall \psi\in C_0^\infty(\mathcal{M}).
	\end{equation} 
	As before, we can show $\phi$ has a $\tau$ derivative by considering the equation obeyed by $\p_{\tau}\phi$ as a distribution. If $F$ is in $H^1$, then $\phi$ has two weak spatial derivatives. It also has a $\tau$ derivative with spacelike weak derivatives. By integrating \eqref{eq:weaksolnperm}, we obtain it also has a second weak time derivative. Thus it is $H^2$, and thus our solution is a weak solution of \eqref{eq:forcedwave}.
	
	We then proceed with the final section in exactly the same way. Note that given our function, $u$, we can take $F'=F-\Box_g u$ smooth. Thus our solution in $H^2$, and therefore is a solution in a weak sense. However, this sense is sufficient for the applications listed in later sections.
\end{proof}

The final thing we need in order to complete existence of solutions is the following: we need to show that initial data matching our condition $(\phi_0,\phi_1)=(u,\p_\tau u)$ on $\Sigma_{\tau_0}$ is dense in $H^1(\Sigma_{\tau_0})$:
\begin{Proposition}\label{Prop:PermDense}
	Let $(\phi_0,\phi_1)$ be any pair of functions in $C^\infty_0(\Sigma_{\tau_0}) \times C^\infty_0(\Sigma_{\tau_0})$. Then there exists a sequence of globally defined functions $u_n\in C^\infty_0(\mathcal{M})\cap\Box_g^{-1}(H^1(\mathcal{M}))$ such that
	\begin{equation}
	(u_n\vert_{\Sigma_{\tau_0}},\p_\tau u_n\vert_{\Sigma_{\tau_0}})\xrightarrow{H^1(\Sigma_{\tau_0})}(\phi_0, \phi_1).
	\end{equation}
\end{Proposition}
\begin{proof}
	We first remove the region over which $g$ is not smooth. Define a smooth sequence $(\phi_{0,n},\phi_{1,n})$ such that
	\begin{equation}
	(\phi_{0,n},\phi_{1,n})\xrightarrow{H^1(\Sigma_{\tau_0})}(\phi_0, \phi_1)
	\end{equation}
	and $\p_{r}\phi_{0,n}=\phi_{1,n}=\p_{r}\phi_{1,n}=0$ for the region $[r_b-1/n,r_b+1/n]$.
	
	Let $\chi$ be a smooth cut-off function which is $1$ outside $[-2,2]$ and $0$ inside $[-1,1]$.
	
	We first construct $\phi_{1,n}$: Let $\phi_{1,n}=\phi_{1}\chi(n(r-r_b))$. It is clear that this tends to $\phi_1$ in the $H^0$ norm. It is also clear that it and its $r$ derivative are $0$ in the required region.
		
	Then we choose $\phi_{0,n}=\chi(n(r-r_b))\phi_0+(1-\chi(n(r-r_b)))\phi_0(r_b)$.
	
	It is clear to see that the $r$ derivative vanishes while $\chi=0$. All that remains is to show that 
	\begin{equation}
	\Vert \left((1-\chi(n(r-r_b)))(\phi_0-\phi_0(r_b)),0\right)\Vert_{H^1(\Sigma_{\tau_0})}\to 0.
	\end{equation}
	It is easy to see that the $L^2$ norm of this tends to $0$. Similarly the angular derivatives tend to $0$. Then all that is left to prove is that the $r$ derivative tends to 0.
	\begin{align}\nonumber
	\Vert\p_r(\phi_0-\phi_{0,n})\Vert_{L^2(\Sigma_{\tau_0})}&=\Vert(1-\chi(n(r-r_b)))\p_r\phi_0-n\chi'(n(r-r_b))(\phi_0-\phi_0(r_b))\Vert_{L^2(\Sigma_{\tau_0})}\\\nonumber
	&\leq\Vert(1-\chi(n(r-r_b)))\p_r\phi_0\Vert_{L^2(\Sigma_{\tau_0})}+\Vert n\chi'(n(r-r_b))(\phi_0-\phi_0(r_b))\Vert_{L^2(\Sigma_{\tau_0})}\\
	&\leq \Vert 1-\chi(n(r-r_b))\Vert_{L^2(\Sigma_{\tau_0})}\sup\vert\p_r\phi_0\vert+\sup_{r\in[r_b-1/n,r_b+1/n]}\vert n(\phi_0-\phi_0(r_b))\vert\Vert\chi'(n(r-r_b))\Vert_{L^2(\Sigma_{\tau_0})}.
	\end{align}
	
	The first term in the RHS tends to $0$, as $(1-\chi(n(r-r_b)))\in [0,1]$ is only supported in $r\in[r_b-2/n,r_b+2/n]$. The supremum in the second terms tends to $\vert\p_r\phi_0(r_b)\vert$, so is bounded. The $\chi'$ in the second term is bounded and only non-zero in a region whose volume tends to $0$. Therefore the whole second term also tends to $0$.
	
	Now, given the pair $(\phi_{0,n},\phi_{1,n})$, we define $u_n:=(\phi_{0,n}+\tau \phi_{1,n})(1-\chi((2n\tau)))$.
	
	As $\p_\tau r_b\leq 1$, we have that $\p_{r}u_n=\p_\tau u_n=\p_{r}\p_\tau u_n=0$ for all $r\in [r_b-1/2n,r_b+1/2n]$. We also have $(u_n,\p_\tau u_n)=(\phi_{0,n},\phi_{1,n})$ at $\tau=0$. Thus we can see
	\begin{equation}
	\Box_g u_n=\frac{1}{\sqrt{-g}}\p_a(\sqrt{-g}g^{ab}\p_b u_n)=H^1\text{-terms}+\sum_{a,b\in \{r,\tau\}}(\p_a g^{ab})\p_b u_n.
	\end{equation}
	The only terms in the sum where $\p_a g^{ab}\notin H^1_{loc}$ is at $r=r_b$. However, in that region we have $\p_r u_n=\p_{0}\tau u_n=0$. Thus $\Box_g u_n\in H^1_{loc}$. As $\phi_0$ and $\phi_1$ are compactly supported, we then obtain that $\Box_g u_n\in H^1$.
\end{proof}

Combining Propositions \ref{Prop:PermExist} and \ref{Prop:PermDense} gives us the following Theorem:

\begin{Theorem}[Existence of Permeating Solutions]\label{Thm:PermExist}
	There exists a dense subset $D$ of $H^1(\Sigma_{\tau_0})$ with the following property. Given initial data $(\phi_0,\phi_1)\in D$, there exists an $H^2_{c\forall\tau}$ solution to \eqref{eq:wave} and \eqref{eq:initial2} in the permeating case.
\end{Theorem}
\begin{proof}
	We use the subset given by
	\begin{equation}
	D=\{(u,\p_\tau u), u\in C_0^\infty(\mathcal{M})\cap\Box_g^{-1}(H^1(\mathcal{M}))\}.
	\end{equation}
	
	This is dense, by Proposition \ref{Prop:PermDense}, and has an $H^2_{c\forall \tau}$ solution by Proposition \ref{Prop:PermExist}.
\end{proof}

\begin{Remark}\label{Rmk:H1PermExist}
	As previously, Theorem \ref{Thm:PermExist} allows us to extend other results. Suppose we obtain a result on boundedness between times slices $\Sigma_{\tau}$ in the $H^1(\Sigma_{\tau})$ norm (not necessarily uniform in time). Then we can use a density argument to obtain the following: given initial data in $H^1(\Sigma_{\tau_0}))$, there exists an $H^1(\Sigma_\tau)\text{ }\forall \tau$ solution, in the sense of distributions (see already Theorem \ref{Thm:H1PermExist}).
\end{Remark}

\section{Boundedness of Solutions}\label{Sec:Bound}

We now look at showing boundedness of solutions. In this section we show that there exists a constant $C=C(M)>0$ such that for any $t^*_0\leq t^*_1\leq t^*_c$, or $\tau_0\leq\tau_1\leq \tau_{c^-}$,
\begin{align}\label{eq:reflbound}
C^{-1}\Vert\phi\Vert^2_{\dot{H}^1(\Sigma_{t^*_1})}\leq\Vert\phi\Vert^2_{\dot{H}^1(\Sigma_{t^*_0})}\leq C\Vert\phi\Vert^2_{\dot{H}^1(\Sigma_{t^*_1})}\\\label{eq:permbound}
C^{-1}\Vert\phi\Vert^2_{\dot{H}^1(\Sigma_{\tau_1})}\leq\Vert\phi\Vert^2_{\dot{H}^1(\Sigma_{\tau_0})}\leq C\Vert\phi\Vert^2_{\dot{H}^1(\Sigma_{\tau_1})}.
\end{align}
See Theorems \ref{Thm:ForwardReflBound} and \ref{Thm:BackReflBound} for the reflective case, and Theorems \ref{Thm:ForwardPermBound} and \ref{Thm:BackPermBound} for the permeating case.

Here, the first inequality in either line is \emph{forward} boundedness, i.e.~showing $\phi$ cannot grow arbitrarily large in the $\dot{H}^1$ norm as we go forward in time. This is done in the reflective/permeating case by Theorem \ref{Thm:ForwardReflBound}/\ref{Thm:ForwardPermBound}. The second inequality is \emph{backwards} boundedness, i.e.~showing $\phi$ cannot grow arbitrarily large in the $\dot{H}^1$ norm as we go backwards in time. This is done in the reflective/permeating case by Theorems \ref{Thm:BackReflBound}/\ref{Thm:BackPermBound}. These four theorems give us Theorem \ref{Thm:VagueBound} from the overview.

Statements \eqref{eq:reflbound} and \eqref{eq:permbound} can also be written  in terms of the maps, $\mathcal{F}_{t^*_0,t^*_1}$ and $\mathcal{F}_{\tau_0,\tau_1}$. These maps take Cauchy data on $\Sigma_{t^*_0,\tau_0}$ to data on $\Sigma_{t^*_1,\tau_1}$. Let $X$ be an everywhere timelike vector field which coincides with the timelike Killing field in a neighbourhood of $\mathcal{I}^-$. We then have boundedness of $\mathcal{F}_{t^*_0,t^*_1}$ and $\mathcal{F}_{\tau_0,\tau_1}$ with respect to the $X$ norm. Here the bounds do not depend on the choice of $t^*_0$, $t^*_1$ or $\tau_0$, $\tau_1$.

In the reflective case, we will show boundedness with respect to $n^{th}$ order energy (Theorems \ref{Thm:NForwardBound} and \ref{Thm:NBackwardBound}). In the permeating case, we show boundedness with respect to $2^{nd}$ order energy (Theorem \ref{Thm:2nd order energy}). Remember any higher order energy would not make sense in the permeating case, as solutions themselves do not necessarily remain in $H^n$ for arbitrary $n>2$.

Theorems \ref{Thm:H1ReflExist} and \ref{Thm:H1PermExist} show existence in the general case of $H^1(\Sigma_{t^*})$ initial data. This is as briefly discussed in Remarks \ref{Rmk:H1ReflExist} and \ref{Rmk:H1PermExist}.

\subsection{Reflective Case}

Unless stated otherwise, all theorems in this subsection refer to solutions of the wave equation \eqref{eq:wave} with reflective boundary conditions (existence shown by Theorem $\ref{Thm:ReflExist}$).

To show boundedness of our solution with respect to the non-degenerate energy, we first discuss boundedness of our solution in terms of the standard $T$-energy. Here
\begin{equation}
T=\p_{t^*}
\end{equation}
in $(t^*,r,\theta,\varphi)$ coordinates. Note that this $T$-energy becomes degenerate near $r=2M$. Once we have boundedness for this energy, we then choose a vector with non-degenerate energy near $r=2M$. This section closely follows the red shift section (section 3.3) in \cite{BHL}.

\begin{Proposition}[Forward $T$-energy Boundedness for the Reflective Case]\label{prop:Ebound}
Let $\phi\in C^\infty_{c\forall t^*}$ be  a solution of the wave equation \eqref{eq:wave} with reflective boundary conditions. We have that
\begin{equation}\label{eq:cons}
E(t^*_0)=E(t^*_1)+\frac{1}{2}\int_{S_{[t^*_0,t^*_1]}}\left[{\dot{R}^*}\left(\left(1+\frac{2M}{r}\right)\dot{R}^{*2}+\frac{4M}{r}{\dot{R}^*}-\left(1-\frac{2M}{r}\right)\right)(\p_r\phi)^2\right] r^2dt^*d\omega^2
\end{equation}
where
\begin{equation}
E(t^*)=\frac{1}{2}\int_{\Sigma_{t^*}}\left[\left(1+\frac{2M}{r}\right)(\p_{t^*}\phi)^2+\left(1-\frac{2M}{r}\right)(\p_r\phi)^2+\frac{1}{r^2}\vert\mathring{\slashed\nabla}\phi\vert^2 \right]r^2drd\omega^2\leq \Vert\phi\Vert^2_{\dot{H}^1(\Sigma_{t^*})}.
\end{equation}

The second term on the right hand side of equation \eqref{eq:cons} is non-negative. We therefore have that $E(t^*)$ is a non-increasing function of $t^*$.
\end{Proposition}	
\begin{proof}
We consider the standard $T$ energy current, $J^T$. Note that if $\phi$ is constant on $r=R^*(t^*)$, then $\p_{t^*}\phi+{\dot{R}^*}\p_r\phi=0$ and $\vert\mathring{\slashed\nabla}\phi\vert^2=0$ in any $S_{[t^*_0,t^*_1]}$ boundary terms. We then integrate $K^T$, the divergence of $J^T$, in a region bounded by $\Sigma_{t^*_0}$, $\Sigma_{t^*_1}$, and $S_{[t^*_0,t^*_1]}$ (the surface of the star between $t^*_0$ and $t^*_1$). Note this divergence is $0$, as $\p_{t^*}$ is a Killing vector field. We thus obtain exactly \eqref{eq:cons} from the resulting boundary terms. Given that $R^*$ is a strictly decreasing function, ${\dot{R}^*}<0$. As the surface of the star is timelike, we have that
\begin{equation*}
\left\vert\left(\begin{array}{c}1\\{\dot{R}^*}\\0\\0\end{array}\right)\frac{dt^*}{d\tau}\right\vert^2=\left(\left(1+\frac{2M}{r}\right)\dot{R}^{*2}+\frac{4M}{r}{\dot{R}^*}-\left(1-\frac{2M}{r}\right)\right)\left(\frac{dt^*}{d\tau}\right)^2=-1.
\end{equation*}
This implies that the $S_{[t^*_0,t^*_1]}$ boundary term in \eqref{eq:cons}  is non-negative. Therefore $E$ is a non-increasing function in $t^*$.

(Note that this method would work if the surface is given by $\{r=r_s(\tau)\}$ for any $r_s(\tau)\in H^1_{loc}$ non-increasing, timelike.)
\end{proof}

\begin{wrapfigure}{r}{5cm}
	\vspace{20mm}
	\begin{tikzpicture}[scale=0.8]
	\node (I)    at ( 0,0) {};
	
	\path
	(I) +(-90:2) coordinate (RPast)
	+(180:1) coordinate (t*)
	+(-30:6) coordinate (C)
	;
	\draw (C) to[out=130, in=0] node[midway, above, sloped] {\tiny $t^*=t^*_c$} (t*) to[out=-40, in=110] node[midway, below, sloped] {\tiny $r=R^*(t^*)$} (RPast);
	\draw (C) to[out=170, in=10] node[midway, below, sloped] {\tiny $t^*=t^*_c-\delta$} (RPast);
	\end{tikzpicture}
\end{wrapfigure}

Next we look at bounding the non-degenerate energy:

\begin{Theorem}[Forward Non-degenerate Energy Boundedness for the Reflective Case]\label{Thm:ForwardReflBound}		
Let $\phi\in C^\infty_{c\forall t^*}$ be  a solution of the wave equation \eqref{eq:wave} with reflective boundary conditions. There exists a constant $C=C(M)>0$ such that
\begin{equation}
\Vert\phi\Vert_{\dot{H}^1(\Sigma_{t^*_1})}\leq C\Vert\phi\Vert_{\dot{H}^1(\Sigma_{t^*_0})}\qquad \forall t^*_0<t^*_1.
\end{equation}
\end{Theorem}

\begin{proof}
We first note that $t^*_0\geq t^*_c$ is in the Schwarzschild exterior region. ($t^*_c$ is the time at which the surface of the star crosses $r=2M$.) Therefore \cite{BHL} gives us the result for this case. Thus if we can prove boundedness for $t^*_1\leq t^*_c$, then the result follows.

We start by choosing a suitable vector field.  Let $X=f(r)\p_{t^*}+h(r)\p_r$. Then integrate $K^X:=\nabla^\nu J^X_\nu=\nabla^\nu(X^\mu)T_{\mu\nu}$ in the region between $\Sigma_{t^*_0}$ and $\Sigma_{t^*_1}$. We proceed to look at coefficients of $\phi$'s derivatives in $K$, $dt^*(J^X)$, $d\rho(J^X)$. If $f$ and $h$ are $C^1$ functions, then the coefficients of derivatives of $\phi$ are given in the table below:

\begin{equation}\label{tbl:coeffs}
{\arraycolsep=1.4pt\def\arraystretch{2.2}
\begin{array}{ |c|c|c|c|c| } 
	\hline
 \text{Coefficients} & (\p_{t^*}\phi)^2 & (\p_{t^*}\phi)(\p_r\phi) & (\p_r\phi)^2 & \vert\mathring{\slashed\nabla}\phi\vert^2\\ 
 \hline
 K^X & \frac{2(M+r)h+4Mrf'+r(2M+r)h'}{2r^2} & \frac{r(r-2M)f'-2Mh}{r^2} & -\frac{2(r-M)h+r(r-2M)h'}{2r^2} & -\frac{h'}{2r^2} \\ 
 dt^*(J^X) & -\frac{2M+r}{2r}f & -\frac{2M+r}{r}h & \frac{4Mh-(r-2M)f}{2r} & -\frac{f}{2r^2} \\ 
 d\rho(J^X) & 0 & 0 & \frac{(1+{\dot{R}^*})({\dot{R}^*}(r+2M)-(r-2M))({\dot{R}^*}f-h)}{2r} & -\\
 \hline
\end{array}.
}
\end{equation}

Again, $\phi$ being constant on $r=R^*(t^*)$ means $\p_{t^*}\phi+{\dot{R}^*}\p_r\phi=0$ and $\vert\mathring{\slashed\nabla}\phi\vert^2=0$ on $S_{[t^*_0,t^*_1]}$. This allows us to ignore coefficients of $\vert\mathring{\slashed\nabla}\phi\vert^2$ in $d\rho(J^X)$.
Thus we can choose $f=1$ and $h$ such that:
\begin{align}
h&\leq 0\\
h&(R^*(t^*))\in [\dot{R}^*(t^*)/2,0]\\
\label{eq:hchoice}h&'\geq 0\\
\label{eq:hchoice1}h&(2M+\epsilon/2)<0\\
h&(2M+\epsilon)=0.
\end{align}

With these choices, $K^X$ is only non-zero in the compact region $\big([t^*_c-\delta,t^*_c]\times[2M,2M+\epsilon]\big)\cap\{r\geq R^*(t^*)\}$. All the coefficients of $K^X$ in table \eqref{tbl:coeffs} thus have a finite supremum. We also obtain that $-dt^*(J^X)$ is strictly positive definite, i.e.~there exists a time independent constant $\epsilon>0$ such that
\begin{equation}\label{eq:posdef}
\epsilon(-dt^*(J^X))\leq (\p_{t^*}\phi)^2+(\p_{r}\phi)^2+\frac{1}{r^2}\vert\mathring{\slashed\nabla}\phi\vert^2 \leq \epsilon^{-1}(-dt^*(J^X)).
\end{equation} 

Thus there exists an $A>0$ such that $\vert K^X\vert\leq -A\mathbbm{1}_{[t^*_c-\delta,t^*_c]} dt^*(J^X)$. Here $\mathbbm{1}_{[t^*_c-\delta,t^*_c]}$ is the indicator function for $t^*\in[t^*_c-\delta,t^*_c]$. Courtesy of our choice of $h$, we also have that $d\rho(J^X)\geq 0$.

Then, by generalised Stokes' theorem, we have that:
\begin{equation}
\int_{t^*=t^*_0}^{t^*_1}K^X=\int_{\Sigma_{t^*_0}}(-dt^*(J^X))-\int_{\Sigma_{t^*_1}}(-dt^*(J^X))-\int_{S_{[t^*_0,t^*_1]}}d\rho(J^X).
\end{equation}
Rearranging and using inequalities \eqref{eq:hchoice} and \eqref{eq:hchoice1}, we obtain
\begin{equation}
g(t^*_1):=\int_{\Sigma_{t^*_1}}(-dt^*(J^X))=\int_{\Sigma_{t^*_0}}(-dt^*(J^X))-\int_{S_{[t^*_0,t^*_1]}}d\rho(J^X)-\int_{t^*=t^*_0}^{t^*_1}K^X
\leq g(t^*_0)+\int_{t^*=t^*_0}^{t^*_1}A\mathbbm{1}_{[t^*_c-\delta,t^*_c]}g(t^*)dt^*.
\end{equation}
Thus Gronwall's inequality gives us that
\begin{equation}
\int_{\Sigma_{t^*_1}}(-dt^*(J^X))\leq\int_{\Sigma_{t^*_0}}(-dt^*(J^X))e^{A\delta}.
\end{equation}

Equation \eqref{eq:posdef} then gives us our result.
\end{proof}

We now consider the map in the backwards direction, going from $\Sigma_{t^*_c}$ down to $\Sigma_{t^*_c-s_{max}}$.

\begin{Lemma}[Finite in Time Backwards Bound in the Reflecting Case]\label{lm:backwardbound}
Let $\phi\in C^\infty_{c\forall t^*}$ be a solution to the wave equation \eqref{eq:wave} with reflective boundary conditions. Let $s_{max}\geq 0$ be any positive constant. Then there exists a constant $C=C(M,s_{max})>0$ such that:
\begin{equation}
\Vert\phi\Vert_{\dot{H}^1(\Sigma_{t^*_c-s})}\leq C\Vert\phi\Vert_{\dot{H}^1(\Sigma_{t^*_c})} \qquad\forall s\in[0,s_{max}].
\end{equation}
\end{Lemma}

\begin{proof}
We again start by letting $X=f(r)\p_{t^*}+h(r)\p_r$, and considering at table \eqref{tbl:coeffs}. As we are now going backwards in time, we require that $d\rho(J^X)\leq 0$. However, we still require $-\int_{\Sigma_{t^*}}dt^*(J^X)\sim\Vert\phi\Vert_{\dot{H}^1(\Sigma_{t^*})}^2$, and that the coefficients of $K^X$ are bounded. For this we pick $h=\inf_{t^*\in[t^*_c-s_{max},t^*_c]}{\dot{R}^*}$, $f=1$. Note $f'=0=h'$. Then we have that the following coefficients:
\begin{equation}
{\arraycolsep=1.4pt\def\arraystretch{2.2}
	\begin{array}{ |c|c|c|c|c| } 
		\hline
		\text{Coefficients} & (\p_{t^*}\phi)^2 & (\p_{t^*}\phi)(\p_r\phi) & (\p_r\phi)^2 & \vert\mathring{\slashed\nabla}\phi\vert^2\\ 
		\hline
		K^X & \frac{2(M+r)h}{2r^2} & \frac{-2Mh}{r^2} & -\frac{2(r-M)h}{2r^2} & 0 \\ 
		dt^*(J^X) & -\frac{2M+r}{2r} & -\frac{2M+r}{r}h & \frac{4Mh-(r-2M)}{2r} & -\frac{1}{2r^2} \\ 
		d\rho(J^X) & 0 & 0 & \frac{(1+{\dot{R}^*})({\dot{R}^*}(r+2M)-(r-2M))({\dot{R}^*}-h)}{2r} & 0\\
		\hline
	\end{array}.
}
\end{equation}

Then $-dt^*(J^X)$ is again strictly positive definite, so obeys equation \eqref{eq:posdef}. The coefficients of $K^X$ are again bounded, so there exists a $A$ such that $\vert K^X\vert\leq-Adt^*(J^X)$. Equation \eqref{eq:timelike} gives us that $h\in(0,1)$. Thus we have that $-\int_{\Sigma_{t^*}}dt^*(J^X)\sim \Vert\phi\Vert_{\dot{H}^1(\Sigma_{t^*})}$, where the constants are only dependent on $s_{max}$. Then let\begin{equation}
g(s):=-\int_{\Sigma_{t^*_c-s}}dt^*(J^X).
\end{equation}

Integrating $K^X$ over the area $t\in[t^*_c-s_1,t^*_c-s_0]$, we have: 
\begin{equation}
g(s_1)= g(s_0)+\int_{S_{[t^*_c-s_1,t^*_c-s_0]}}d\rho(J^X)+\int_{t\in[t^*_c-s_1,t^*_c-s_0]}K^X\leq g(s_0)+A\int_{s'=s_0}^{s_1}g(s)ds.
\end{equation}
Then by Gronwall's Inequality,
\begin{equation}
g(s)\leq g(0)e^{As_{max}}.
\end{equation}
Again, as $-\int_{\Sigma_{t^*}}dt^*(J^X)\sim\Vert\phi\Vert_{\dot{H}^1(\Sigma_{t^*})}^2$, we are done.
\end{proof}

Lemma \ref{Lem:BackExpBound} and Theorem \ref{Thm:ForwardReflBound} give us the conditions mentioned in Remark \ref{Rmk:H1ReflExist}, so we have the following Theorem:
\begin{Theorem}[$H^1$ Existence of Reflective Solutions]\label{Thm:H1ReflExist}
	Let $(\phi_0,\phi_1)\in H^1(\Sigma_{t^*_0})$, where $t^*_0\leq t^*_c$. There exists a solution $\phi$ to the wave equation \eqref{eq:wave} with reflective boundary conditions such that
	\begin{equation}
	(\phi\vert_{\Sigma_{t^*_0}},\p_{t^*}\phi\vert_{\Sigma_{t^*_0}})=(\phi_0,\phi_1).
	\end{equation}
	Here this restriction holds in a trace sense, and $\phi$ is a solution in the sense of distributions. Finally, $\phi\in \dot{H}^1(\Sigma_{t^*})$ for all $t^*\leq t^*_c$.
\end{Theorem}
\begin{proof}
	This is a result of Theorem \ref{Thm:ReflExist} and a density argument. This density argument relies on the bounds given by Lemma \ref{Lem:BackExpBound} and Theorem \ref{Thm:ForwardReflBound}.
\end{proof}
\begin{Remark}
	Note that our existence result, Theorem \ref{Thm:H1ReflExist} allows us to define the forwards map:
	\begin{align}
	\mathcal{F}_{(t_0^*,t_1^*)}&:\mathcal{E}^X_{\Sigma_{t^*_0}}\to \mathcal{E}^X_{\Sigma_{t^*_1}}\\
	\mathcal{F}_{(t_0^*,t_1^*)}\left(\phi_0,\phi_1\right)&:=\left(\phi\vert_{\Sigma_{t^*_1}},\p_{t^*}\phi\vert_{\Sigma_{t^*_1}}\right) \quad \text{where $\phi$ is the solution to \eqref{eq:wave} and \eqref{eq:initial}}.
	\end{align}
	Then Theorem \ref{Thm:ForwardReflBound} gives boundedness of $\mathcal{F}_{(t_0^*,t_1^*)}$:
	\begin{equation}
	\Vert\mathcal{F}_{(t^*_0,t^*_1)}\left(\phi_0,\phi_1\right)\Vert_X\leq C\Vert(\phi_0,\phi_1)\Vert_X\quad \forall t^*_0\leq t^*_1\leq t^*_c
	\end{equation}
	for some $C=C(M)>0$.
\end{Remark}
Now we wish to obtain a bound for our solution which does not depend on the time interval we are looking in.

\begin{Theorem}[Backwards Non-degenerate Energy Boundedness for the Reflective Case]\label{Thm:BackReflBound}
Let $\phi$ be a solution, to the wave equation \eqref{eq:wave} with reflective boundary conditions (as given by Theorem \ref{Thm:H1ReflExist}). There exists a constant $C=C(M)>0$ such that
	\begin{equation}
	\Vert\phi\Vert_{\dot{H}^1(\Sigma_{t^*_0})}^2\leq C\Vert\phi\Vert_{\dot{H}^1(\Sigma_{t^*_1})}^2\qquad \forall t^*_0<t^*_1<t^*_c.
	\end{equation}
\end{Theorem}
\begin{Remark}
	The main properties of $R^*$ that will be used in this proof are $R^*\to\infty$ and $\dot{R}^*\to 0$ as $t^*\to-\infty$.
\end{Remark}
\begin{Remark}
	This Theorem can also be written that there exists $C=C(M)>0$ s.t.
	\begin{equation}
	C\Vert\mathcal{F}_{(t^*_0,t^*_1)}\left(\phi_0,\phi_1\right)\Vert_X\geq \Vert(\phi_0,\phi_1)\Vert_X\quad \forall t^*_0\leq t^*_1\leq t^*_c
	\end{equation}
\end{Remark}
\begin{proof}
	The previous theorem gives our result for any finite distance back in time. Thus we only need to show uniform boundedness for all $t^*<t^*_2$ for any sufficiently far back $t^*_2$. Let $X=f(t^*)\p_{t^*}+h\p_r$, for $h$ constant. We then consider the modified current $J^{X,w}$, given by \eqref{eq:modcurrent}. If we choose $w=h/2r$, then we have:
	\begin{align}\nonumber
	-dt^*(J^{X,w})=&\left(1+\frac{2M}{r}\right)\frac{f}{2}(\p_{t^*}\phi)^2+\left(1+\frac{2M}{r}\right)h\p_{t^*}\phi\p_r\phi+\left(\left(1-\frac{2M}{r}\right)\frac{f}{2}-\frac{2M}{r}h\right)(\p_r\phi)^2+\frac{f}{2r^2}\vert\mathring{\slashed\nabla}\phi\vert^2\\\label{eq:Generalt^*}
	&+\left(1+\frac{2M}{r}\right)\frac{h}{r}\phi\p_{t^*}\phi-\frac{2M}{r^2}h\phi\p_r\phi-\Bigg(\left(1+\frac{2M}{r}\right)\p_{t^*}\left(\frac{h}{2r}\right)-\frac{2M}{r}\p_r\left(\frac{h}{2r}\right)\Bigg)\phi^2.
	\end{align}
	\begin{equation}\label{eq:GeneralBoundary}
	d\rho(J^{X,w})=\frac{1}{2r}(1+{\dot{R}^*})\left((r+2M){\dot{R}^*}-(r-2M)\right)({\dot{R}^*}f-h)(\p_r\phi)^2.
	\end{equation}
	\begin{align}\nonumber
	K^{X,w}=&-\left(\left(1+\frac{2M}{r}\right)\frac{\dot{f}}{2}+\frac{Mh}{r^2}\right)(\p_{t^*}\phi)^2-\left(\left(1-\frac{2M}{r}\right)\frac{\dot{f}}{2}+\frac{Mh}{r^2}\right)(\p_r\phi)^2\\\label{eq:GeneralK}
	&-\frac{1}{r^2}\left(\frac{\dot{f}}{2}-\frac{h}{r}\right)\vert\mathring{\slashed\nabla}\phi\vert^2-\Box_g\left(\frac{h}{2r}\right)\phi^2.
	\end{align}
	
	We choose
	\begin{equation}
	f(t^*)=1+\frac{1}{\log\left(\frac{R^*}{2M}\right)}\qquad h(t^*,r)=-\epsilon,\qquad \epsilon\in(0,1).
	\end{equation}
	Given $\Box_g(h/2r)=\frac{M\epsilon}{r^4}$, we then obtain
	\begin{align}\nonumber
	\int_{S^2}K^{X,w}=&-\int_{S^2}\Bigg(\left(\left(1+\frac{2M}{r}\right)\frac{-{\dot{R}^*}}{2R^*(\log\left(\frac{R^*}{2M}\right))^2}-\frac{M\epsilon}{r^2}\right)(\p_{t^*}\phi)^2+\left(\left(1-\frac{2M}{r}\right)\frac{-{\dot{R}^*}}{2R^*(\log\left(\frac{R^*}{2M}\right))^2}-\frac{M\epsilon}{r^2}\right)(\p_r\phi)^2\\
	&+\frac{1}{r^2}\left(\frac{-{\dot{R}^*}}{2R^*(\log\left(\frac{R^*}{2M}\right))^2}+\frac{\epsilon}{r}\right)\vert\mathring{\slashed\nabla}\phi\vert^2+\frac{M\epsilon}{r^4}\phi^2\Bigg).
	\end{align}

	By choosing $t^*$ large and negative enough, we have $R^*(t^*)$ is arbitrarily large. Thus we have that $\int_{S^2}K^{X,w}\leq0$ for sufficiently negative $t^*$. For large enough negative $t^*$, we have $-{\dot{R}^*}<\epsilon/2$. As $f$ is bounded above (by $2$, for example), then ${\dot{R}^*}f-h>0$. Thus from equation \eqref{eq:GeneralBoundary}, we obtain that $d\rho(J^{X,w})\leq 0$. 
	
	Finally, we look at $dt^*(J^{X,w})$:
	\begin{align}\nonumber
	-\int_{\Sigma_{t^*}}dt^*(J^{X,w})=&\int_{\Sigma_{t^*}}\left(1+\frac{2M}{r}\right)\frac{f}{2}(\p_{t^*}\phi)^2-\left(1+\frac{2M}{r}\right)\epsilon\p_{t^*}\phi\p_r\phi+\left(\left(1-\frac{2M}{r}\right)\frac{f}{2}+\frac{2M}{r}\epsilon\right)(\p_r\phi)^2\\\nonumber
	&+\frac{f}{2r^2}\vert\mathring{\slashed\nabla}\phi\vert^2-\left(1+\frac{2M}{r}\right)\frac{\epsilon}{r}\phi\p_{t^*}\phi+\frac{2M}{r^2}\epsilon\phi\p_r\phi+\left(\frac{M\epsilon}{r^3}\right)\phi^2\\\nonumber
	=&\int_{\Sigma_{t^*}}\left(1+\frac{2M}{r}\right)\frac{f}{2}(\p_{t^*}\phi)^2-\left(1+\frac{2M}{r}\right)\epsilon\p_{t^*}\phi\p_r\phi+\left(\left(1-\frac{2M}{r}\right)\frac{f}{2}+\frac{M}{r}\epsilon\right)(\p_r\phi)^2\\\nonumber
	&+\frac{f}{2r^2}\vert\mathring{\slashed\nabla}\phi\vert^2-\left(1+\frac{2M}{r}\right)\frac{\epsilon}{r}\phi\p_{t^*}\phi+\frac{M\epsilon}{r^3}(r\p_r\phi+\phi)^2\\\nonumber
	=&\int_{\Sigma_{t^*}}\left(1+\frac{2M}{r}\right)\frac{\epsilon}{2}\left(\p_{t^*}\phi-\p_r\phi-\frac{\phi}{r}\right)^2+\left(1+\frac{2M}{r}\right)\frac{f-\epsilon}{2}(\p_{t^*}\phi)^2+\left(\left(1-\frac{2M}{r}\right)\frac{f}{2}-\frac{\epsilon}{2}\right)(\p_r\phi)^2\\\nonumber
	&+\frac{f}{2r^2}\vert\mathring{\slashed\nabla}\phi\vert^2-\left(1+\frac{2M}{r}\right)\frac{\epsilon}{r}\phi\p_r\phi+\frac{M\epsilon}{r^3}(r\p_r\phi+\phi)^2\\\nonumber
	=&\int_{\Sigma_{t^*}}\left(1+\frac{2M}{r}\right)\frac{\epsilon}{2}\left(\p_{t^*}\phi-\p_r\phi-\frac{\phi}{r}\right)^2+\left(1+\frac{2M}{r}\right)\frac{f-\epsilon}{2}(\p_{t^*}\phi)^2+\left(\left(1-\frac{2M}{r}\right)\frac{f}{2}-\frac{\epsilon}{2}\right)(\p_r\phi)^2\\
	&+\frac{f}{2r^2}\vert\mathring{\slashed\nabla}\phi\vert^2+\frac{\epsilon}{2r^2}\phi^2+\frac{M\epsilon}{r^3}(r\p_r\phi+\phi)^2.
	\end{align}
	Here we have integrated the $\phi\p_r\phi$ term by parts (remembering that there is an $r^2$ term in the volume form).
	
	We can bound the $\phi$ terms by using the following version of Hardy's inequality:
	\begin{equation}\label{eq:Hardy}
	\exists C>0\quad s.t.\quad\int_{\Sigma_{t^*}}\left(\frac{\phi}{r}\right)^2\leq C\int_{\Sigma_{t^*}}(\p_r\phi)^2,
	\end{equation}
	where $C$ is independent of $t^*$.
	
	Using \eqref{eq:Hardy}, we have
	\begin{equation}
	0\leq\int_{\Sigma_{t^*}}\left(1+\frac{2M}{r}\right)\frac{\epsilon}{2}\left(\p_{t^*}\phi-\p_r\phi-\frac{\phi}{r}\right)^2
	+\frac{\epsilon}{2r^2}\phi^2+\frac{M\epsilon}{r^3}(r\p_r\phi+/\phi)^2\leq C\Vert\phi\Vert_{\dot{H}^{1}(\Sigma_{t^*})}^2
	\end{equation}
	for a $t^*$ independent constant $C$.
	
	Since $f>1$ and $\epsilon<1$, we have $-dt^*(J^{X,w})\sim\Vert\phi\Vert_{\dot{H}^{1}(\Sigma_{t^*})}^2$. Thus by generalised Stokes' theorem, we have boundedness of the solution:
	\begin{align}\nonumber
	\Vert\phi\Vert_{\dot{H}^{1}(\Sigma_{t^*_0})}^2\leq A\int_{\Sigma_{t^*_0}}(-dt^*(J^{X,w}))&=A\left(\int_{\Sigma_{t^*_0}}(-dt^*(J^{X,w}))+\int_{S_{[t^*_0,t^*_1]}}d\rho(J^X)+\int_{t^*\in[t^*_0,t^*_1]}K^{X,w}\right)\\
	&\leq A\int_{\Sigma_{t^*_1}}(-dt^*(J^{X,w}))\leq C\Vert\phi\Vert_{\dot{H}^{1}(\Sigma_{t^*_1})}^2.
	\end{align}
\end{proof}

\begin{Corollary}\label{Cor:vreflbound}
	Let $\phi$ be a solution, to the wave equation \eqref{eq:wave} with reflective boundary conditions (as given by Theorem \ref{Thm:H1ReflExist}). There exists a constant $C=C(M)>0$, such that
	\begin{equation}
	-\int_{\Sigma_{v_0}}dv(J^{X,w})\leq-\int_{\Sigma_{v_1}}dv(J^{X,w})\leq C\Vert\phi\Vert_{\dot{H}^1(\Sigma_{t^*_0})}^2\quad\forall v_0\leq v_1\leq t^*_0.
	\end{equation}
\end{Corollary}
\begin{proof}
	This is done with the exact same currents as in Theorem \ref{Thm:BackReflBound}. For the second inequality (and to show that the integral over $\Sigma_{v_1}$ exists), we integrate between $\Sigma_{t^*_0}$, $\Sigma_{v_1}$, and $\Sigma_{u}$, and then allow $u\to\infty$. Note that the integral on $\Sigma_u$ always has the correct sign, and that $\Sigma_{v_0}$ is entirely in the past of $\Sigma_{t^*}$ for $v_1\leq t^*_0$. For the first inequality, we just integrate over the region between $\Sigma_{v_0}$, $\Sigma_{v_1}$, and $\Sigma_{u}$, and again allow $u\to\infty$.
\end{proof}
We now try to extend Theorems \ref{Thm:ForwardReflBound} and \ref{Thm:BackReflBound} to $\dot{H}^n$ norms. To do this, we will need the following 3-part Lemma:
\begin{Lemma}\label{BigLem}	
	Let $\phi\in C^\infty_{c\forall t^*}$ be a solution to the wave equation \eqref{eq:wave} with reflective boundary conditions. We have the following results:
	\begin{enumerate}
		 \item Let $\Omega_i$ be the angular Killing vector fields earlier (see \eqref{eq:AngularKillingFields}). Then $\Box_g\left(\frac{1}{r^{\vert p\vert}}\Omega^p\p_r^m\p_{t^*}^{n-1-m-\vert p\vert}\phi\right)$ only contains at most $n^{th}$ order derivatives. Furthermore, all coefficients of these derivatives are smooth and have all their derivatives bounded. Thus there exists a constant $D=D(M, n)>0$ such that 
		\begin{equation}
		\left\Vert\Box_g\left(\frac{1}{r^{\vert p\vert}}\Omega^p\p_r^m\p_{t^*}^{n-1-m-\vert p\vert}\phi\right)\right\Vert_{L^2(\Sigma_{t^*})}\leq D\Vert\phi\Vert_{\dot{H}^{n}(\Sigma_{t^*})}.
		\end{equation}
		\item There exists a $t^*_0\leq t^*_c$ and a constant $C=C(M,t^*_0)>0$ such that 
		\begin{equation}
		C\left(\Vert\bar{\p}_{t^*}^n\phi\Vert^2_{\dot{H}^1(\Sigma_{t^*})}+\Vert\phi\Vert^2_{\dot{H}^{n-1}(\Sigma_{t^*})}\right)\geq\Vert\phi\Vert^2_{\dot{H}^n(\Sigma_{t^*})} \quad t^*\leq t^*_0.
		\end{equation}
		Here $\bar{\p}_{t^*}$ is the $t^*$ derivative with respect to $(t^*,\rho=r-R^*(t^*)+2M,\theta,\varphi)$ coordinates, as given in \eqref{eq:RhoCoords}.
		\item Given any finite time $t^*_0\leq t^*_1\leq t^*_c$, there exists a constant $A=A(n,t^*_0,t^*_1,M)$ such that 
		\begin{equation}
		\frac{1}{A} \Vert\phi\Vert_{\dot{H}^n(\Sigma_{t^*_0})}^2\leq \Vert\phi\Vert^2_{\dot{H}^n(\Sigma_{t^*_1})}\leq A \Vert\phi\Vert_{\dot{H}^n(\Sigma_{t^*_0})}^2.
		\end{equation}
	\end{enumerate}
\end{Lemma}
\begin{Remark}
	Note that when calculating $\Vert\psi\Vert_{\dot{H}^1(\Sigma_{t^*})}^2$, we can use the $\bar{\p}_{t^*}$ derivative in place of the $\p_{t^*}$ derivative. This is due to the fact that these norms can differ by at most a factor of $2$, since
	\begin{equation}
	\Vert\p_{t^*}\psi\Vert_{L^2(\Sigma_{t^*})}-\Vert\p_{r}\psi\Vert_{L^2(\Sigma_{t^*})}\leq\Vert(\p_{t^*}-\dot{R}^*\p_r)\psi\Vert_{L^2(\Sigma_{t^*})}=\Vert\bar{\p}_{t^*}\psi\Vert_{L^2(\Sigma_{t^*})}\leq\Vert\p_{t^*}\psi\Vert_{L^2(\Sigma_{t^*})}+\Vert\p_{r}\psi\Vert_{L^2(\Sigma_{t^*})}.
	\end{equation}
	This in turn implies
	\begin{equation}
	\frac{1}{2}\left(\Vert\p_{t^*}\psi\Vert_{L^2(\Sigma_{t^*})}^2+\Vert\p_{r}\psi\Vert_{L^2(\Sigma_{t^*})}^2\right)\leq\Vert\bar{\p}_{t^*}\psi\Vert_{L^2(\Sigma_{t^*})}^2+\Vert\p_{r}\psi\Vert_{L^2(\Sigma_{t^*})}^2\leq 2\left(\Vert\p_{t^*}\psi\Vert_{L^2(\Sigma_{t^*})}^2+\Vert\p_{r}\psi\Vert_{L^2(\Sigma_{t^*})}^2\right).
	\end{equation}
\end{Remark}
\begin{proof}
	\begin{enumerate}
	\item Note $\Omega_i$ and $\p_{t^*}$ commute with $\Box_g$. Thus for this part we only need to check $\Box_g\left(\frac{1}{r^{\vert p\vert}}\p_r^{n-1}\phi\right)$ explicitly. Using the fact that $\Box_g\phi=0$, we obtain:
	\begin{align}
	\Box_g\left(\frac{1}{r^{\vert p\vert}}\p_r^{n-1}\phi\right)=\frac{n-1}{r^{2+\vert p \vert}}\Big((n-2)\p_{t^*}^2\p_r^{n-3}\phi+2(r-M)\p_{t^*}^2\p_r^{n-2}\phi-4M\p_{t^*}\p_r^{n-1}\phi&-n\p_r^{n-1}\phi-2(r-M)\p_r^n\phi\Big)\\\nonumber
	&-\frac{\vert p\vert}{r^{1+\vert p \vert}}\nabla^\mu r\nabla_\mu\p_r^{n-1}\phi.
	\end{align}
	
	Given in the above case, $\vert p \vert \leq n-1$, then we have our result.
	
	\item We first look at how the wave operator commutes with $\bar{\p}_{t^*}$: 
	\begin{align}\label{eq:BoxRho}
	\Box_g\left(\frac{1}{r^{\vert p\vert}}\Omega^p\bar{\p}_{t^*}^{n-\vert p\vert}(\phi)\right)= \Box_g\left(\frac{1}{r^{\vert p\vert}}\Omega^p\left(\p_{t^*}+\dot{R}^*\p_r\right)^{n-\vert p\vert}\phi\right)=\sum_{m=0}^{n-\vert p\vert}&\left(\begin{matrix} {n-\vert p\vert}\\m	\end{matrix}\right)\frac{\dot{R}^{*m}}{r^{\vert p\vert}}\Box_g\left(\Omega^p\p_{t^*}^{{n-\vert p\vert}-m}\p_r^m\phi\right)\\\nonumber&+\left(\text{Bounded lower order terms}\right).
	\end{align}
	
	As $\p_{t^*}$ and $\Omega$ commute with $\Box_g$, we can ignore the $m=0$ term in the sum. Then, by the first part of the Lemma, we can bound the right hand side of \eqref{eq:BoxRho}. It is bounded by $\vert\dot{R}^*\vert$ times a constant multiple of the $\dot{H}^{n+1}$ norm, plus lower order terms:
	\begin{equation}
	\left\Vert\Box_g\left(\frac{1}{r^{\vert p\vert}}\Omega^p\bar{\p}_{t^*}^{n-\vert p\vert}(\phi)\right)\right\Vert_{L^2(\Sigma_{t^*})}^2\leq D\left(\Vert\phi\Vert_{\dot{H}^{n}(\Sigma_{t^*})}^2+\dot{R}^{*2}\Vert\phi\Vert_{\dot{H}^{n+1}(\Sigma_{t^*})}^2\right).
	\end{equation}
	
	We also have that $\Omega^p\bar{\p}_{t^*}^n(\phi)=0$ on the boundary of the star. Thus we can then proceed by using an elliptic estimate (such as in \cite{LecturesOnGemoetryOfManifolds}) on $\bar{\p}_{t^*}^n\phi$. 
	
	We consider the elliptic operator, $L$, given by
	\begin{equation}
	L\psi:=\left(1-\frac{2M}{r}-\frac{4M\dot{R}^*}{r}-\left(1+\frac{2M}{r}\right)\dot{R}^{*2}\right)\p_r^2\psi+\frac{1}{r^2}\mathring{\slashed\triangle}\psi=f(t^*,r)\p_r^2\psi+\frac{1}{r^2}\mathring{\slashed\triangle}\psi.
	\end{equation}
	
	Thus we have
	\begin{align}\nonumber
	\int_{\Sigma_{t^*}}(L\psi)^2=\int_{\Sigma_{t^*}}&f^2(\p_r\psi)^2+\frac{2f}{r^2}\p_r^2\psi\mathring{\slashed\triangle}\psi+\frac{1}{r^4}(\mathring{\slashed\triangle}\psi)^2\\\label{eq:LLowerBound}
	=\int_{\Sigma_{t^*}}&f^2(\p_r^2\psi)^2+\frac{2f}{r^2}\vert\p_r\mathring{\slashed\nabla}\psi\vert^2+\frac{2\p_rf}{r^2}\mathring{\slashed\nabla}\psi.\p_r\mathring{\slashed\nabla}\psi+\frac{1}{r^4}\vert\mathring{\slashed\nabla}\mathring{\slashed\nabla}\psi\vert^2-\int_{S_{t^*}}\frac{2f}{r^2}\mathring{\slashed\nabla}\psi.\p_r\mathring{\slashed\nabla}\psi\\\nonumber
	\geq\int_{\Sigma_{t^*}}&\frac{1}{2}(\p_r^2\psi)^2+\frac{1}{2r^2}\vert\p_r\mathring{\slashed\nabla}\psi\vert^2-\frac{C}{r^2}\vert\mathring{\slashed\nabla}\psi\vert^2+\frac{1}{r^4}\vert\mathring{\slashed\nabla}\mathring{\slashed\nabla}\psi\vert^2-\int_{S_{t^*}}\frac{2f}{r^2}\mathring{\slashed\nabla}\psi.\p_r\mathring{\slashed\nabla}\psi\\\nonumber
	\geq\frac{1}{2}\Vert\psi&\Vert_{\dot{H}^2(\Sigma_{t^*})}^2-\frac{1}{2}\Vert\p_{t^*}\psi\Vert_{\dot{H}^1(\Sigma_{t^*})}^2-C\Vert\psi\Vert_{\dot{H}^1(\Sigma_{t^*})}^2-\int_{S_{t^*}}\frac{2f}{r^2}\mathring{\slashed\nabla}\psi.\p_r\mathring{\slashed\nabla}\psi.
	\end{align}
	
	By rearranging equation \eqref{eq:wave} in coordinates given by \eqref{eq:RhoCoords}, we have that
	\begin{equation}\label{eq:LUpperBound}
	\Vert L\psi\Vert_{L^2(\Sigma_{t^*})}^2\leq C\left(\Vert\bar{\p}_{t^*}\psi\Vert_{\dot{H}^{1}(\Sigma_{t^*})}^2+\Vert\Box_g\psi\Vert_{L^2(\Sigma_{t^*})}^2+\Vert\psi\Vert_{\dot{H}^{1}(\Sigma_{t^*})}^2\right).
	\end{equation}
	
	Combining \eqref{eq:LLowerBound} and \eqref{eq:LUpperBound} with $\psi=\bar{\p}_{t^*}^n\phi$ ($=0$ on $S_{t^*}$), and noting that
	\begin{equation}
	\Vert\p_{t^*}\psi\Vert_{\dot{H}^{1}(\Sigma_{t^*})}^2\leq C\left(\Vert\bar{\p}_{t^*}\psi\Vert_{\dot{H}^{1}(\Sigma_{t^*})}^2+\Vert\psi\Vert_{\dot{H}^{1}(\Sigma_{t^*})}^2\right),
	\end{equation}
	we obtain
	\begin{equation}\label{eq:1stEllipticBound}
	\Vert \bar{\p}_{t^*}^n\phi\Vert^2_{\dot{H}^2(\Sigma_{t^*})}\leq C\left(\Vert\bar{\p}_{t^*}^{n+1}\phi\Vert^2_{\dot{H}^1(\Sigma_{t^*})}+\Vert\phi\Vert^2_{\dot{H}^{n}(\Sigma_{t^*})}+\dot{R}^{*2}\Vert\phi\Vert_{\dot{H}^{n+1}(\Sigma_{t^*})}^2\right).
	\end{equation}
	
	We then look at $\psi=\frac{1}{r^{\vert p \vert}}\Omega^p\bar{\p}_{t^*}^{n-1}(\phi)$, where $p$ is a multi-index of size $1$ (as this also vanishes on $S_{t^*}$). As the $L^2$ norms of $\bar{\p}_{t^*}^2\psi$ and $\bar{\p}_{t^*}\p_{r}\psi$ are bounded by the left hand side of \eqref{eq:1stEllipticBound}, we repeat the above argument to get that
	\begin{equation}\label{2ndEllipticBound}
	\Vert \bar{\p}_{t^*}^n\phi\Vert^2_{\dot{H}^2(\Sigma_{t^*})}+\Vert \frac{1}{r^{\vert p \vert}}\Omega^p\bar{\p}_{t^*}^{n-\vert p\vert}(\phi)\Vert^2_{\dot{H}^2(\Sigma_{t^*})}\leq C\left(\Vert\bar{\p}_{t^*}^{n+1}\phi\Vert^2_{\dot{H}^1(\Sigma_{t^*})}+\Vert\phi\Vert^2_{\dot{H}^{n}(\Sigma_{t^*})}+\dot{R}^{*2}\Vert\phi\Vert_{\dot{H}^{n+1}(\Sigma_{t^*})}^2\right),
	\end{equation}
	for $\vert p\vert=1$. 
	
	We repeat this argument $n$ times to obtain that \eqref{2ndEllipticBound} is true for all $\vert p\vert\leq n$. The coefficient of $\p^2_r$ in \eqref{eq:wave} (with respect to the coordinates in \eqref{eq:RhoCoords}) is bounded above and away from $0$. This means, we can rearrange \eqref{eq:wave}, to bound all $r$ derivatives to obtain:
	\begin{equation}
	\Vert\phi\Vert_{\dot{H}^{n+2}(\Sigma_{t^*})}^2\leq C\left(\Vert\bar{\p}_{t^*}^{n+1}\phi\Vert^2_{\dot{H}^1(\Sigma_{t^*})}+\Vert\phi\Vert^2_{\dot{H}^{n}(\Sigma_{t^*})}+\dot{R}^{*2}\Vert\phi\Vert_{\dot{H}^{n+1}(\Sigma_{t^*})}^2\right).
	\end{equation}
	
	If we then choose $t^*_0$ such that $\dot{R}^{*2}C<1$, then we can rearrange the above to get the required result.
	
	\item We proceed in a very similar way to our previous results for finite-in-time boundedness; we use energy currents, Stokes' theorem, and then Gronwall's inequality. For this case, our energy currents will be
	\begin{equation}
	\sum_{n=1}^{n=N}\sum_{\vert p\vert=0}^{n-1}J^X\left(\frac{1}{r^{\vert p\vert}}\Omega^p\bar{\p}_{t^*}^{n-1-\vert p\vert}\phi\right)
	\end{equation}
	where we choose $X=\p_{t^*}+\dot{R}^*\p_r$ (timelike), so $-\int_{\Sigma_{t^*}}dt^*(J^X)\sim\Vert.\Vert_{\dot{H}^1(\Sigma_{t^*})}^2$. Note here that $\Omega$ are our angular Killing vector fields, and $p$ is a multi-index. Now, as $\Box_g\frac{1}{r^{\vert p\vert}}\Omega^p\bar{\p}_{t^*}^{n-1-\vert p \vert}\phi\neq 0$, we obtain an extra term in our bulk integral:
	\begin{multline}\label{eq:Stokes'dr}
	\int_{t^*=t^*_0}^{t^*_1}\left(K^X+X\left(\frac{1}{r^{\vert p\vert}}\Omega^p\bar{\p}_{t^*}^{n-1-\vert p \vert}\phi\right)\Box_g\left(\frac{1}{r^{\vert p\vert}}\Omega^p\bar{\p}_{t^*}^{n-1-\vert p \vert}\phi\right)\right)\\=\int_{\Sigma_{t^*_0}}(-dt^*(J^X))-\int_{\Sigma_{t^*_1}}(-dt^*(J^X))-\int_{S_{[t^*_0,t^*_1]}}d\rho(J^X).
	\end{multline}
	
	As in part 2, we have that the coefficients of $\p_r^2$ in \eqref{eq:wave} are bounded above and away from $0$. Suppose we have bounded the $L^2$ norms of all derivatives up to $N^{th}$ order that have fewer that $2$ $r$ derivatives. We can then use \eqref{eq:wave} to bound the remaining derivatives up to $N^{th}$ order.
	
	Now we consider the new second term in \eqref{eq:Stokes'dr}. The first part of this Lemma gives us that the sum of these additional term can be bounded by
	\begin{align}\label{eq:BigCurrentBound}
	\left\vert\sum_{n=1}^{n=N}\sum_{\vert p\vert=0}^{n-1}X\left(\frac{1}{r^{\vert p\vert}}\Omega^p\bar{\p}_{t^*}^{n-1-\vert p \vert}\phi\right)\Box_g\left(\frac{1}{r^{\vert p\vert}}\Omega^p\bar{\p}_{t^*}^{n-1-\vert p \vert}\phi\right)\right\vert\leq C\int_{t^*=t^*_0}^{t^*_1}\Vert\phi\Vert^2_{\dot{H}^N(\Sigma_{t^*})}dt^*\\\nonumber\leq -C'\sum_{n=1}^{n=N}\sum_{\vert p\vert=0}^{n-1}\int_{t^*=t^*_0}^{t^*_1}\int_{\Sigma_{t^*}}dt^*\left(J^X\left(\frac{1}{r^{\vert p\vert}}\Omega^p\bar{\p}_{t^*}^{n-1-\vert p \vert}\phi\right)\right),
	\end{align}
	where we have used that $\Box_g$ commutes with $\p_{t^*}$ and each $\Omega_i$.
	As usual, we can then bound the $K^X$ terms by a multiple of this.
	
	Finally, we note that $\Omega^p\bar{\p}_{t^*}^{n-1-\vert p \vert}\phi=0$ on $S_{[t^*_0,t^*_1]}$, and $X$ is tangent to this surface. Therefore  $d\rho(J^X)$ also vanishes. Thus from equation \eqref{eq:Stokes'dr}, we obtain
	\begin{equation}
	g(t^*_1):=\sum_{n=1}^{n=N}\sum_{\vert p\vert=0}^{n-1}\int_{\Sigma_{t^*_0}}(-dt^*(J^X(\Omega^p\bar{\p}_{t^*}^{n-1-\vert p\vert}\phi))\leq g(t^*_0)+C\int_{t^*=t^*_0}^{t^*_1}g(t^*)dt^*,
	\end{equation}
	\begin{equation}
	g(t^*_0)\leq g(t^*_1)+C\int_{t^*=t^*_0}^{t^*_1}g(t^*)dt^*.
	\end{equation}
	Then, in a similar manner to Gronwall's inequality, we will show $g(t^*) \leq e^{c(t^*-t^*_0)}g(t^*_0)$ for $t^*\geq t^*_0$.
	
	The $g(t^*_0)=0$ is trivial. We proceed to prove that if $g(t^*_0)$ non-zero, then $g(t^*) < (1+\delta)e^{c(t^*-t^*_0)}g(t^*_0)$ for all $\delta>0$. Suppose that there exists a $t^*_2$ such that $g(t^*_2)=(1+\delta)e^{C(t^*_2-t^*_0)}g(t^*_0)$, but up to this point, $g(t^*_2)<(1+\delta)e^{C(t^*_2-t^*_0)}g(t^*_0)$. Then we obtain
	\begin{align}\nonumber
	g(t^*_2)<(1+\delta)g(t^*_0)+C\int_{t^*=t^*_0}^{t^*_2}g(t^*)dt^*&\leq(1+\delta)g(t^*_0)+C\int_{t^*=t^*_0}^{t^*_2}(1+\delta)e^{C(t^*-t^*_0)}g(t^*_0)dt^*\\
	&=(1+\delta)g(t^*_0)+[(1+\delta)e^{C(t^*-t^*_0)}g(t^*_0)]_{t^*_0}^{t^*_2}=(1+\delta)e^{C(t^*_2-t^*_0)}g(t^*_0)=g(t^*_2),
	\end{align}
	which gives us a contradiction.
	
	We similarly have $g(t^*_0)\leq e^{C(t^*_1-t^*_0)}g(t^*_1)$.
	
	Thus by letting $A=e^{C(t^*_1-t^*_0)}$ in the statement of the Lemma, we are done.
\end{enumerate}
\end{proof}

The above lemma then allows us to come to our $n^{th}$ energy uniform boundedness results:
\begin{Theorem}[Forward $n^{th}$ order Non-degenerate Energy Boundedness for the Reflective Case]\label{Thm:NForwardBound}
Let $\phi\in C^\infty_{c\forall t^*}$ be a solution to the wave equation \eqref{eq:wave} with reflective boundary conditions. There exists a constant $E=E(n,M)$ such that 
\begin{equation}\label{eq:falsebound}
\Vert\phi\Vert_{\dot{H}^n(\Sigma_{t^*_1})}\leq E\Vert\phi\Vert_{\dot{H}^{n}(\Sigma_{t^*_0})} \qquad \forall \phi\in C^\infty_{c\forall t^*} \quad \forall t^*_1\in [t^*_0,t^*_c].
\end{equation}	
\end{Theorem}	
\begin{proof}
As with previous uniform boundedness results, we look at bounding the energy uniformly for sufficiently far back in time. Then we use our local result (part 3 of Lemma \ref{BigLem}) to obtain a uniform bound for all $t^*$.

We proceed inductively, by considering $\bar{\p}_{t^*}^n(\phi)$. Here $\bar{\p}_{t^*}=\p_{t^*}+\dot{R}^*\p_r$ is the partial $t^*$ derivative with respect to the coordinates given in \eqref{eq:RhoCoords}.
\begin{align}\nonumber
\Box_g\left(\bar{\p}_{t^*}^n(\phi)\right)=&\Box_g\left(\left(\p_{t^*}+\dot{R}^*\p_r\right)^n\phi\right)=\sum_{m=0}^{n}\left(\begin{matrix} n\\m	\end{matrix}\right)\dot{R}^{*m}\Box_g\left(\p_{t^*}^{n-m}\p_r^m\phi\right)+\ddot{R}^*\left(\text{Lower order terms with bounded coefficients}\right)\\
=&\sum_{m=0}^n\left(\begin{matrix} n\\m	\end{matrix}\right)\dot{R}^{*m}m\Bigg(\frac{2}{r}\left(1-\frac{M}{r}\right)\left(\p_{t^*}^{n-m+2}\p_{r}^{m-1}\phi-\p_{t^*}^{n-m}\p_r^{m+1}\phi\right)-\frac{4M}{r^2}\p_{t^*}^{m-n+1}\p_r^m\phi\\\nonumber&
+\frac{m-1}{r^2}\p_{t^*}^{n-m+2}\p_r^{m-2}\phi-\frac{m+1}{r^2}\p_{t^*}^{n-m}\p_r^m\phi\Bigg)+\ddot{R}^*(\text{Lower order terms with bounded coefficients}).
\end{align}

In Oppenheimer--Snyder, $\ddot{R}<0$. We have, by the induction hypothesis, that for some $A=A(M,n)>0$
\begin{equation}
\int_{t^*=t^*_0}^{t^*_1}\int_{\Sigma_{t^*}}\left\vert\ddot{R}(\text{Lower order terms with bounded coefficients})\right\vert^2 dt^*\leq \int_{t^*_0}^{t^*_1}A\Vert\phi\Vert_{\dot{H}^n(\Sigma_{t^*_0})}^2\vert\ddot{R}^*\vert^2 dt^*
\end{equation}

We also have that if we fix $t^*_-$ to be large and negative enough, $R^*(t^*)\geq A\vert t^*\vert^{2/3}\geq 0$, $0\leq-\dot{R}^*\leq B\vert t^*\vert^{-1/3}$ for all $t^*\leq t^*_-$. This means that if $t^*_0, t^*_1\leq t^*_-$,
\begin{align}
\left\vert\int_{\Sigma_{t^*}}\sum_{m=0}^n\left(\begin{matrix} n\\m	\end{matrix}\right)\dot{R}^{*m}m\left(\frac{m-1}{r^2}\p_{t^*}^{n-m+2}\p_r^{m-2}\phi-\frac{m+1}{r^2}\p_{t^*}^{n-m}\p_r^m\phi\right)\right\vert^2\leq \frac{1}{{R^*}^4}C'\Vert\phi\Vert_{\dot{H}^n(\Sigma_{t^*_0})}^2\\\nonumber\leq \frac{C'^2}{3A^4}\vert t^*\vert^{-8/3}\Vert\phi\Vert_{\dot{H}^n(\Sigma_{t^*_0})}^2\leq C\vert t^*\vert^{-8/3}\Vert\phi\Vert_{\dot{H}^n(\Sigma_{t^*_0})}^2,
\end{align}
for some $C=C(M,n,t^*_-)>0$.

So now we consider the modified current $J^{X,\epsilon/2r}(\bar{\p}_{t^*}^n(\phi))$, as given by \eqref{eq:modcurrent}. Here $X=\p_{t^*}+\epsilon\p_{r}$ and $0<\epsilon\ll 1$ is a fixed, small constant. Given we are already restricting ourselves to $t^*_0,t^*_1\leq t^*_-$, from \eqref{eq:Generalt^*} we have that
\begin{equation}
-\int_{\Sigma_{t^*}}dt^*(J^{X,\epsilon/2r}(\bar{\p}_{t^*}^n(\phi)))\geq c\Vert\bar{\p}_{t^*}^n\phi\Vert^2_{\dot{H}^1(\Sigma_{t^*})}
\end{equation}
for some positive constant, $c=c(M,n,t^*_-)>0$. From \eqref{eq:GeneralBoundary}, we have
\begin{equation}
d\rho(J^T(X^n(\phi)))\geq 0.
\end{equation}

Thus applying generalised Stokes' theorem, we obtain
\begin{align}\nonumber
-\int_{\Sigma_{t^*_1}}dt^*(J^{X,\epsilon/2r}(\bar{\p}_{t^*}^n(\phi)))\leq&-\int_{\Sigma_{t^*_0}}dt^*(J^{X,\epsilon/2r}(\bar{\p}_{t^*}^n(\phi)))-\int_{t^*_0}^{t^*_1}\int_{\Sigma_{t^*}}\p_{t^*}\left(\bar{\p}_{t^*}^n(\phi)\right)\Box_g(\bar{\p}_{t^*}^n(\phi))+K^{X,\epsilon/2r}dt^*\\
\leq& -\int_{\Sigma_{t^*_0}}dt^*(J^{X,\epsilon/2r}(\bar{\p}_{t^*}^n(\phi)))+\int_{t^*_0}^{t^*_1}\Vert \p_{t^*}\bar{\p}_{t^*}^n\phi\Vert_{L^2(\Sigma_{t^*})}\left(A\vert\ddot{R}^*\vert+C\vert t^*\vert^{-4/3}\right)\Vert\phi\Vert_{\dot{H}^n(\Sigma_{t^*_0})}-K^{X,\epsilon/2r}dt^*\\\nonumber
-\int_{t^*_0}^{t^*_1}\int_{\Sigma_{t^*}}\p_{t^*}\bar{\p}_{t^*}^n(\phi)&\left(\sum_{m=0}^n\left(\begin{matrix} n\\m	\end{matrix}\right)\dot{R}^{*m}m\left(\frac{2}{r}\left(1-\frac{M}{r}\right)\left(\p_{t^*}^{n-m+2}\p_{r}^{m-1}\phi-\p_{t^*}^{n-m}\p_r^{m+1}\phi\right)-\frac{4M}{r^2}\p_{t^*}^{m-n+1}\p_r^m\phi\right)\right)dt^*.
\end{align}

We now note that in the case $m\geq 2$, every term has a coefficient which can be bounded by $A\dot{R}^{*2}/R^*\leq B\vert t^*\vert^{-4/3}$. We can similarly bound any terms with a $1/r^2$ coefficient. Thus we have that
\begin{align}\label{eq:horriblebound}
-\int_{t^*_0}^{t^*_1}\int_{\Sigma_{t^*}}\p_{t^*}\bar{\p}_{t^*}^n(\phi)\left(\sum_{m=0}^n\left(\begin{matrix} n\\m	\end{matrix}\right)\dot{R}^{*m}m\left(\frac{2}{r}\left(1-\frac{M}{r}\right)\left(\p_{t^*}^{n-m+2}\p_{r}^{m-1}\phi-\p_{t^*}^{n-m}\p_r^{m+1}\phi\right)-\frac{4M}{r^2}\p_{t^*}^{m-n+1}\p_r^m\phi\right)\right)dt^*\\\nonumber
\leq \int_{t^*_0}^{t^*_1} B\vert t^*\vert^{-4/3}\Vert\bar{\p}_{t^*}^n\phi\Vert^2_{\dot{H}^1(\Sigma_{t^*})}dt^*-\int_{t^*_0}^{t^*_1}\int_{\Sigma_{t^*}}\frac{n\dot{R}^*}{r}\left(\p_{t^*}^{n+1}\phi-\p_{t^*}^{n-1}\p_r^{2}\right)\p_{t^*}\bar{\p}_{t^*}^n(\phi) dt^*,
\end{align}
Here, we have used part 2 of Lemma \ref{BigLem} to bound $\p_{t^*}\bar{\p}_{t^*}^n(\phi)$ by $\Vert\bar{\p}_{t^*}^n\phi\Vert_{\dot{H}^1(\Sigma_{t^*})}$.

In order to bound the final term, it is useful to note that swapping between $\p_{t^*}$ and $\bar{\p}_{t^*}$ introduces terms with a factor of $\dot{R}^*$. Any derivative that now has an extra factor of $\dot{R}^*$ can be absorbed into the $B$ term in \eqref{eq:horriblebound}. Thus we can freely swap between the two derivatives when bounding this final term. We can similarly ignore any $\bar{\p}_{t^*}r$ terms.
\begin{align}\nonumber
\int_{t^*_0}^{t^*_1}\int_{\Sigma_{t^*}}\frac{n\dot{R}^*}{r}\left(\p_{t^*}^{n+1}\phi-\p_{t^*}^{n-1}\p_r^{2}\right)\p_{t^*}\bar{\p}_{t^*}^n(\phi) dt^*\geq&\int_{t^*_0}^{t^*_1}\int_{\Sigma_{t^*}}\frac{n\dot{R}^*}{r}\bar{\p}_{t^*}^{n-1}\left(\p_{t^*}^2\phi-\p_r^{2}\phi\right)\bar{\p}_{t^*}^{n+1}\phi\text{ } dt^*-\int_{t^*_0}^{t^*_1} B\vert t^*\vert^{-4/3}\Vert\bar{\p}_{t^*}^n\phi\Vert^2_{\dot{H}^1(\Sigma_{t^*})}dt^*\\\nonumber
\geq&\int_{t^*_0}^{t^*_1}\int_{\Sigma_{t^*}}\frac{n\dot{R}^*}{r^3}\bar{\p}_{t^*}^{n-1}\mathring{\slashed\triangle}\phi\bar{\p}_{t^*}^{n+1}\phi\text{ } dt^*-\int_{t^*_0}^{t^*_1} B\vert t^*\vert^{-4/3}\Vert\bar{\p}_{t^*}^n\phi\Vert^2_{\dot{H}^1(\Sigma_{t^*})}dt^*\\
\geq&-\int_{t^*_0}^{t^*_1}\int_{\Sigma_{t^*}}\frac{n\dot{R}^*}{r^3}\bar{\p}_{t^*}^{n-1}\mathring{\slashed\nabla}\phi.\bar{\p}_{t^*}^{n+1}\mathring{\slashed\nabla}\phi\text{ } dt^*-\int_{t^*_0}^{t^*_1} B\vert t^*\vert^{-4/3}\Vert\bar{\p}_{t^*}^n\phi\Vert^2_{\dot{H}^1(\Sigma_{t^*})}dt^*\\\nonumber
\geq&\int_{t^*_0}^{t^*_1}\int_{\Sigma_{t^*}}\frac{n\dot{R}^*}{r^3}\vert\bar{\p}_{t^*}^{n}\mathring{\slashed\nabla}\phi\vert^2dt^*-\int_{t^*_0}^{t^*_1} B\vert t^*\vert^{-4/3}\Vert\bar{\p}_{t^*}^n\phi\Vert^2_{\dot{H}^1(\Sigma_{t^*})}dt^*\\\nonumber
&-\left\vert\frac{n\dot{R}^*}{R^*}\right\vert_{t^*_0}\Vert\phi\Vert_{\dot{H}^{n+1}(\Sigma_{t^*_0})}^2-\left\vert\frac{n\dot{R}^*}{R^*}\right\vert_{t^*_1}\Vert\phi\Vert_{\dot{H}^{n+1}(\Sigma_{t^*_1})}^2.
\end{align}

We then have, from \eqref{eq:GeneralK}, that
\begin{equation}
\int_{t^*_0}^{t^*_1}\int_{\Sigma_{t^*}}K^{X,\epsilon/2r}dt^*\geq\int_{t^*_0}^{t^*_1}\int_{\Sigma_{t^*}}\frac{\epsilon}{r^3}\vert\bar{\p}_{t^*}^{n}\mathring{\slashed\nabla}\phi\vert^2dt^*-\int_{t^*_0}^{t^*_1} D\vert t^*\vert^{-4/3}\Vert\bar{\p}_{t^*}^n\phi\Vert^2_{\dot{H}^1(\Sigma_{t^*})}dt^*,
\end{equation}
for some fixed constant $D=D(M,n,t^*_-)>0$.

Finally, we note that
\begin{align}\nonumber
\int_{t^*_0}^{t^*_1}\Vert \p_{t^*}\bar{\p}_{t^*}^n\phi\Vert_{L^2(\Sigma_{t^*})}&\left(A\vert\ddot{R}^*\vert+C\vert t^*\vert^{-4/3}\right)\Vert\phi\Vert_{\dot{H}^n(\Sigma_{t^*_0})}dt^*\\\nonumber&\leq A\int_{t^*_0}^{t^*_1}\left(\vert\ddot{R}^*\vert+\vert t^*\vert^{-4/3}\right)\Vert \bar{\p}_{t^*}^n\phi\Vert_{\dot{H}^1(\Sigma_{t^*})}^2+\left(\vert\ddot{R}^*\vert+\vert t^*\vert^{-4/3}\right)\Vert\phi\Vert_{\dot{H}^n(\Sigma_{t^*_0})}^2dt^*\\
&\quad-\int_{t^*_0}^{t^*_1}\int_{\Sigma_{t^*}}\frac{\epsilon+n\dot{R}^*}{r^3}\vert\bar{\p}_{t^*}\mathring{\slashed\nabla}\phi\vert^2dt^*\\\nonumber
&\leq A\int_{t^*_0}^{t^*_1}\left(\vert\ddot{R}^*\vert+\vert t^*\vert^{-4/3}\right)\Vert \bar{\p}_{t^*}^n\phi\Vert_{\dot{H}^1(\Sigma_{t^*})}^2dt^*+E\Vert\phi\Vert_{\dot{H}^n(\Sigma_{t^*_0})}^2,
\end{align}
for $E=E(M,n,t^*_-)>0$. This is given $t^*_-$ negative enough that $\vert\dot{R}^*\vert\leq\epsilon/n$ and that we can apply part 2 of Lemma \ref{BigLem}.

Adding these all together, we get
\begin{align}
\Vert\phi\Vert_{\dot{H}^{n+1}(\Sigma_{t^*_1})}^2&\leq C\Vert\phi\Vert_{\dot{H}^{n+1}(\Sigma_{t^*_0})}^2+D\Vert\phi\Vert_{\dot{H}^{n}(\Sigma_{t^*_0})}+A\int_{t^*_0}^{t^*_1}\left(\vert\ddot{R}^*\vert+\vert t^*\vert^{-4/3}\right)\Vert \bar{\p}_{t^*}^n\phi\Vert_{\dot{H}^1(\Sigma_{t^*})}^2dt^*\\\nonumber&
\leq  C\Vert\phi\Vert_{\dot{H}^{n+1}(\Sigma_{t^*_0})}^2+A\int_{t^*_0}^{t^*_1}\left(\vert\ddot{R}^*\vert+\vert t^*\vert^{-4/3}\right)\Vert \bar{\p}_{t^*}^n\phi\Vert_{\dot{H}^1(\Sigma_{t^*})}^2dt^*
\end{align}
where constants $C,D,A$ all only depend on $M$, $n$ and $t^*_-$.

Thus by Gronwall's inequality, we have
\begin{equation}
\Vert\phi\Vert_{\dot{H}^{n+1}(\Sigma_{t^*_1})}^2\leq C\Vert\phi\Vert_{\dot{H}^{n+1}(\Sigma_{t^*_0})}^2 \exp\left(A\int_{t^*_0}^{t^*_1}\left(\vert\ddot{R}^*\vert+\vert t^*\vert^{-4/3}\right)\right)dt^*\leq C\Vert\phi\Vert_{\dot{H}^{n+1}(\Sigma_{t^*_0})}^2 \exp\left(A\left(\vert\dot{R}(t^*_-)\vert+\vert t^*_-\vert^{-1/3}\right)\right),
\end{equation}
for all $t^*_0\leq t^*_1\leq t^*_-$. We can then proceed to cover the interval $[t^*_-,t^*_c]$ by using part 3 of Lemma \ref{BigLem}. Thus we obtain our result.
\end{proof}

The last theorem we then prove in this section is backwards $n^{th}$ order energy boundedness.

\begin{Theorem}[Backwards $n^{th}$ order Non-degenerate Energy Boundedness for the Reflective Case]\label{Thm:NBackwardBound}
	Let $\phi\in C^\infty_{c\forall t^*}$ be a solution to the wave equation \eqref{eq:wave} with reflective boundary conditions. There exists a constant $E=E(n,M)$ such that 
	\begin{equation}\label{eq:falsebound2}
	\Vert\phi\Vert_{\dot{H}^n(\Sigma_{t^*_0})}\leq E\Vert\phi\Vert_{\dot{H}^{n}(\Sigma_{t^*_1})} \qquad \forall \phi\in C^\infty_{c\forall t^*} \quad \forall t^*_1\in [t^*_0,t^*_c].
	\end{equation}	
\end{Theorem}
\begin{proof}
	This is proved identically to Theorem \ref{Thm:NForwardBound}, but let $X=\p_{t^*}-\epsilon\p_{r}$, and we are done (for positive definiteness of the surface terms, see Theorem \ref{Thm:BackReflBound}).
\end{proof}

\subsection{Permeating Case}
We now look at the permeating case. In the Oppenheimer--Snyder model for the interior of the star, we have that our metric is $C^0$, but piecewise smooth. Thus, as given by Theorem \ref{Thm:PermExist}, we are dealing with a weak solution rather than a classical solution, i.e.~$\phi\in H^1(\Sigma_\tau)$ a solution to 
\begin{equation}
\int_{\tau=-\infty}^{\infty}\int_{\Sigma_\tau}\sqrt{-g}g^{ab}\p_a\phi\p_b\left(\frac{\psi}{\sqrt{-g}}\right) dVd\tau=0 \qquad \forall \psi\in C^\infty_0(\mathcal{M}).
\end{equation} 
(Note that in the coordinates chosen below, the determinant of $\sqrt{-g}$ is $r^2\sin\theta$)

The metric in the interior of our star has the form (see Section \ref{sec:int}):
\begin{equation}
ds^2=\begin{cases}
-\left(1-\frac{2Mr^2}{r_b^3}\right)d\tau^2+2\sqrt{\frac{2Mr^2}{r_b^3}}d\tau dr+dr^2+r^2g_{S^2} & r<r_b:=(R_b^\frac{3}{2}-\frac{3\tau}{2}\sqrt{2M})^\frac{2}{3}\\-\left(1-\frac{2M}{r}\right)d\tau^2+2\sqrt{\frac{2M}{r}}d\tau dr+dr^2+r^2g_{S^2} & r\geq r_b
\end{cases}
\end{equation}
for constants $R_b$ and $M$. Here, $r_b$ is the boundary of the star.

We note that the null hypersurface given by $r=r_b\left(3-2\sqrt{\frac{r_b}{2M}}\right)$ is part of our event horizon. This means when we construct the backwards scattering map, we will require data on this as well as the $r=2M$, $\tau>\tau_c$ surface.

We begin our study of boundedness by noticing that our usual $\p_\tau$-energy does not give the same bound as before. This is due to the fact that $\p_\tau$ is no longer a Killing vector. We therefore obtain a term arising from $K^{\p_\tau}$ inside the star. We can still obtain a bound from integrating $K^{\p_\tau}$, however it is now exponentially growing in $\tau$:

\begin{Lemma}[Finite in Time Forwards Bound in the Permeating Case]\label{lem:TimeBound}
	Let $\phi\in H^2_{c\forall \tau}$ be a weak solution to the wave equation \eqref{eq:wave} with permeating boundary conditions. There exists a constant $B=B(M)>0$ such that
	\begin{equation}
	\Vert\phi\Vert_{\dot{H}^1(\Sigma_{\tau_1})}^2\leq-2\int_{\Sigma_{\tau_1}}d\tau(J^X[\phi])\leq -2\int_{\Sigma_{\tau_0}}d\tau(J^X[\phi])e^{B(\tau_1-{\tau_0})}\leq 4\Vert\phi\Vert_{\dot{H}^1(\Sigma_{\tau_0})}^2e^{B(\tau_1-{\tau_0})} \qquad\forall\tau_1\geq\tau_0
	\end{equation}
	for suitably chosen future directed timelike $X$.
\end{Lemma}
\begin{proof}
	Choose $f(r)$ to be a smooth cut off function
	\begin{align}
	f(r)\begin{cases}
	=-\frac{1}{2} & r\in[\frac{3M}{2},\frac{5M}{2}]\\
	=0 &r\notin[M,3M]\\
	\in[-\frac{1}{2},0] &r\in (M,\frac{3M}{2})\cup(\frac{5M}{2},3M)
	\end{cases}.
	\end{align}
	Note $f$ has bounded derivatives. Then if we let $X=\p_{\tau}+f(r)\p_r$ we have that:
	\begin{equation}
	-d\tau(J^X)=\begin{cases}
	\frac{1}{2}\left(\frac{1}{r^2}\vert\mathring{\slashed\nabla}\phi\vert^2+(\p_{\tau}\phi)^2+2f\p_{\tau}\phi\p_r\phi+\left(1-\frac{2M r^2}{r_b^3}-2f\sqrt{\frac{2M r^2}{r_b^3}}\right)(\p_{r}\phi)^2\right) & r<r_b\\\frac{1}{2}\left(\frac{1}{r^2}\vert\mathring{\slashed\nabla}\phi\vert^2+(\p_{\tau}\phi)^2+2f\p_{\tau}\phi\p_r\phi+\left (1-\frac{2M}{r}-2f \sqrt{\frac{2M}{r}}\right)(\p_r\phi)^2\right) & r\geq r_b	\end{cases}.
	\end{equation}
	\begin{equation}
	K^X=\begin{cases}\frac{1}{2}\Bigg(-\frac{f'}{r^2}\vert\mathring{\slashed\nabla}\phi\vert^2+(\frac{2f}{r}+f')(\p_{\tau}\phi)^2-\left(\frac{6Mr}{r_b^3}+\frac{6\sqrt{2M}f}{\sqrt{r_b^3}}\right)\p_{\tau}\phi\p_r\phi\\\qquad+\left((1-\frac{2Mr^2}{r_b^3})f'-(1-\frac{4Mr^2}{r_b^3})\frac{2f}{r}+\frac{3\sqrt{2M}^3r^2}{\sqrt{r_b^9}}\right)(\p_r\phi)^2\Bigg)&r<r_b\\\frac{1}{2}\Bigg(-\frac{f'}{r^2}\vert\mathring{\slashed\nabla}\phi\vert^2+(\frac{2f}{r}+f')(\p_{\tau}\phi)^2-\frac{3f}{r}\sqrt{\frac{2M}{r}}\p_{\tau}\phi\p_r\phi\\\qquad+\left((1-\frac{2M}{r})f'-(1-\frac{M}{r})\frac{2f}{r}\right)(\p_r\phi)^2\Bigg) & r\geq r_b
	\end{cases}.
	\end{equation}
	
Thus $K^X$ can always be bounded by multiples of $-d\tau(J^X)$.
\begin{equation}\label{eq:PermStokes}
-\int_{\Sigma_{\tau_0}}d\tau(J^X)=-\int_{\Sigma_{\tau_1}}d\tau(J^X)+\int_{\tau=\tau_0}^{\tau_1}\int_{\Sigma_\tau}K^X-\int_{\mathcal{H}^+\cap \{ r < 2M \} }dn(J^X).
\end{equation}

We can also note that the contribution from the part of the horizon in \eqref{eq:PermStokes} is of the form $-T_{ab}X^an^b$ for future directed normal $n$. By the dominant energy condition, we have that this term has the correct sign. Thus letting $g(\tau)=-\int_{\Sigma_{\tau}}d\tau(J^X)$, we have that \begin{equation}
g(\tau)\leq g(\tau_0)+A\int_{s^*=\tau_0}^{\tau}g(s)ds,
\end{equation}
which gives us our result by Gronwall's Inequality.
\end{proof}

\begin{Remark}
	For the purposes of the scattering map however, we will not want to disregard the surface term from the event horizon. Instead we will want to consider a norm on the horizon such that the map from a surface $\Sigma_{\tau}$ to $\Sigma_{\tau_c}\cup(\mathcal{H}\cap\{r< 2M\})$ is bounded in both directions. Letting $X=\p_\tau+f(r)\p_r$, then we have
	\begin{equation}
	-dn(J^X)=\frac{1}{2}\left(3\sqrt{\frac{2M}{r_b}}-1+f(r)\right)\left(\p_\tau\phi+3\Bigg(1-\sqrt{\frac{2M}{r_b}}\Bigg)\p_r\phi\right)^2+\left(3\left(1-\sqrt{\frac{2M}{r_b}}\right)-f\right)\frac{1}{2r^2}\vert\mathring{\slashed\nabla}\phi\vert^2.
	\end{equation}
	If we then use the $f$ from Lemma \ref{lem:TimeBound}, we have all these terms being positive definite. Therefore the norm we will consider on the surface contains only the $L^2$ norms of the angular derivatives and the derivative with respect to the vector $\p_\tau+3\Bigg(1-\sqrt{\frac{2M}{r_b}}\Bigg)\p_r$.
\end{Remark}

\begin{Lemma}[Finite-in-Time Backward Non-degenerate Energy Boundedness for the Permeating Case]\label{Lem:BackExpBound}
	Let $\phi\in H^2_{c\forall \tau}$ be  a weak solution to the wave equation \eqref{eq:wave} with permeating boundary conditions. There exists a constant $B=B(M)>0$ such that for all $\tau_0\leq\tau_1\leq\tau_{c^-}$, we have
	\begin{equation}
		\Vert\phi\Vert_{\dot{H}^1(\Sigma_{\tau_0})}^2\leq 4\Vert\phi\Vert_{\dot{H}^1(\Sigma_{\tau_1})}^2e^{B(\tau_1-\tau_0)}.
	\end{equation}
	Here $(\tau_{c^-},r=0)$ is defined by equation \eqref{eq:tauc-}.
\end{Lemma}
\begin{proof}
	This is proved identically to the previous lemma. We bounding $K^X$ below instead of above, and we ignore the boundary term, as $\mathcal{H}^+\cap\{\tau<\tau_{c^-}\}=0$.
\end{proof}

Lemmas \ref{lem:TimeBound} and \ref{Lem:BackExpBound} give us the conditions mentioned in Remark \ref{Rmk:H1PermExist}. Thus we have the following Theorem:
\begin{Theorem}[$H^1$ Existence of Permeating Solutions]\label{Thm:H1PermExist}
	Let $(\phi_0,\phi_1)\in H^1(\Sigma_{\tau_0})$, where $\tau_0\leq\tau_{c^-}$. There exists a solution $\phi$ to the wave equation \eqref{eq:wave} with permeating boundary conditions, such that
	\begin{equation}
	(\phi\vert_{\Sigma_{\tau_0}},\p_{\tau}\phi\vert_{\Sigma_{\tau_0}})=(\phi_0,\phi_1)
	\end{equation}
	Here this restriction holds in a trace sense, and $\phi$ is a solution in the sense of distributions. Finally, $\phi\in \dot{H}^1(\Sigma_{\tau})$ for all $\tau\leq\tau_c$.
\end{Theorem}
\begin{proof}
	This is a result of Theorem \ref{Thm:PermExist} and a density argument. This density argument relies on the bounds given by Lemmas \ref{lem:TimeBound} and \ref{Lem:BackExpBound}.
\end{proof}

\begin{Remark}
	As in the permeating case, our existence result Theorem \ref{Thm:H1PermExist} allows us to define the forwards map:
	\begin{align}
	\mathcal{F}_{(\tau_0,\tau_1)}&:\mathcal{E}^X_{\Sigma_{\tau_0}}\to \mathcal{E}^X_{\Sigma_{\tau_1}}\\
	\mathcal{F}_{(\tau_0,\tau_1)}\left(\phi_0,\phi_1\right)&:=\left(\phi\vert_{\Sigma_{\tau_1}},\p_{t^*}\phi\vert_{\Sigma_{\tau_1}}\right) \quad \text{where $\phi$ is the solution to \eqref{eq:wave} and \eqref{eq:initial2}}.
	\end{align}
\end{Remark}
We can use Lemmas \ref{lem:TimeBound} and \ref{Lem:BackExpBound} to bound the solution over any finite time interval. Thus we can now consider only the case where $r_b\gg 2M$, i.e.~$\frac{2M}{r_b}<\epsilon$ for some small, fixed epsilon. Once we have a uniform bound for $\frac{2M}{r_b}<\epsilon$, we can bound solutions of the wave equation for $\tau\leq\tau_c$ using  Lemmas \ref{lem:TimeBound} and \ref{Lem:BackExpBound}. Previous work on the external Schwarzschild space-time gives us the required bounds for $\tau>\tau_c$.

This brings us to our next result:
\begin{Proposition}[Forward Non-degenerate Energy Boundedness for the Permeating Case, Sufficiently Far Back]\label{Prop:UniBound}
	Let $\phi$ be a solution to the wave equation \eqref{eq:wave} with permeating boundary conditions (as given by Theorem \ref{Thm:H1PermExist}). There exists a constant, $A=A(M)>0$, and a time, $\tau^*$ such that
	\begin{equation}
	\Vert\phi\Vert_{\dot{H}^1(\Sigma_{\tau_0})}\leq A\Vert\phi\Vert_{\dot{H}^1(\Sigma_{\tau_1})}\qquad \forall\tau_0<\tau_1\in(-\infty,\tau^*].
	\end{equation}
\end{Proposition}
\begin{proof}
	For this proof, we choose a time dependent vector field. Let $Y=h(\tau)\p_\tau$. Then we have that 
	\begin{equation}
	-d\tau(J^Y)=\begin{cases}
	\frac{h}{2}\left(\frac{1}{r^2}\vert\mathring{\slashed\nabla}\phi\vert^2+(\p_{\tau}\phi)^2+\left(1-\frac{2M r^2}{r_b^3}\right)(\p_{r}\phi)^2\right) & r<r_b\\\frac{h}{2}\left(\frac{1}{r^2}\vert\mathring{\slashed\nabla}\phi\vert^2+(\p_{\tau}\phi)^2+\left (1-\frac{2M}{r}\right)(\p_r\phi)^2\right) & r\geq r_b	\end{cases}.
	\end{equation}
	\begin{equation}\label{eq:KY}
	K^Y=\begin{cases}
	-\frac{1}{2}\Bigg(\frac{h'}{r^2}\vert\mathring{\slashed\nabla}\phi\vert^2+h'(\p_{\tau}\phi)^2-\frac{2Mr}{r_b^3}3h\p_{\tau}\phi\p_{r}\phi & \\\qquad+\left(h'\left(1-\frac{2M r^2}{r_b^3}\right)-\frac{2M r^2}{r_b^3}\frac{3h}{r}\right)(\p_{r}\phi)^2\Bigg) & r<r_b \\\frac{h'}{2}\left(\frac{1}{r^2}\vert\mathring{\slashed\nabla}\phi\vert^2+(\p_{\tau}\phi)^2+\left (1-\frac{2M}{r}\right)(\p_r\phi)^2\right) & r\geq r_b	\end{cases}.
\end{equation}

Now, we would like both of these to be everywhere positive definite. For this, we need to pick $h>0$ and bounded. We also need $h'<0$, with $-h'>\frac{3M}{r_b^2}h$. Thus we can choose, for example,
\begin{equation}
h(\tau)=1-\left(\frac{2M}{r_b}\right)^{1/4}\in [1-\left(\frac{2M}{r_b(\tau^*)}\right)^{1/4},1]
\end{equation}
\begin{equation}
h'(\tau)=-\left(\frac{r_b}{2M}\right)^{1/4} \frac{2M}{4r_b^2}\leq-\left(\frac{r_b(\tau^*)}{2M}\right)^{1/4} \frac{2M}{4r_b^2}<-\frac{3M}{r_b^2}h
\end{equation}
where we have chosen $\tau^*$ s.t. $\left(\frac{r_b(\tau^*)}{2M}\right)^{1/4}>6$. This choice also gives us 
\begin{equation}
\left(1-\frac{2M}{r_b(\tau^*)}\right)\left(1-\left(\frac{2M}{r_b(\tau^*)}\right)^\frac{1}{4}\right)\Vert\phi\Vert_{\dot{H}^1(\Sigma_{\tau})}^2\leq-2\int_{\Sigma_{\tau}}d\tau(J^X)\leq\Vert\phi\Vert_{\dot{H}^1(\Sigma_{\tau})}^2.
\end{equation}

Finally, using these inequalities in Stokes' Theorem gives
\begin{equation}
\Vert\phi\Vert_{\dot{H}^1(\Sigma_{\tau_0})}^2\geq-2\int_{\Sigma_{\tau_0}}d\tau(J^Y)=-2\int_{\Sigma_{\tau_1}}d\tau(J^Y)+2\int_{\tau=\tau_0}^{\tau_1}\int_{\Sigma_\tau}K^Y\geq-2\int_{\Sigma_{\tau_1}}d\tau(J^Y)\geq A\Vert\phi\Vert_{\dot{H}^1(\Sigma_{\tau_1})}^2
\end{equation}
as required.
\end{proof}

\begin{Theorem}[Forward Non-degenerate Energy Boundedness for the Permeating Case]\label{Thm:ForwardPermBound}
	Let $\phi$ be a solution to the wave equation \eqref{eq:wave} with permeating boundary conditions (as given by Theorem \ref{Thm:H1PermExist}). There exists a constant $A=A(M)$ such that
	\begin{equation}\label{eq:forwardpermbound}
	\Vert\phi\Vert_{\dot{H}^1(\Sigma_{\tau_1})}\leq A\Vert\phi\Vert_{\dot{H}^1(\Sigma_{\tau_0})}\qquad \forall\tau_0\leq\tau_1.
	\end{equation}
\end{Theorem}
\begin{proof}
	Previous works on Schwarzschild exterior space time (e.g.~\cite{BHL}), gives us that \eqref{eq:forwardpermbound} holds for $\tau_c\leq\tau_0\leq\tau_1$. Thus if we prove the result for the case $\tau_1\leq\tau_c$, we can combine these results to obtain \eqref{eq:forwardpermbound} for all $\tau_0\leq\tau_1$.
	
	Let $A$ and $\tau^*$ be defined as in Proposition \ref{Prop:UniBound}. Let $B$ be as defined in Lemma \ref{lem:TimeBound}. Remember $\tau_c$ is defined to be the time at which the surface of the star crosses $r=2M$. We then have that
	\begin{equation}
	\Vert\phi\Vert_{\dot{H}^1(\Sigma_{\tau_1})}^2\leq 4Ae^{B(\tau_c-{\tau^*})}\Vert\phi\Vert_{\dot{H}^1(\Sigma_{\tau_0})}^2 \qquad \forall\tau_0\leq\tau_1\leq\tau_c.
	\end{equation}
\end{proof}

In a similar way, we can obtain a boundedness statement for the reverse direction:

\begin{Theorem}[Backwards Non-degenerate Energy Boundedness for the Permeating Case]\label{Thm:BackPermBound}
	Let $\phi$ be a solution to the wave equation \eqref{eq:wave} with permeating boundary conditions (as given by Theorem \ref{Thm:H1PermExist}). There exists a constant $A=A(M)$, and a time $\tau^{*-}$ such that
	\begin{equation}
	\Vert\phi\Vert_{\dot{H}^1(\Sigma_{\tau_0})}\leq A\Vert\phi\Vert_{\dot{H}^1(\Sigma_{\tau_1})}\qquad \forall\tau_0\leq \tau_1\leq \tau^{*-}.
	\end{equation}
\end{Theorem}

\begin{proof}
	As before, we consider $Y=h(\tau)\p_{\tau}$. However, this time we require the sign of $K^Y$ to be non-positive. Thus we choose
	\begin{equation}
	h(\tau)=1+\left(\frac{2M}{r_b}\right)^\frac{1}{4}\in [1,2].
	\end{equation}
	\begin{equation}
	h'(\tau)=\left(\frac{r_b}{2M}\right)^{1/4} \frac{2M}{4r_b^2}\geq\left(\frac{r_b(\tau^{*-})}{2M}\right)^{1/4} \frac{2M}{4r_b^2}>\frac{3M}{r_b^2}h.
	\end{equation}
	
	Thus if we choose $\tau^{*-}$ s.t. $\left(\frac{r_b(\tau^{*-})}{2M}\right)^{1/4}>12$, we have $K^Y\leq0$ (see equation \eqref{eq:KY}). Then, as before we obtain:
	\begin{equation}
\Vert\phi\Vert_{\dot{H}^1(\Sigma_{\tau_1})}^2\geq-\int_{\Sigma_{\tau_1}}d\tau(J^Y)=-\int_{\Sigma_{\tau_0}}d\tau(J^Y)-\int_{\tau=\tau_0}^{\tau_1}\int_{\Sigma_\tau}K^Y\geq-\int_{\Sigma_{\tau_0}}d\tau(J^Y)\geq \frac{1}{2}\left(1-\frac{2M}{r_b(\tau^{*-})}\right)\Vert\phi\Vert_{\dot{H}^1(\Sigma_{\tau_0})}^2.
	\end{equation}
\end{proof}
\begin{Corollary}[Uniform Boundedness for the Permeating Case]
	Let $\phi$ be a solution to the wave equation \eqref{eq:wave} with permeating boundary conditions (as given by Theorem \ref{Thm:H1PermExist}). There exist constants $B=B(M)>0,b=b(M)>0$ such that
	\begin{equation}
	b\Vert\phi\Vert_{\dot{H}^1(\Sigma_{\tau_1})}\leq\Vert\phi\Vert_{\dot{H}^1(\Sigma_{\tau_0})}\leq B\Vert\phi\Vert_{\dot{H}^1(\Sigma_{\tau_1})}\qquad\forall\tau_0,\tau_1\leq\tau_{c^-}.
	\end{equation}
\end{Corollary}
	\begin{proof}
		We have the forward bound due to Theorem \ref{Thm:ForwardPermBound}. The backwards bound is done by combining Theorem \ref{Thm:BackPermBound} and Corollary \ref{Lem:BackExpBound} over the finite time interval $[\tau^{*-},\tau_{c^-}]$.
	\end{proof}
	
\begin{Corollary}\label{Cor:vpermbound}
	Let $\phi$ be a solution to the wave equation \eqref{eq:wave} with permeating boundary conditions (as given by Theorem \ref{Thm:H1PermExist}). There exists a constant, $C=C(M)>0$, such that
	\begin{equation}
	-\int_{\tilde\Sigma_{v_0}}dn(J^Y)\leq-\int_{\tilde\Sigma_{v_1}}dn(J^Y)\leq C\Vert\phi\Vert_{\dot{H}^1(\Sigma_{\tau_0})}^2\quad\forall v_0\leq v_1\leq \tau_0+r_b(\tau_0)
	\end{equation}
	where n is the normal to $\tilde\Sigma_{v}$.
\end{Corollary}
\begin{proof}
	We integrate $K^Y$ between the relevant surfaces and use Stokes' theorem to obtain these bounds.
\end{proof}

We finally look to extend the result to $\dot{H}^2$ norms.
\begin{Theorem}\label{Thm:2nd order energy}
	Let $\phi\in H_c^2$ be a solution to the wave equation \eqref{eq:wave} (as given by Theorem \ref{Thm:PermExist}). There exists a constant $C=C(M)>0$ such that
	\begin{align}
	\Vert\phi\Vert_{\dot{H}^2(\Sigma_{\tau_1})}^2&\leq C\Vert\phi\Vert_{\dot{H}^2(\Sigma_{\tau_0})}^2\qquad \forall \tau_0<\tau_1<\tau_c\\
	\Vert\phi\Vert_{\dot{H}^2(\Sigma_{\tau_0})}^2&\leq C\Vert\phi\Vert_{\dot{H}^2(\Sigma_{\tau_1})}^2\qquad \forall \tau_0<\tau_1<\tau_{c^-}.
	\end{align}
\end{Theorem}
\begin{proof}
	As in the reflective case, we first prove the local in time case. Let
	\begin{equation}
	X=\p_{\tau}+\chi\left(\frac{r}{r_b}\right)\dot{r_b}\p_r,
	\end{equation} 
	where $\chi$ is a smooth cut-off function which vanishes outside $[1/2,3/2]$ and is identically $1$ inside $[2/3,4/3]$.
	
	Note that $X$ is tangent to the boundary over which irregularities of $g$ occur. This means that derivatives of components of $g$ in the $X$ direction are still $H^1_{loc}$.
	
	We then proceed to write \eqref{eq:wave} in terms of this $X$:
	\begin{align}\label{eq:timelikewave}
	\Box_g\phi=\begin{cases}-\p_{\tau}(X\phi)+\left(2\sqrt{\frac{2Mr^2}{r_b^3}}+\chi\dot{r_b}\right)\p_r(X\phi)+\left(1-\frac{2Mr^2}{r_b^3}-\dot{r_b}\left(2\sqrt{\frac{2Mr^2}{r_b^3}}+\chi\dot{r_b}\right)\right)\p_r^2\phi+\frac{1}{r^2}\mathring{\slashed\triangle}\phi+\text{Lower Order Terms}\\-\p_{\tau}(X\phi)+\left(2\sqrt{\frac{2M}{r}}+\chi\dot{r_b}\right)\p_r(X\phi)+\left(1-\frac{2M}{r}-\dot{r_b}\left(2\sqrt{\frac{2M}{r}}+\chi\dot{r_b}\right)\right)\p_r^2\phi+\frac{1}{r^2}\mathring{\slashed\triangle}\phi+\text{Lower Order Terms}.\end{cases}
	\end{align}
	
	Here note:
	\begin{align}
	1-\frac{2Mr^2}{r_b^3}-\dot{r_b}\left(2\sqrt{\frac{2Mr^2}{r_b^3}}+\chi\dot{r_b}\right)=1-\frac{2Mr^2}{r_b^3}+\sqrt{\frac{2M}{r_b}}\left(2\sqrt{\frac{2Mr^2}{r_b^3}}-\chi\sqrt{\frac{2M}{r_b}}\right)=1-\frac{2Mr^2}{r_b^3}+\frac{2Mr}{r_b^2}\left(2-\chi\frac{r_b}{r}\right)>0\\
	1-\frac{2M}{r}-\dot{r_b}\left(2\sqrt{\frac{2M}{r}}+\chi\dot{r_b}\right)=1-\frac{2M}{r}+\sqrt{\frac{2M}{r_b}}\left(2\sqrt{\frac{2M}{r}}-\chi\sqrt{\frac{2M}{r_b}}\right)=1-\frac{2M}{r}+\frac{2M}{\sqrt{rr_b}}\left(2-\chi\sqrt{\frac{r}{r_b}}\right)>0.
	\end{align}
	We can approximate $\phi$ on $\Sigma_{\tau}$ by smooth functions, and then manipulate \eqref{eq:timelikewave} in an identical way to \eqref{eq:LLowerBound} to obtain:
	\begin{equation}\label{eq:H2elliptic}
	\int_{\Sigma_{\tau}}(\p_{\tau}X\phi)^2\geq\epsilon\Vert \phi\Vert_{\dot{H}^2(\Sigma_{t^*})}^2-C\Vert X\phi\Vert_{\dot{H}^1(\Sigma_{t^*})}^2-C\Vert  \phi\Vert_{\dot{H}^1(\Sigma_{t^*})}^2-\int_{S_{\tau,r=0}}2f\mathring{\slashed\nabla}\phi.\p_r\mathring{\slashed\nabla}\phi d\omega^2.
	\end{equation}
	Here $\epsilon>0$ and $C>0$ may depend on the time interval we are considering. One can then show that for smooth approximations to $\phi$, the final term in \eqref{eq:H2elliptic} vanishes.
	
	Thus to prove the local in time result, we can consider the following:
	\begin{equation}
	f(\tau):=\Vert\phi\Vert_{\dot{H}^1(\Sigma_{\tau})}^2-\int_{\Sigma_{\tau}}d\tau\left(J^X(X\phi)\right)\sim\Vert\phi\Vert_{\dot{H}^2(\Sigma_{\tau})}^2.
	\end{equation}
	
	Looking at $\Box_g(X\phi)$ as a distribution, we obtain:
	\begin{equation}
	\int_{\Sigma_{\tau}}\vert\left(\Box_g(X\phi)\right)XX\phi\vert=\int_{\Sigma_{\tau}}\vert\left(\Box_g(X\phi)-X(\Box_g\phi)\right)XX\phi\vert\leq C\Vert\phi\Vert_{\dot{H}^2(\Sigma_{\tau})}^2.
	\end{equation}
	
	As in previous cases, we have that $\vert K^X(X\phi)\vert\leq-C d\tau\left(J^X(X\phi)\right)$ for some $C>0$. Applying Stokes theorem and boundedness of $\dot{H}^1(\Sigma)$ norms, we obtain:
	\begin{equation}
	\frac{1}{C}f(\tau_1)-\int_{\tau_0}^{\tau_1}f(\tau)d\tau\leq f(\tau_0)\leq Cf(\tau_1)+\int_{\tau_0}^{\tau_1}Cf(\tau)d\tau
	\end{equation}
	for some $C=C(M,\tau_0,\tau_1)\geq 0$.	Then an application of Gronwall's lemma gives the local result, in either direction.
	
	Once we are sufficiently far back in time, we can consider
	\begin{equation}
	Y^\pm=\left(1\pm\left(\frac{2M}{r_b}\right)^{1/4}\right)\p_{\tau},
	\end{equation}
	as given in the proof of Proposition \ref{Prop:UniBound} and Theorem \ref{Thm:BackPermBound}. Define
	\begin{equation}
	g(\tau)^\pm:=\Vert\phi\Vert_{\dot{H}^1(\Sigma_{\tau})}^2-\int_{\Sigma_{\tau}}d\tau\left(J^{Y\pm}(T\phi)\right)\sim\Vert\phi\Vert_{\dot{H}^2(\Sigma_{\tau})}^2.
	\end{equation}
	
	\begin{equation}
	\int_{\Sigma_{\tau}}\vert\left(\Box_g(T\phi)\right)Y^\pm T\phi\vert=\int_{\Sigma_{\tau}}\vert\left(\Box_g(T\phi)-T(\Box_g\phi)\right)Y^\pm T\phi\vert\leq \frac{C}{r_b^2}\Vert\phi\Vert_{\dot{H}^2(\Sigma_{\tau})}^2\leq\frac{C'}{r_b^2}g(\tau)^\pm.
	\end{equation}
	
	Now we can choose $\tau\leq\tau^*$ negative enough that $\mp K^{Y^\pm}(T\phi)\geq 0$. Thus
	\begin{equation}
	\mp K^{Y\pm}(T\phi)+\vert\left(\Box_g(T\phi)\right)Y^\pm T\phi\vert\leq \frac{C'}{r_b^2}g(\tau)^\pm,
	\end{equation}
	In this region we can use the boundedness of $\dot{H}^1$ norms given by Theorems \ref{Thm:ForwardPermBound} and \ref{Thm:BackPermBound} to obtain
	\begin{align}
	g(\tau_1)^+\leq C g(\tau_0)^++\int_{\tau_0}^{\tau_1}\frac{C}{r_b^2}g(\tau)^+d\tau\\
	g(\tau_0)^-\leq C g(\tau_1)^-+\int_{\tau_0}^{\tau_1}\frac{C}{r_b^2}g(\tau)^-d\tau.
	\end{align}
	
	An application of Gronwall's lemma, completes the proof, on noting that $\int_{-\infty}^{\tau^*}r_b(\tau)^{-2}d\tau\leq\infty$.
\end{proof}

\begin{Remark}
The first order energy results from this section can be given using the forwards map and the energy space notation from \eqref{eq:energydef} and \eqref{eq:energyspacedef}. Let $X$ be strictly timelike everywhere (for example, as in Theorem \ref{Thm:ForwardReflBound} or Lemma \ref{lem:TimeBound}). We then have that there exist $A_1=A_1(M)>0$ and $A_2=A_2(M)>0$ such that
\begin{align}
A_1^{-1}\Vert(\phi_0,\phi_1)\Vert_X\leq \Vert\mathcal{F}_{t^*_0,t^*_1}(\phi_0,\phi_1)\Vert_X&\leq A_1\Vert(\phi_0,\phi_1)\Vert_X\qquad\qquad \forall t^*_0\leq t^*_1\leq{t^*_c}\\
A_2^{-1}\Vert(\phi_0,\phi_1)\Vert_X\leq \Vert\mathcal{F}_{\tau_0,\tau_1}(\phi_0,\phi_1)\Vert_X&\leq A_2\Vert(\phi_0,\phi_1)\Vert_X\qquad\qquad \forall \tau_0\leq \tau_1\leq\tau_{c^-}\\
\Vert\mathcal{F}_{\tau_0,\tau_2}(\phi_0,\phi_1)\Vert_X&\leq A_2\Vert(\phi_0,\phi_1)\Vert_X\qquad\qquad \forall \tau_0\leq\tau_{c^-}\leq\tau_2\leq\tau_c.
\end{align}
\end{Remark}

\section{The Scattering Map}\label{Sec:TheScatMap}

We now look at the radiation fields. More specifically, we will look at the maps between data on Cauchy surfaces and past/future radiation fields. The maps $\mathcal{G}^+$ and $\mathcal{F}^-$ will take data on Cauchy surfaces to future and past radiation fields respectively. The maps $\mathcal{G}^-$ and $\mathcal{F}^+$ will take future and past radiation fields to data on Cauchy surfaces respectively. We look at obtaining boundedness or non-boundedness results for these. Then we will look at combining them to obtain the scattering map itself, $\mathcal{S}^+$, and boundedness results for this.

\begin{wrapfigure}{r}{7cm}
	\begin{tikzpicture}[scale=1]
	\node (I)    at ( 0,0) {};
	
	\path 
	(I) +(0:6)   coordinate[label=0:$i^0$] (Iright)
	+(0:0) coordinate (Ileft)
	+(0:2) coordinate (tc)
	+(-90:6) coordinate[label=0:$i^-$] (Ibot);
	\draw (Ileft) -- 
	node[midway, above, sloped] {$\Sigma_{t^*_0}$}
	(Iright) -- 
	node[midway, below, sloped] {$\mathcal{I}^-$}
	(Ibot) --
	node[midway, below, sloped]    {\small $r=0$}    
	(Ileft) -- cycle;
	\draw (Ibot) to[out=60, in=-100]
	node[midway, above, sloped] {\tiny $r=R^*(t^*)$} (tc);
	\draw (4,-1) -- node[midway, above, sloped] {$v_0$} (4.5,-1.5);
	\draw (3,-2) -- node[midway, below, sloped] {$v_1$} (3.5,-2.5);
	\draw (4,-1) -- node[midway, above, sloped] {$I_{u_0}$} (3,-2);
	
	\end{tikzpicture}\label{fig:IutoI-}
\end{wrapfigure}

\subsection{Radiation Field}

In order to discuss the idea of a scattering map, we must first ensure that the radiation field actually exists. To do this, we look at the limiting process when we take $r\phi$ towards $\mathcal{I}^{+/-}$ to .

\begin{Proposition}[Existence of the Past Radiation Field]\label{Prop:radexist}
Let $\phi$ be a solution of \eqref{eq:wave} in $H^2_{c\forall \tau}$. with either reflective or permeating boundary conditions. Then there exists a function $\psi(v,\theta,\varphi)$ such that as $u\to-\infty$, $r\phi (u,v,\theta,\phi)\to\psi(v,\theta,\phi)$ in $H^1_{loc}$ (i.e.~ $\mathring{\slashed\nabla}(r\phi)(u,v,\theta,\phi)\to\mathring{\slashed\nabla}\psi(v,\theta,\phi)$ and $\p_v(r\phi)(u,v,\theta,\phi)\to\p_v\psi(v,\theta,\phi)$ in the $L^2_{loc}$ norm). Also, $\psi=0$ for sufficiently large $v$. Here, $v$ is the outgoing radial null coordinate and $u$ is the ingoing radial null coordinate, as defined in section \ref{sec:ext} and \ref{sec:int}. (We view $\psi$ as a function
$\psi:\mathcal{I}^-\to\R$.)
\end{Proposition}

\begin{Remark}
Which boundary conditions we use does not matter for this proof, as they coincide near $\mathcal{I}^\pm$. For all fixed $u$, there exists a $v_0$ such that $(u,v)$ is in the exterior of the star $\forall v\geq v_0$. We will do this proof for the reflective case, but it can be easily changed via coordinate transform to that of the permeating case.
\end{Remark}

\begin{proof}
	Let $\phi$ be a solution of \eqref{eq:wave}, which is in $H^2_{c\forall \tau}$. We will first consider a region local to an interval of $\mathcal{I}^-$. Let $D_{u_0,v_0,v_1}:=(-\infty,u_0]\times[v_1,v_0]\times S^2\subset \mathcal{M}$. Here we have chosen $v_0$ such that $\phi=0$ for all $v\geq v_0$ (as $\phi$ compactly supported on each $\Sigma_{\tau}$). We consider the surfaces of constant $u$, $I_u:=\Sigma_{u_0}\cap[v_1,v_0]$, for $u\in(-\infty,u_0]$. We will then let $u\to-\infty$. 
	
	Given $u_0$ negative and large enough, we have that $D_{u_0,v_0,v_1}$ is outside the star. Therefore $\p_{t^*}$ (or $\p_\tau$) is a Killing vector field, and $K^{\p_{t^*}}=0$. Thus we can integrate $K^{\p_{t^*}}$ in the region $[v_1,v_0] \times [u_1,u_0]\times S^2$ and apply Stokes' theorem. Now, as $\phi=0$ for all $v\geq v_0$, the surface term from $\Sigma_{v_0}\cap[u_1,u_0]$ is $0$. If we let $F_{[v_1,v_0]}$ be the surface term from $I_{u_0}$, we obtain
	\begin{equation}\label{eq:rphibound}
	F_{[v_1,v_0]}\geq c \int_{I_{u_1}}(\p_v\phi)^2+c\int_{\Sigma_{v_1}\cap(-\infty,u_0]}\left(\frac{1}{2r^2}\vert\mathring{\slashed\nabla}\phi\vert^2+\frac{2}{1-\frac{2M}{r}}(\p_u\phi)^2\right)\geq c'(v_0,v_1)\int_{I_{u_1}}(r\phi)^2 dvd\omega^2
	\end{equation}
	\begin{equation}\label{eq:pvrphibound}
	F_{[v_1,v_0]}\geq c \int_{I_{u_1}}(\p_v(r\phi))^2dvd\omega^2.
	\end{equation}
	Here $c=c(M)>0, c'=c'(M,v_0,v_1)>0$. For the first inequality, we have used the fact that $(\p_vr)/r\to0$ as $u\to-\infty$. We have also used Poincar\'e's inequality to bound $\Vert r\phi\Vert_{L^2(I_{u_1})}$ by $\Vert\p_u(r\phi)\Vert_{L^2(I_{u_1})}$. Note this is only possible as $\phi=0$ when $v=v_0$.
	
	The above integral bounds only required $\phi$ to be compactly supported and in $H^1(\Sigma_{t^*_0})$. As $\phi\in H^2_{c\forall\tau}$, we have $\Omega_i\phi\in H^2_{c\forall\tau}$. (Here $\Omega_i$ are the angular Killing fields given by equation \eqref{eq:AngularKillingFields}). By applying the above bounds to $\Omega_i\phi$, we can similarly obtain a uniform bound on $\mathring{\slashed\nabla}(r\phi)$ restricted to $I_{u_1}$. Thus we have $r\phi$ restricted to $I_u$ for $u\in\{u_0,u_0-1,u_0-2,...\}$ form a bounded sequence in $L^2$. They also have bounded angular derivatives and $v$ derivative (in $L^2$). This means there exists a subsequence with a limiting function $\psi\in H^1_{loc}(\mathcal{I}^-)$ such that:
	\begin{align}
	r\phi(u_i)&\xrightarrow[i\to\infty]{L^2(\{v\in[v_0,v_1]\}\times S^2)}\psi\\
	\p_v(r\phi)(u_i)&\xrightarrow[i\to\infty]{\text{weakly in } L^2(\{v\in[v_0,v_1]\}\times S^2)}\p_v\psi\\\label{eq:angularradconv}\mathring{\slashed\nabla}\left(r\phi\right)(u_i)&\xrightarrow[i\to\infty]{ L^2(\{v\in[v_0,v_1]\}\times S^2)}\mathring{\slashed\nabla}\psi.
	\end{align}
	Here strong convergence of the angular derivatives comes from commuting with the angular Killing vector fields: given any of the angular Killing fields in \eqref{eq:AngularKillingFields}, we can look at $\Omega_i\phi$. This is also a solution of the wave equation \eqref{eq:wave} for which \eqref{eq:rphibound} and \eqref{eq:pvrphibound} apply, so we can obtain a (strong) limit function $\Omega_i(r\phi)\to\psi_i\in L^2$ as above. By uniqueness of weak limits, we obtain that $\psi_i=\Omega_i\psi$, and thus the convergence is strong.
	
Next, we wish to show strong convergence of the $v$ derivative. We will show that the sequence $\p_v(r\phi)(u_i)$ is a Cauchy sequence in $L^2$, and therefore converges.

By integrating $K^{\p_{t^*}}$ between $I_{u_i}$ and $I_{u_j}$, we obtain the following equality:
\begin{equation}
\int_{I_{u_i}}\frac{1}{2r^2}\vert\mathring{\slashed\nabla}\phi\vert^2+\frac{2}{1-\frac{2M}{r}}(\p_v\phi)^2=\int_{I_{u_j}}\frac{1}{2r^2}\vert\mathring{\slashed\nabla}\phi\vert^2+\frac{2}{1-\frac{2M}{r}}(\p_v\phi)^2+\int_{\Sigma_{v_1}\cap[u_j,u_i]}\frac{1}{2r^2}\vert\mathring{\slashed\nabla}\phi\vert^2+\frac{2}{1-\frac{2M}{r}}(\p_u\phi)^2.
\end{equation}

We then rearrange this, and use equation \eqref{eq:rphibound} (with angular derivatives) and equation \eqref{eq:pvrphibound}. Introducing the volume form in the $I_u$ integrals, we obtain:
\begin{align}\label{eq:pvrphiCauchy}
\left\vert\int_{I_{u_i}}2(\p_v(r\phi))^2dvd\omega^2-\int_{I_{u_j}}2(\p_v(r\phi))^2dvd\omega^2\right\vert\leq&\left\vert\int_{\Sigma_{v_1}\cap[u_j,u_i]}\frac{1}{2r^2}\vert\mathring{\slashed\nabla}\phi\vert^2+\frac{2}{1-\frac{2M}{r}}(\p_u\phi)^2\right\vert\\\nonumber&+\frac{1}{2\min_{I_{u_i}}(r^2)}\left(\int_{I_{u_i}}\frac{1}{2r^2}\vert\mathring{\slashed\nabla}r\phi\vert^2dvd\omega^2+\int_{I_{u_j}}\frac{1}{2r^2}\vert\mathring{\slashed\nabla}r\phi\vert^2dvd\omega^2\right)
\end{align}

Thanks to equation \eqref{eq:angularradconv}, we know that the bracketed section on the right hand side of \eqref{eq:pvrphiCauchy} is bounded. As $\min_{I_{u_i}}(r^2)\to \infty$, we have this term tends to $0$ as $u_i\to-\infty$. From \eqref{eq:rphibound}, the integral on the surface $v=v_1$ in \eqref{eq:pvrphiCauchy} converges as we let $u_j\to-\infty$. This means that
\begin{equation}
\left\vert\int_{\Sigma_{v_1}\cap[u_j,u_i]}\frac{1}{2r^2}\vert\mathring{\slashed\nabla}\phi\vert^2+\frac{2}{1-\frac{2M}{r}}(\p_u\phi)^2\right\vert\leq\left\vert\int_{\Sigma_{v_1}\cap(-\infty,u_i]}\frac{1}{2r^2}\vert\mathring{\slashed\nabla}\phi\vert^2+\frac{2}{1-\frac{2M}{r}}(\p_u\phi)^2\right\vert\to 0\qquad \text{as }u_i\to -\infty
\end{equation}

Thus as $u_i\to-\infty$, we have that the right hand side of \eqref{eq:pvrphiCauchy} tends to $0$. Therefore the left hand side tends to $0$. This implies the sequence of $\p_v\phi$s converges strongly in $L^2$. By uniqueness of limits, they converge to the same limit as the weak $L^2$ limit above.

Finally, we wish to show that the limit of this sequence does not depend on the choice of $u_i$, i.e.~we wish to show

\begin{align}
r\phi(u)&\xrightarrow[u\to-\infty]{L^2(\{v\in[v_0,v_1]\}\times S^2)}\psi\\
\p_v(r\phi)(u)&\xrightarrow[u\to-\infty]{L^2(\{v\in[v_0,v_1]\}\times S^2)}\p_v\psi\\\mathring{\slashed\nabla}\left(r\phi\right)(u)&\xrightarrow[u\to-\infty]{ L^2(\{v\in[v_0,v_1]\}\times S^2)}\mathring{\slashed\nabla}\psi.
\end{align}

We obtain this by noting our argument for the $L^2$ convergence of $\p_v(r\phi)$ works for all sequences $u_i\to-\infty$. By Poincar\'e's inequality, this gives us convergence of $r\phi$ in the $L^2$ norm. Then, as before we can apply the angular Killing vector fields to obtain $L^2$ convergence of the angular derivatives.
\end{proof}
\begin{wrapfigure}{r}{7cm}
	\vspace{-5mm}
	\begin{tikzpicture}[scale=0.8]
	\node (I)    at ( 0,0) {};
	
	\path 
	(I) +(0:6)   coordinate[label=0:$i^0$] (Iright)
	+(0:0) coordinate (Ileft)
	+(0:2) coordinate (tc)
	+(-90:6) coordinate[label=0:$i^-$] (Ibot)
	+(0:5) coordinate (Iuright)
	+(-90:5) coordinate (Iubot)
	;
	\draw (Ileft) -- 
	node[midway, above, sloped] {$\Sigma_{t^*_0}$}
	(Iright) -- 
	node[midway, below, sloped] {$\mathcal{I}^-$}
	(Ibot) --
	node[midway, below, sloped]    {\small $r=0$}    
	(Ileft) -- cycle;
	\draw (Ibot) to[out=60, in=-100]
	node[midway, above, sloped] {\tiny $r=R^*(t^*)$} (tc);
	\draw (0,-4.4) -- node[midway, above, sloped] {\tiny $\{v=v_1\}$} (0.8,-5.2);
	\end{tikzpicture}
\end{wrapfigure}

\begin{Remark}\label{Rmk:ForwardRadExist}
	We could repeat this argument to show existence of the radiation field at $\mathcal{I}^+$. However, as the region $t^*\geq t^*_c$, or similarly $\tau\geq\tau_c$, is a region of Schwarzschild space-time, we will refer directly to results previously obtained on Schwarzschild. For example, existence of the radiation field is shown in \cite{Moschidis}. These results will be discussed in more detail in Section \ref{Sec:Schw}.
	
	One should also note that in the Schwarzschild case, the radiation field on $\mathcal{H}^+$ also plays a role. This will have to be considered for the forward scattering map.
\end{Remark}
\subsection{The Backwards Map $\mathcal{F}^-_{r,p}$}

Now that we have existence of the radiation field, we make the following definition. Given a solution, $\phi$ to the wave equation \eqref{eq:wave} smooth in a neighbourhood of $\mathcal{I}^-$, we define
\begin{equation}
\mathcal{F}^-_{r,p}(\phi_0,\phi_1):=\left(\lim_{u\to-\infty}r\phi\right)(v,\theta,\phi)
\end{equation}
where $\phi$ is the solution to \eqref{eq:wave} and \eqref{eq:initial} or \eqref{eq:initial2}. Here the subscript $r,p$ refers to the reflective and permeating cases respectively. This solution exists for finite energy $(\phi_0,\phi_1)$, by Theorems \ref{Thm:ReflExist} and \ref{Thm:PermExist}. This limit exists by Proposition \ref{Prop:radexist} as an $H^1_{loc}$ function. However, we do not yet know what function space this map $\mathcal{F}^-$ maps into. This motivates the following Theorem:

\begin{Theorem}[Non-degenerate Energy Boundedness of the Radiation Field]\label{Thm:RadBound}
	Let $X$ be the vector given in the proof of Theorem \ref{Thm:ForwardReflBound} (reflective) or in Lemma \ref{lem:TimeBound} (permeating). We then have that $\mathcal{F}^-_{r,p}$ is bounded with respect to $X$ energy on ${\Sigma_{t^*_c}}$ or ${\Sigma_{\tau_{c^-}}}$. Therefore
	\begin{align}
		\mathcal{F}^-_r:\mathcal{E}^X_{\Sigma_{t^*_c}}\to\mathcal{E}^{\p_{t^*}}_{\mathcal{I}^-}\\
		\mathcal{F}^-_p:\mathcal{E}^X_{\Sigma_{\tau_{c^-}}}\to\mathcal{E}^{\p_\tau}_{\mathcal{I}^-}.
	\end{align} 
	Furthermore, let
	\begin{equation}
	\mathcal{F}^+_{r,p}:\Ima(\mathcal{F}^-_{r,p})\to\mathcal{E}^X_{\Sigma_{t^*_c,\tau_{c^-}}}
	\end{equation}
	be the inverse of $\mathcal{F}^-_{r,p}$ where it is defined. Then this inverse is also bounded.
\end{Theorem}
\begin{proof}
	For the reflective case, we prove this result for data in $C^\infty_{c\forall t^*}$. For the permeating case,we prove this result for data in $H^2_{c\forall \tau}\cap\Omega_i^{-1}(H^2_{c\forall \tau})$, i.~e.~ $\phi\in H^2_{c\forall \tau}$ and $\Omega_i\phi\in H^2_{c\forall \tau}$ for each $i$. Here $\Omega_i$ are the angular Killing fields defined in \eqref{eq:AngularKillingFields}. This result then applies for all functions in $\mathcal{E}^X_{\Sigma_{t^*,\tau}}$ by an easy density argument. We have already established the existence of the radiation field in $H^1_{loc}(\mathcal{I}^-)$ for a dense subset of $\mathcal{E}^X_{\Sigma_{t^*,\tau}}$ by Proposition \ref{Prop:radexist}.
	
	We now establish boundedness of $\mathcal{F}^-$. We will use the vector fields and energy currents used to prove Theorem \ref{Thm:BackReflBound} and \ref{Thm:BackPermBound}. We will integrate these energy currents in a region bounded by a time slice, say $\Sigma_{t^*_0}$ or $\Sigma_{\tau_0}$, and by a surface ${v=v_1}$. We will also need to include the surface of the star for the reflective case. We then have that the boundary term from the surface of the star has the correct sign, as does the term from $\{v=v_1\}$. We also know that the integral over $\Sigma_{t^*_0}$/$\Sigma_{\tau_0}$ converges, and finally that the bulk term $K^X$ is non-positive. Using these, we obtain:
	\begin{equation}
	\int_{\Sigma_{t^*_0}}(-dt^*(J^X))=\int_{\{v=v_1\}}(-dv(J^X))-\int_{t^*=t^*_0}^{t^*_1}K^X+\int_{\mathcal{I}^-\cap\{v>v_1\}}(-du(J^X))\geq\int_{\mathcal{I}^-\cap\{v>v_1\}}(-du(J^X)).
	\end{equation} 
	
	Thus we have uniform bounds on the integral over $\mathcal{I}^-\cap\{v\geq v_1\}$ for all $v_1$. By taking the limit $v\to-\infty$, we obtain a bound on the integral over $\mathcal{I}^-$.
	
	We then choose $R^*$ sufficiently large, and $J^{X,w}$ as in the proof of Theorem \ref{Thm:BackReflBound}. In the reflective case, we have
	\begin{equation}
	-du(J^{X,w})=2\left(1-\frac{2M}{r}\right)^{-1}\left(1+\frac{1}{\log(R^*/2M)}-\frac{r+2M}{r-2M}\epsilon\right)(\p_v\phi)^2+\frac{1}{2r^2}\left(1+\frac{1}{\log(R^*/2M)}+\epsilon\right)\vert\mathring{\slashed\nabla}\phi\vert^2+\phi-terms.
	\end{equation}
	
	Now, we know that $r\phi$ tends to a $\dot{H}^1$ limit in $[v_1,v_0]$. This means that the $\phi$ terms tend to $0$ (as $r\to\infty$ for $u\to-\infty$, $v\in[v_1,v_0]$). Thus we obtain 
	
	\begin{align}\nonumber
	-\int_{\Sigma_{t^*}}dt^*(J^{X,w})&\geq-\int_{\mathcal{I}^-\cap[v_1,v_0]}du(J^{X,w})\\\nonumber
	&=\int_{\mathcal{I}^-\cap[v_1,v_0]}2\left(1-\frac{2M}{r}\right)^{-1}\left(1+\frac{1}{\log(R^*/2M)}-\frac{r+2M}{r-2M}\epsilon\right)(\p_v\phi)^2
	+\frac{1}{2r^2}\left(1+\frac{1}{\log(R^*/2M)}+\epsilon\right)\vert\mathring{\slashed\nabla}\phi\vert^2\\\nonumber
	&\geq\int_{\mathcal{I}^-\cap[v_1,v_0]}\left(1+\frac{1}{\log(R^*/2M)}-\frac{r+2M}{r-2M}\epsilon\right)\left(\p_v (r\phi)-\phi\p_v r\right)^2dvd\omega^2\\\label{eq:ScatBackReflBound}
	&\geq C\int_{\mathcal{I}^-\cap[v_1,v_0]}\left(1+\frac{1}{\log(R^*/2M)}-\epsilon\right)(\p_v(r\phi))^2dvd\omega^2\sim\int_{\mathcal{I}^-\cap[v_1,v_0]}(\p_v(r\phi))^2dvd\omega^2.
	\end{align}
	
	As the upper bound in \eqref{eq:ScatBackReflBound} is independent of $v_1$, we can let $v_1\to-\infty$ to obtain a bound on the integral over all of $\mathcal{I}^-$.

	Thus $\mathcal{F}^-$ is bounded. Next we show that the inverse of $\mathcal{F}^-$ is bounded. For this, we first need to prove a Proposition:
	
	\begin{Proposition}\label{prop:decay}
		Let $\phi\in C^\infty_{c\forall t^*}$ be any solution of \eqref{eq:wave} with reflective boundary conditions. Then there exists a sequence $v_i\to-\infty$ such that 
		\begin{equation}\label{eq:Refldecay}
		\int_{\Sigma_{v_i}}dn(J^{T})\to 0 \quad as\quad v\to-\infty,
		\end{equation}
		for $J^T$ as in Theorem \ref{Thm:ForwardReflBound}.
		
		Similarly, let $\phi\in H^2_{c\forall \tau}$ a solution to \eqref{eq:wave} with permeating boundary conditions and $\Omega_i\phi\in H^2_{c\forall \tau}$. We then have that there exists a sequence $v_i\to-\infty$ such that
		\begin{equation}\label{eq:Permdecay}
		\int_{\tilde\Sigma_{v_i}}dn(J^{Y})\to 0 \quad as\quad v\to-\infty,
		\end{equation}
		for $J^Y$ as in Theorem \ref{Prop:UniBound}.
	\end{Proposition}
	
	\begin{proof}

		We will look at the reflective case first, and calculate $-dv(J^T)$:
	\begin{equation}
	-\int_{\Sigma_{v_0}}dv(J^{T})=\int_{\Sigma_{v_0}}\frac{1}{2r^2}\vert\mathring{\slashed\nabla}\phi\vert^2+\frac{2(\p_u\phi)^2}{1-\frac{2M}{r}}=\int_{\Sigma_{v_0}}\left(\frac{1-\frac{2M}{r}}{4r^2}+(\p_u\phi)^2\right)r^2dud\omega^2
	\end{equation}

	We then proceed by using the $r^p$ method \cite{NewPhysSpace}. We consider $\psi:=r\phi$. As $\phi$ is a solution of the wave equation \eqref{eq:wave}, we have:
	\begin{equation}
		\frac{4}{1-\frac{2M}{r}}\p_v\p_u\psi=\frac{1}{r^2}\mathring{\slashed\triangle}\psi-\frac{2M}{r^3}\psi.
	\end{equation}

	We then integrate from $\Sigma_{v_1}$ back to $\Sigma_{v_0}$. We are using the volume form $dudvd\omega^2$, so there is \emph{no factor of $r^2$}.
	\begin{align}\nonumber
	\int_{\Sigma_{v_1}}\frac{4r}{1-\frac{2M}{r}}(\p_u\psi)^2dud\omega^2\geq&\left(\int_{\Sigma_{v_1}}-\int_{\Sigma_{v_0}}-\int_{S_{[v_0,v_1]}}\right)\left(\frac{4r}{1-\frac{2M}{r}}(\p_u\psi)^2dud\omega^2\right)\\\nonumber
	=&\int_{v_0}^{v_1}\p_{v}\left(\int_{\Sigma_{v}}\frac{4r}{1-\frac{2M}{r}}(\p_u\psi)^2dud\omega^2\right)dv\\\nonumber
	=&\iint_{v_0}^{v_1}\p_u\psi\left(\frac{8r}{1-\frac{2M}{r}}\p_v\p_u\psi+\frac{1-\frac{4M}{r}}{1-\frac{2M}{r}}4\p_u\psi\right)dudvd\omega^2\\\label{eq:ReflDecay}
	=&\iint_{v_0}^{v_1}2r\p_u\psi\left(\frac{1}{r^2}\mathring{\slashed\triangle}\psi-\frac{2M}{r^3}\psi\right)+\frac{1-\frac{4M}{r}}{1-\frac{2M}{r}}4(\p_u\psi)^2dudvd\omega^2\\\nonumber
	=&\iint_{v_0}^{v_1}-\frac{1}{r}\p_u(\vert\mathring{\slashed\nabla}\psi\vert^2)-\frac{2M}{r^2}\p_u(\psi^2)+\frac{1-\frac{4M}{r}}{1-\frac{2M}{r}}4(\p_u\psi)^2dudvd\omega^2\\\nonumber
	=&\int_{\mathcal{I}^-}\left(\frac{1}{r}\vert\mathring{\slashed\nabla}\psi\vert^2-\frac{2M}{r^2}\psi^2\right)dvd\omega^2\\\nonumber
	&+\iint_{v_0}^{v_1}\p_u\left(\frac{1}{r}\right)\vert\mathring{\slashed\nabla}\psi\vert^2+\p_u\left(\frac{2M}{r^2}\right)\psi^2+\frac{1-\frac{4M}{r}}{1-\frac{2M}{r}}4(\p_u\psi)^2dudvd\omega^2\\\nonumber
	\geq&\iint_{v_0}^{v_1}\left(1-\frac{2M}{r}\right)\left(\frac{1}{2r^2}\vert\mathring{\slashed\nabla}\psi\vert^2+\frac{2M}{r^3}\psi^2\right)+\frac{1-\frac{4M}{r}}{1-\frac{2M}{r}}4(\p_u\psi)^2dudvd\omega^2dudvd\omega^2\\\nonumber
	\geq&\iint_{v_0}^{v_1}\left(\frac{1-\frac{2M}{r}}{2r^2}\vert\mathring{\slashed\nabla}\phi\vert^2+(\p_u\phi)^2\right)r^2dudvd\omega^2\geq \int_{v_0}^{v_1}\left(\int_{\Sigma_{v}}(-dv(J^T))\right)dv.
	\end{align}
	Here we have integrated by parts to go from the third to fourth line and from the fourth to fifth line. As $\phi=\vert\mathring{\slashed\nabla}\phi\vert=0$ on the surface of the star, this boundary term from vanishes.

	To go from the penultimate to the final line, we have used that
	\begin{align}\nonumber
	\int_{\Sigma_{v}}\frac{1-\frac{4M}{r}}{1-\frac{2M}{r}}4(\p_u\psi)^2dud\omega^2&\geq\int_{\Sigma_{v}}(\p_u\psi)^2dud\omega^2=\int_{\Sigma_{v}}r^2(\p_u\phi)^2+\p_u\left(r\p_u r \phi^2\right)-\p_u^2 r\frac{\psi^2}{r}dud\omega^2\\
	&\geq\int_{\Sigma_{v}}r^2(\p_u\phi)^2-\left(1-\frac{2M}{r}\right)\left(1-\frac{e^2M}{r}\right)\frac{M}{2r^3}\psi^2dud\omega^2.
	\end{align}

	The upper bound that we started with in inequality \eqref{eq:ReflDecay} is independent of $v_0$. As we let $v_0\to-\infty$, we obtain
	\begin{equation}
	\int_{-\infty}^{v_1}\left(\int_{\Sigma_{v}}(-dv(J^T))\right)dv\leq \int_{\Sigma_{v_1}}\frac{4r}{1-\frac{2M}{r}}(\p_u\psi)^2dud\omega^2<\infty.
	\end{equation}

	Therefore there exists a sequence $v_i\to-\infty$ such that $\int_{\Sigma_{v_i}}(-dn(J^T))\to 0$, as required. This gives the result in the reflective case, \eqref{eq:Refldecay}.

	Next we look at the permeating case. For this case, let $Y=h(\tau)\p_\tau$. We then have
	\begin{equation}
	-dn(J^Y)=\begin{cases}\frac{h}{2}\left(\frac{1}{r^2}\vert\mathring{\slashed\nabla}\phi\vert^2+\left(1-\sqrt{\frac{2M}{r}}\right)\left(1+\sqrt{\frac{2M}{r}}\right)^{-1}\left(\p_\tau\phi-\left(1+\sqrt{\frac{2M}{r}}\right)\p_r\phi\right)^2\right)&r\geq r_b\\
	\frac{h}{2}\left(\frac{1}{r^2}\vert\mathring{\slashed\nabla}\phi\vert^2+(\p_{\tau}\phi)^2+\left(1-\frac{2M r^2}{r_b^3}\right)(\p_{r}\phi)^2\right) & r<r_b
	\end{cases}
	\end{equation}
	where $n$ is the normal to $\tilde{\Sigma}_{\tau}$.

	We then perform something similar to the reflective case above. However, as we do not have an explicit coordinate system $u, v$, we will use $\p_\tau$ and $\p_r$. Let $f$ be given by
	\begin{equation}
	f(\tau,r)=\begin{cases}
	\sqrt{\frac{2Mr^2}{r_b^3}}&r<r_b\\
	\sqrt{\frac{2M}{r}}&r\geq r_b\end{cases}
	\end{equation}
	so our metric is of the form
	\begin{equation}
	g=-(1-f^2)d\tau^2+2fd\tau dr+dr^2+r^2g_{S^2}.
	\end{equation}
	Again, let $\psi:=r\phi$. We obtain
	\begin{equation}
	(\p_\tau+(1-f)\p_r)(\p_\tau-(1+f)\p_r)\psi=f'(\p_\tau-(1+f)\p_r)\psi+(2ff'-\dot{f})\frac{\psi}{r}+\frac{1}{r^2}\mathring{\slashed\triangle}\psi.
	\end{equation}
	Note that $(2ff'-\dot{f})$ is not continuous over $r=r_b$.

	Then, in a similar way to the reflective case, we obtain that:
	\begin{align}\nonumber
	C\int_{\Sigma_{v_1}}r((\p_\tau-(1+f)\p_r)\psi)^2dud\omega^2\geq&\iint_{v_0}^{v_1}(\p_\tau+(1-f)\p_r)(r(\p_\tau-(1+f)\p_r)\psi)^2dudvd\omega^2\\\nonumber
	=&\iint_{v_0}^{v_1}\Bigg(-(\p_\tau-(1+f)\p_r)\left(\frac{1}{r}\vert\mathring{\slashed\nabla}\psi\vert^2-(2ff'-\dot{f})\psi^2\right)+(1+f)\frac{1}{r^2}\vert\mathring{\slashed\nabla}\psi\vert^2\\\nonumber
	&-[(\p_\tau-(1+f)\p_r)(2ff'-\dot{f})]\psi^2\\\nonumber
	&+(1+2rf'-f)((\p_\tau-(1+f)\p_r)\psi)^2 \Bigg)dudvd\omega^2\\\nonumber
	\geq&\iint_{v_0}^{v_1}\frac{1}{r^2}\vert\mathring{\slashed\nabla}\psi\vert^2+\frac{1}{2}((\p_\tau-(1+f)\p_r)\psi)^2 dudvd\omega^2\\\label{eq:PermDecay}
	&+\iint_{r< r_b}\frac{M}{2r_b^3}\psi^2dudvd\omega^2+\iint_{r\geq r_b}\frac{2M}{r^3}\psi^2dudvd\omega^2\\\nonumber
	&+\int_{r=r_b}(2ff'-\dot{f})\vert_{r_b^-}^{{r_b}^+}\psi^2dvd\omega^2+\int_{r=0}(2ff'-\dot{f})\psi^2dvd\omega^2\\\nonumber
	\geq&\iint_{v_0}^{v_1}\frac{1}{r^2}\vert\mathring{\slashed\nabla}\psi\vert^2+\frac{1}{2}((\p_\tau-(1+f)\p_r)\psi)^2 dudvd\omega^2\\\nonumber
	&+\iint_{r< r_b}\frac{M}{2r_b^3}\psi^2dudvd\omega^2+\iint_{r\geq r_b}\frac{2M}{r^3}\psi^2dudvd\omega^2-\int_{r=r_b}\frac{3M}{r_b^2}\psi^2dvd\omega^2.
	\end{align}

	We will only be using this $r^p$ method to bound the $Y$-energy for the exterior, so we note
	\begin{align}\nonumber
	\int_{\Sigma_{v}\cap\{r\geq r_b\}}(\p_u\psi)^2dud\omega^2&=\int_{\Sigma_{v}\cap\{r\geq r_b\}}r^2(\p_u\phi)^2+\p_u\left(r\p_u r \phi^2\right)-\p_u^2 r\frac{\psi^2}{r}dud\omega^2\\\nonumber
	&\geq\int_{\Sigma_{v}\cap\{r\geq r_b\}}r^2(\p_u\phi)^2dud\omega^2-\int_{r=r_b}\frac{2}{r_b}\psi^2dud\omega^2\\
	&=\int_{\Sigma_{v}\cap\{r\geq r_b\}}r^2(\p_u\phi)^2-\int_{r=r_b}\frac{2}{r_b}\psi^2dud\omega^2.
	\end{align}

	Now, we need to bound the $\psi/{r_b}$ surface term, and therefore also bound the $3M\psi/r_b^2$ surface term in \eqref{eq:PermDecay}.

	We proceed by noting
	\begin{align}\nonumber
	\int_{r=r_b}\frac{1}{r}\psi^2dud\omega^2&=-\int_{r< r_b}(\p_\tau-(1+f)\p_r)\left(\frac{1}{r}\chi\left(\frac{2Mr^2}{r_b^3}\right)\psi^2\right)dudvd\omega^2\\\nonumber
	&=\int_{r<r_b}-\frac{2}{r}\psi(\p_\tau-(1+f)\p_r)\psi-(1+f)\frac{1}{r^2}\psi^2+\left((1+f)\frac{4M}{r_b^3}+\frac{2Mr}{r_b^4}\dot{r_b}\right)\chi'\psi^2dudvd\omega^2\\\nonumber
	&\leq \int_{r<r_b}((\p_\tau-(1+f)\p_r)\psi)^2-\left(\frac{1}{r}\psi+(\p_\tau-(1+f)\p_r)\psi\right)^2+\frac{CM}{r_b^3}\psi^2dudvd\omega^2\\\label{eq:SurfaceDecayBound}
	&\leq\int_{r<r_b}((\p_\tau-(1+f)\p_r)\psi)^2+\frac{CM}{r_b^3}\psi^2dudvd\omega^2.
	\end{align}

	Dividing equations \eqref{eq:PermDecay} and \eqref{eq:SurfaceDecayBound} by $4C$, say, we can absorb the $\psi^2$ term in \eqref{eq:SurfaceDecayBound} by our $\psi^2$ bulk term in \eqref{eq:PermDecay}. This gives us
	\begin{equation}
	\int_{\Sigma_{v_1}}r((\p_\tau-(1+f)\p_r)\psi)^2dud\omega^2\geq c\int_{v_0}^{v_1}\int_{\Sigma_{v}\cap\{r\geq r_b\}}\left(\frac{1}{r^2}\vert\mathring{\slashed\nabla}\phi\vert^2+(\p_u\phi)^2\right)\geq c\int_{v_0}^{v_1}\int_{\Sigma_{v}\cap\{r\geq r_b\}}(-dn(J^T)).
	\end{equation}
		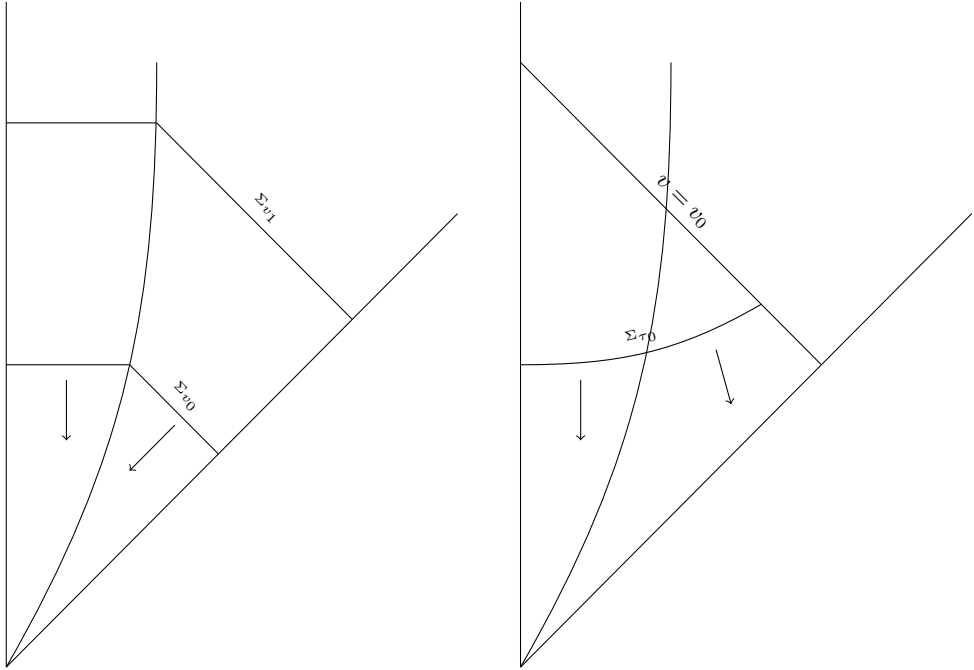
\begin{figure}[H]
			\centering
			\begin{tikzpicture}[scale=4]
			\draw (0,0.2) to (0,-2);
			\draw (0,-2) to[in=-90,out=60] (0.5,0);
			\draw (0,-2) to (1.5,-0.5);
			\draw (0,-0.2) to (0.5,-0.2);
			\draw (0.5,-0.2) to node[midway, above, sloped] {\tiny $\Sigma_{v_1}$} (1.15,-0.85);
			\draw (0,-1) to (0.41,-1);
			\draw (0.41,-1) to node[midway, above, sloped] {\tiny $\Sigma_{v_0}$} (0.705,-1.295);
			\draw[->] (0.2,-1.05) to (0.2,-1.25);
			\draw[->] (0.56,-1.2) to (0.41,-1.35);
			\end{tikzpicture}$\qquad$
			\begin{tikzpicture}[scale=4]
			\draw (0,0.2) to (0,-2);
			\draw (0,-2) to[in=-90,out=60] (0.5,0);
			\draw (0,-2) to (1.5,-0.5);
			\draw (0,0) to node[midway, above, sloped] {\small $v=v_0$} (1,-1);
			\draw (0,-1) to[in=-150,out=0] node[midway, above, sloped] {\tiny $\Sigma_{\tau_0}$} (0.8,-0.8);
			\draw[->] (0.2,-1.05) to (0.2,-1.25);
			\draw[->] (0.65,-0.95) to (0.70,-1.13);
			\end{tikzpicture}
			\caption{Left: Region for integrating in the $r^p$ method as above. Right: Region for integrating for interior decay see below.}
		\end{figure}
	Finally, we need to consider the interior of our star. To do this, we will restrict ourselves to $v\leq v_1<0$. In this region, we will consider
	\begin{equation}
	X=-\sqrt{\frac{2Mr^2}{r_b^3}}\p_r, \qquad w=-\frac{1}{2}\sqrt{\frac{2M}{r_b^3}}.
	\end{equation}
	We will integrate the modified current $J^{X,w}$, as in \eqref{eq:modcurrent}.

	This has the properties that:
	\begin{align}
	-\nabla.J^{X,w}&\geq 0\\
	-\nabla.J^{X,w}\vert_{r< r_b}&\geq c\sqrt{\frac{2M}{r_b^3}}(-d\tau(J^Y))=\p_\tau(\log r_b)c(-d\tau(J^Y))\\
	\vert d\tau(J^{X,w})\vert&\leq A(-d\tau(J^Y))+\sqrt{\frac{2M}{r_b^3}}\psi d\tau(\phi)+\frac{6M}{4{r_b}^3}\phi^2.
	\end{align}
	We now integrate $\nabla.J^{X,w}$ in the region $R_{\tau_0,v_1}:=\{v\leq v_1, \tau\geq\tau_0\}$ to obtain:
	\begin{multline}\label{eq:QILED}
	\int_{\Sigma_{v_1}\cap\{\tau\leq\tau_0\}}-dn(J^{X,w})-\int_{\Sigma_{\tau_0}\cap\{v\leq v_1\}}-d\tau(J^{X,w})=c\int_{R_{\tau_0,v_1}}-\nabla.J^{X,w}\geq\int_{[\tau_0,\tau_1]\cap \{r< r_b\}}\sqrt{\frac{2M}{r_b^3}}(-d\tau(J^Y)).
	\end{multline}
	Here $\tau_1$ is chosen such that $(\tau_1,r_b(\tau_1))$ has $v\leq v_1$.

	The terms on the left hand side of equation \eqref{eq:QILED} can both be bounded by data on $\Sigma_{v_1}$. 

	As $\frac{2M}{r_b^3}\to 0$, by choosing $\tau_1$ sufficiently far back, we can combine equations \eqref{eq:PermDecay} and \eqref{eq:QILED} to see
	\begin{equation}
	C\geq \int_{\tau_0}^{\tau_1}\sqrt{\frac{2M}{r_b^3}}\int_{\tilde{\Sigma}_{\tau}}(-dn(J^Y)).
	\end{equation}
	This $C$ only depends our initial data on $\Sigma_{v_1}$, and is finite for compactly supported data. As $\sqrt{\frac{2M}{r_b^3}}$ is not integrable, there must be a sequence of $v_i\to\infty$ such that the required result holds. Thus we have proven this for the permeating case, \eqref{eq:Permdecay}.

	\end{proof}

	We now return to proving Theorem \ref{Thm:RadBound}. As $\Omega_i\phi$ is also $H^2_{c\forall \tau}$, we can obtain that, from equation \eqref{eq:rphibound},
	\begin{equation}
	F'_{[v_1,v_0]}\geq C \sum_{i=1}^3\int_{I_u\cap [v_1,v_0]}(\p_v\Omega_i\phi)^2\geq C(v_0,v_1)\int_{I_u\cap [v_1,v_0]}\vert \mathring{\slashed\nabla} r\phi\vert^2 dvd\omega^2
	=C(v_0,v_1)\int_{I_u\cap [v_1,v_0]}\vert \mathring{\slashed\nabla}\phi\vert^2 r^2dvd\omega^2.
	\end{equation}
	Thus, we have
	\begin{equation}
	\int_{I_u\cap [v_1,v_0]}\vert \mathring{\slashed\nabla}\phi\vert^2 dvd\omega^2\leq\frac{F'}{C\inf_{I_u\cap [v_1,v_0]}{r}^2}\to 0\ as\ u\to-\infty.
	\end{equation}
	
	Given this, we can simply use $X=\p_{t^*}$, $w=0$ for all angular frequencies. Then $K^X=0$, $-dt^*(J^X)\sim\Vert\phi\Vert_{\dot{H}^1}^2$, and $d\rho(J^X)<0$.
	\begin{multline}
	-\int_{\mathcal{I}^-}du(J^{X})=\lim_{u\to-\infty}\int_{I_u}\Bigg(\left(\p_v (r\phi)-\phi\p_v r\right)^2+\frac{1}{4}\left(1-\frac{2M}{r}\right)\vert\mathring{\slashed\nabla}\phi\vert^2\Bigg)dvd\omega^2
	=\int_{\mathcal{I}^-}(\p_v(r\phi))^2dvd\omega^2.
	\end{multline}
	Here the first equality uses Proposition \ref{prop:decay}.

	Thus, again by Stokes' theorem, we have that $\mathcal{F}^-_r$ is 
	bounded. We also $\mathcal{F}^+_r$ is bounded, where this inverse is defined.
	
	When we look at the permeating case, we use the $\tau,r,\theta,\varphi$ metric. We consider $-du(J^Y)$ where $Y=(1\pm(2M/r_b)^{1/4})\p_\tau$. In these coordinates, the null hypersurfaces (when outside the star) have a normal given by $du=d\tau-(1-\sqrt{2M/r})^{-1}dr$. When inside the star, they need to have a normal given by $du=\alpha(d\tau-(1-\sqrt{2Mr^2/r_b^3})^{-1}dr)$ (the u coordinate exists, thanks to Frobenius' theorem, see for example \cite{ReallBH}, with $\alpha$ bounded above and away from 0). On $I_u$, we have 
	\begin{equation}
	-du(J^Y)=\begin{cases}
	2\left(1\pm\left(\frac{2M}{r_b}\right)^{1/4}\right)\left(1-\frac{2M}{r}\right)^{-1}(\p_v\phi)^2+\frac{1}{2r^2}\left(1\pm\left(\frac{2M}{r_b}\right)^{1/4}\right)\vert\mathring{\slashed\nabla}\phi\vert^2 & r\geq r_b
	\\2\alpha\left(1\pm\left(\frac{2M}{r_b}\right)^{1/4}\right)\left(1-\frac{2Mr^2}{r_b^3}\right)^{-1}(\p_v\phi)^2+\frac{\alpha}{2r^2}\left(1\pm\left(\frac{2Mr^2}{r_b^3}\right)^{1/4}\right)\vert\mathring{\slashed\nabla}\phi\vert^2 & r< r_b
	\end{cases}.
	\end{equation}
	So integrating $K^Y$ and taking the same limits as in the reflective case, we obtain 
	\begin{equation}
	-\int_{\mathcal{I}^-}du(J^Y)=
	\int_{\mathcal{I}^-}\left(1\pm\left(\frac{2M}{r_b}\right)^{1/4}\right)(\p_v(r\phi))^2dvd\omega^2\sim\int_{\mathcal{I}^-}(\p_v(r\phi))^2dvd\omega^2.
	\end{equation}
	This gives, by Stokes', that $\mathcal{F}^-_p$ and $\mathcal{F}^+_p$ are bounded.
\end{proof}

\begin{Proposition}[Density of the Image of the Backwards Scattering Map]\label{Prop:ImageDensity}
	The image of $\mathcal{F}^-_{r,p}$ is dense in $\mathcal{E}^{\p_{t^*,\tau}}_{\mathcal{I}^-}$, with respect to the $\p_{t^*,\tau}$-energy. As the inverse $\mathcal{F}^+_{r,p}$ is bounded on that image, this gives us that 
	\begin{equation}
	\Ima(\mathcal{F}^-_{r,p})=\mathcal{E}^{\p_{t^*,\tau}}_{\mathcal{I}^-}.
	\end{equation}
	Therefore that our scattering map has a well defined inverse on this space.
	
\end{Proposition}
\begin{proof}
	We prove this by showing that the set of all compactly supported smooth functions on $\mathcal{I}^-$ are in the image of $\mathcal{F}^-_{r,p}$. We will consider the reflective case, though the permeating case follows exactly the same logic. 
	Given $\psi:\mathcal{I}^-\to\R$ a compactly supported smooth function, suppose $\psi=0$ for $v<v_0$. Then we can pick $\Sigma_{t^*_0}$ such that $\{t^*>t^*_0\}\cap\{v>v_0\}$ is entirely outside the star. Therefore existence of the solution in this region is entirely in the Schwarzschild exterior space-time. This means we can refer to previous existence proofs in the Schwarzschild case, such as Theorem 4 in \cite{KerrScatter}.
	
	Once we have that the image of $\mathcal{F}^-_{r,p}$ is dense in $\mathcal{E}^{\p_{t^*,\tau}}_{\mathcal{I}^-}$, we proceed as follows. Given any element $\psi\in\mathcal{E}^{\p_{t^*,\tau}}_{\mathcal{I}^-}$, there exists a sequence in the image of $\mathcal{F}^-_{r,p}$, $\mathcal{F}^-_{r,p}(\phi_i)$ such that $\mathcal{F}^-_{r,p}(\phi_i)\to\psi$. As $\mathcal{F}^-_{r,p}(\phi_i)$ is a Cauchy sequence and the inverse of $\mathcal{F}^-_{r,p}$ is bounded, we have that $\phi_i$ is also a Cauchy sequence. $\mathcal{E}^X_{\Sigma_{t^*}}$ is complete, so $\phi_i$ converges to some $\phi\in\mathcal{E}^X_{\Sigma_{t^*}}$. As $\mathcal{F}^-_{r,p}$ is bounded and linear, $\mathcal{F}^-_{r,p}(\phi)=\psi\in \Ima(\mathcal{F}^-_{r,p})$. Therefore $\Ima(\mathcal{F}^-_{r,p})=\mathcal{E}^{\p_{t^*,\tau}}_{\mathcal{I}^-}$.
\end{proof}

This leads us to our final result for this section.

\begin{Theorem}[Existence and Boundedness of the Forward Maps]\label{Thm:RadBackBound}
	The forward maps
	\begin{equation}
	\mathcal{F}^+_{r}:\mathcal{E}^{\p_{t^*_c}}_{\mathcal{I}^-}\to\mathcal{E}^X_{\Sigma_{t^*}}
	\end{equation}
	\begin{equation}
	\mathcal{F}^+_{p}:\mathcal{E}^{\p_{\tau}}_{\mathcal{I}^-}\to\mathcal{E}^X_{\Sigma_{\tau_{c^-}}}
	\end{equation}
	are well defined and bijective, bounded with bounded inverses.
\end{Theorem}
\begin{proof}
	From Theorem \ref{Thm:RadBound} we have that this map (and its inverse) are bounded, where they are defined. We then have from Proposition \ref{Prop:ImageDensity} that $\Ima(\mathcal{F}^-_{r,p})=\mathcal{E}^{\p_{t^*_c}}_{\mathcal{I}^-}$, which then gives us our result.
\end{proof}

\subsection{Schwarzschild Scattering and the Future Radiation Field}\label{Sec:Schw}

If we restrict our space-time to the region $t^*\geq t^*_c$, or similarly $\tau\geq\tau_c$, then this is a subregion of Schwarzschild space-time.

We now define the following map:
\begin{align}\label{eq:Gdef}
\mathcal{G}^+_{r,p}: \mathcal{E}^X_{\Sigma_{t^*_c}}&\to\mathcal{E}^{\p_{t^*,\tau}}_{\mathcal{H}^+}\times\mathcal{E}^{\p_{t^*,\tau}}_{\mathcal{I}^+}\\\nonumber
(\phi_0,\phi_1)&\mapsto\left(\phi|_{\mathcal{H}^+},\left(\lim_{v\to\infty}r\phi\right)(u,\theta,\phi)\right)
\end{align}
where $\phi$ is the solution to \eqref{eq:wave} and \eqref{eq:initial} or \eqref{eq:initial2}.  As this map is in the region $t^*\geq t^*_c$, we have the following Proposition from \cite{KerrScatter}:
\begin{Proposition}\label{prop:SwarzBound}
	The map $\mathcal{G}^+_{r,p}$, as defined by \eqref{eq:Gdef}, exists, is bounded, injective, non-surjective. Its inverse $\mathcal{G}^-_{r,p}$ (considered as a map from $\Ima(\mathcal{G}^+_{r,p})$ to $\mathcal{E}^X_{\Sigma_{t^*_c}}$) is unbounded. We also have that $\Ima(\mathcal{G}^+_{r,p})$ is dense in $\mathcal{E}^{\p_{t^*,\tau}}_{\mathcal{H}^+}\times\mathcal{E}^{\p_{t^*,\tau}}_{\mathcal{I}^+}$.
	
	However, when considered as a map from $\mathcal{E}^{T}_{\Sigma_{t^*_c}}$, $\mathcal{G}^+_{r,p}$ is unitary, injective and surjective. Thus it has unitary inverse, $\mathcal{G}^-_{r,p}$.
\end{Proposition}

See Proposition \ref{prop:Ebound} for a statement of what the degenerate $T$-energy looks like.

The unboundedness of $\mathcal{G}^-_{r,p}$ contrasts with the boundedness of $\mathcal{F}^+_{r,p}$, and is caused by the blue-shift instability along the (time reversed) future horizon $\mathcal{H}^+$.

\subsection{The Scattering Map}

Thus we can consider the scattering map:
\begin{align}
\mathcal{S}^+_{r,p}&:\mathcal{E}^{\p_{t^*,\tau}}_{\mathcal{I}^-}\to\mathcal{E}^{\p_{t^*,\tau}}_{\mathcal{H}^+}\times\mathcal{E}^{\p_{t^*,\tau}}_{\mathcal{I}^+}\\
\mathcal{S}^+_{r,p}&:=\mathcal{G}^+_{r,p}\circ\mathcal{F}^+_{r,p}
\end{align}
to map the radiation field at $\mathcal{I}^-$ to the radiation fields at $\mathcal{I}^+\cup\mathcal{H}^+$. We now consider boundedness of both this map and its inverse, $\mathcal{S}^-_{r,p}$, defined only on $\Ima(\mathcal{S}^+_{r,p})$.

\begin{Theorem}[Boundedness/Non-Boundedness of the Scattering Map Forwards/Backwards Respectively]\label{Thm:O-SScat}
	The scattering maps $\mathcal{S}^+_{r,p}$ are bounded and injective, with respect to the $L^2$ norm of $\p_v(r\phi)$ on $\mathcal{I}^-$ and $\mathcal{H}^+$ and of $\p_u(r\phi)$ on $\mathcal{I}^+$. However, they are not surjective, and $\mathcal{S}^-_{r,p}$ are not bounded.
\end{Theorem}
\begin{proof}
	We have from Proposition \ref{prop:SwarzBound} that the maps $\mathcal{G}^+_{r,p}$ are injective, bounded, and not surjective. The maps $\mathcal{F}^+_{r,p}$ are injective, bounded and surjective (Theorem \ref{Thm:RadBound}). When combined, we obtain that the forwards scattering maps $\mathcal{S}^+_{r,p}$ are bounded, injective but not surjective.
	
	Given $\mathcal{G}^-_{r,p}$ are unbounded and $\mathcal{F}^-_{r,p}$, $\mathcal{F}^+_{r,p}$ are bounded, we also have that $\mathcal{S}^-_{r,p}$ are unbounded.

	This means that there exist solutions with bounded non-degenerate energy on $\mathcal{I}^+\cup\mathcal{H}^+$ which have infinite non-degenerate energy on $\Sigma_{t^*_c}$ (see \cite{Blue}). Then these solutions have infinite energy on $\mathcal{I}^-$, by Theorem \ref{Thm:RadBound}.
\end{proof}

This result is in stark contrast to scattering in the Schwarzschild case. If one considers the backwards scattering map on Schwarzschild:
\begin{equation}
\mathcal{S}^-_{Schw}:\mathcal{E}_{\mathcal{H}^+}^T\times\mathcal{E}_{\mathcal{I}^+}^T\to\mathcal{E}_{\mathcal{H}^-}^T\times\mathcal{E}_{\mathcal{I}^-}^T,
\end{equation} 
this is an isometry (as $T$ is a Killing field throughout all the space-time). Thus if we consider $\mathcal{S}^-_{Schw}$
restricted to $\mathcal{I}^-$, this is bounded with respect to the $T$-energy on $\mathcal{I}^-$ and $\mathcal{I}^+$, and either the $X$ or $T$-energy on $\mathcal{H}^+$. Here, the $T$-energy on $\mathcal{I}^\pm$ is the only canonical choice of energy, as it is the energy generated by the timelike Killing field. In effect, we get a cancelling of the blue-shift effect on $\mathcal{H}^-$ with the (equal and opposite) time reversed effect on $\mathcal{H}^+$.

However, when considering this same $T$-energy on $\mathcal{I}^-$ in Oppenheimer--Snyder, we have that $\mathcal{F}^+_{r,p}$ is an isomorphism between $\mathcal{E}^T_{\mathcal{I}^-}$ and $\mathcal{E}^X_{\Sigma_{t^*_c}}$, or $\mathcal{E}^X_{\Sigma_{\tau_{c^-}}}$. This means that when looking at $\Sigma_{t^*}$ or $\Sigma_{\tau}$ in Oppenheimer--Snyder, we are forced to choose $X$-energy on these timelike slices. This contrasts with Schwarzschild where one can consider the $T$-energy throughout the bulk of the space-time.

In particular, if we consider the backwards reflection map from $\mathcal{I}^+$ to $\mathcal{I}^-$ (where choice of energy is canonical) then this is bounded in Schwarzschild space-time, but unbounded in Oppenheimer--Snyder.
\bibliographystyle{plain}
\bibliography{SES-O}

\end{document}